\documentclass[11pt]{article}
\usepackage[margin=1in]{geometry}
\usepackage{amsmath,amssymb,amsthm,array,booktabs,microtype}
\usepackage{xcolor}
\usepackage{float}
\usepackage{tikz}
\usetikzlibrary{arrows.meta}
\usepackage[colorlinks=true,linkcolor=blue,citecolor=blue,urlcolor=blue]{hyperref}
\usepackage{fec}
\usepackage{mathpazo}
\usepackage{authblk}
\usepackage[backend=bibtex, style=alphabetic, maxbibnames=99, backref=true, backrefstyle=three]{biblatex}
\AtBeginBibliography{\emergencystretch=2em}

\newtheorem{theorem}{Theorem}[section]
\newtheorem{proposition}[theorem]{Proposition}
\newtheorem{lemma}[theorem]{Lemma}
\newtheorem{corollary}[theorem]{Corollary}
\newtheorem{question}[theorem]{Question}
\theoremstyle{definition}
\newtheorem{definition}[theorem]{Definition}
\newtheorem{remark}[theorem]{Remark}
\newcommand{\SQ}{\operatorname{SQ}}
\newcommand{\TV}{\operatorname{TV}}
\newcommand{\Prb}{\mathbb P}
\newcommand{\Samp}{\mathrm{Samp}}
\colorlet{ourbound}{red!70!black}
\newcommand{\ours}[1]{\textcolor{ourbound}{#1}}
\newcommand{\withcite}[2]{{\renewcommand{\arraystretch}{1}\begin{tabular}[t]{@{}c@{}}#1\\\footnotesize#2\end{tabular}}}

\newcount\Comments
\newcommand{\kibitz}[2]{\ifnum\Comments=1{\leavevmode\color{#1}{#2}}\fi}

\title{The Quartic Gap Between Quantum and Dequantized Algorithms for Linear Algebra Is Tight}
\author[1]{Shouvanik Chakrabarti\thanks{Corresponding author: \texttt{shouvanik.chakrabarti@jpmchase.com}}}
\author[1]{Dylan Herman}
\author[1]{Junhyung Lyle Kim}
\author[1]{Guneykan Ozgul}
\affil[1]{Global Technology Applied Research, JPMorganChase, New York, NY 10001, USA}
\date{}

\begin{document}
\maketitle
{\renewcommand{\thefootnote}{}%
 \footnotetext{\copyright~2026 JPMorgan Chase \& Co.}}

\begin{abstract}
 Quantum algorithms for linear algebra that access a matrix $A$ through quantum random access memory (QRAM) prepare quantum states proportional to the solutions of linear systems and related problems. Dequantized classical algorithms, when given sampling and query access to $A$, return sampling and query access to an approximate solution. Both kinds of algorithms use a number of queries that is polynomial in the normalized Frobenius norm $\rho=\|A\|_F/\|A\|$ and does not directly depend on the dimension. For linear systems at fixed condition number and accuracy, quantum algorithms use $\widetilde O(\rho)$ queries, whereas the best dequantized algorithms use $\widetilde O(\rho^4)$. We prove that this quartic gap is tight in the worst case, up to a logarithmic factor. For every constant $\kappa_0>1$, we construct dense linear systems with condition number at most $\kappa_0$ for which sampling from the solution within a sufficiently small constant total variation distance requires $\Omega(\rho^4/\log\rho)$ classical queries but only $\Theta(\rho)$ quantum queries. The same separation holds, for errors below smaller thresholds, for positive-definite systems, least squares, ridge regression, matrix powers and square roots, imaginary-time and real-time Hamiltonian evolution, and sampling from the top principal component. Our results therefore show that many quantum algorithms for linear algebra, achieve a near-quartic speedup in the worst case, over classical algorithms with sampling and query access. All of these classical lower bounds follow from the same framework, which is based on a lower bound for sampling from distributions that depend on a fixed set of Fourier coefficients of an unknown Boolean function.
\end{abstract}

\subsection*{AI Usage Disclosure}
\begingroup
    \small
    The primary theoretical result of this paper, the nearly tight lower bound for dequantized algorithms for dense linear systems, was discovered by GPT-6.1 Sol. We used GPT-6.1 Sol and Claude Opus 5.5 interactively to extend this result to other settings, to abstract the main theoretical framework behind the proofs, and to draft and audit large sections of the text. We set the research program: we posed the initial and follow-up questions and proposed isolating the abstract framework. The proofs that the models first produced were hard to follow and to verify, even for experts, and understanding them took us considerable effort. Much of our work with the models went into rewriting these proofs so that quantum algorithms researchers can follow and verify them and understand the ideas behind them. We have also independently verified all AI-generated results and take responsibility for their correctness.
\par
\endgroup

\section{Introduction}\label{sec:introduction}

\subsection{Motivation}\label{sec:motivation}

Quantum algorithms for linear algebra can solve linear systems, regression problems, and other matrix-function problems in time that grows only polylogarithmically with the dimension, provided that the input can be accessed coherently and that the output is a quantum state. Harrow, Hassidim, and Lloyd gave the first such algorithm, which prepares a state proportional to the solution of a sparse linear system \cite{HHL2009}. Linear systems, regression, Hamiltonian simulation, and general matrix functions were later treated in the same framework, based on block encodings and quantum singular value transformation (QSVT) \cite{ChakrabortyEtAl2019,GilyenEtAl2019}. For dense matrices stored in quantum random access memory (QRAM), these algorithms access the input through a block encoding of $A/\|A\|_F$. Their cost then depends on the normalized Frobenius norm $\rho=\|A\|_F/\|A\|$, on the condition number $\kappa=\|A\|\|A^{-1}\|$, and on the accuracy, but only polylogarithmically on the dimension. Kerenidis and Prakash used this input model for a quantum recommendation system \cite{KerenidisPrakash2017}, which was proposed as an exponential quantum speedup for a practical machine-learning task.

A subsequent line of work on \emph{dequantization} showed that classical algorithms can also avoid a polynomial dependence on the dimension when given a classical analogue of QRAM, called \emph{sampling and query access}. With sampling and query access to a matrix $A$, a classical algorithm can query the entries of $A$, the norms of its rows and columns, and its Frobenius norm; it can also sample a row or a column with probability proportional to its squared norm, and an entry of a given row or column with probability proportional to its squared magnitude. Dequantized algorithms receive sampling and query access to the input and return the same kind of access to an approximate solution, from which samples of the solution can be drawn. Tang used sampling and query access to give a classical recommendation algorithm \cite{Tang2019}, building on the sampling techniques of the low-rank approximation algorithm of Frieze, Kannan, and Vempala \cite{FriezeEtAl2004}. The cost of the classical recommendation algorithm is polynomial in $\|A\|_F/\sigma=\rho\,\|A\|/\sigma$, where $\sigma$ is a threshold on the singular values of $A$, in the inverse accuracy, and in further parameters of the recommendation problem. Further work dequantized principal component analysis, low-rank linear systems, and the QSVT framework as a whole \cite{Tang2021,ChiaEtAl2020Systems,JethwaniEtAl2020,ChiaEtAl2022}. Since the cost of these classical algorithms is polynomial in $\rho$, in the condition number, and in the inverse accuracy, and does not directly depend on the dimension, quantum algorithms with QRAM have no exponential speedup over them when these three parameters are at most polylogarithmic in the dimension. However, polynomial gaps remain between the numbers of queries of the classical and the quantum algorithms.

Later work has narrowed these gaps. At fixed condition number and accuracy, the regression algorithm of Gily\'en, Song, and Tang uses $\widetilde O(\rho^6)$ queries \cite{GilyenSongTang2022}. This dependence on $\rho$ was reduced to $\widetilde O(\rho^4)$ by the analysis of the randomized Kaczmarz method in the April 2023 arXiv revision of the work of Shao and Montanaro, by the classical singular value transformation of Bakshi and Tang, and by the analysis of a quantum-inspired linear-system algorithm by Chen et al.\ \cite{ShaoMontanaro2022,BakshiTang2024,ChenEtAl2025}. Since the corresponding quantum algorithms use $\widetilde O(\rho)$ queries \cite{ChakrabortyEtAl2019,GilyenEtAl2019}, the best known dequantized algorithms use a number of queries that is, up to logarithmic factors, the fourth power of the number used by the quantum algorithms. We study the following question.

\begin{question}\label{q:main}
Fix the condition number and the accuracy. In the worst case over matrices with normalized Frobenius norm $\rho$, how many queries does a classical algorithm with sampling and query access require to sample from the solution of a linear system, or from the output of a related task in quantum linear algebra? In particular, are $\widetilde\Omega(\rho^4)$ queries necessary, or can dequantized algorithms be improved to use $\widetilde O(\rho^2)$ queries?
\end{question}

The answer to Question~\ref{q:main} determines the largest worst-case speedup of quantum algorithms for linear algebra over dequantized algorithms in these input models. If $\widetilde O(\rho^2)$ classical queries sufficed, the quantum algorithms based on block encodings, which use $\widetilde O(\rho)$ queries, would offer at most a quadratic speedup. Babbush et al.\ estimate that, unless the overheads of quantum error correction improve substantially, algorithms with a quadratic speedup will not achieve a quantum advantage on early generations of error-corrected quantum computers, whereas algorithms with a quartic speedup are considerably more promising \cite{BabbushEtAl2021}. Hoefler, H\"aner, and Troyer reach a similar conclusion by comparing projected quantum and classical hardware: they argue that speedups beyond quadratic are essential for a practical quantum advantage \cite{HoeflerEtAl2023}. A tight lower bound also matters for classical algorithm design, because it shows that the fourth-power dependence of the known dequantized algorithms cannot be improved in the worst case.

Question~\ref{q:main} is not resolved by existing lower bounds, which are at most quadratic in $\rho$ for well-conditioned linear systems or concern problems other than linear algebra. Montanaro and Shao prove that sampling from the solution of a well-conditioned linear system requires classical communication linear in the dimension when one player holds the matrix and another player holds the right-hand side \cite{MontanaroShao2024Communication}. Since each input query of a classical algorithm with sampling and query access can be simulated by a constant number of messages of logarithmic length, their communication bound implies a query lower bound linear in the dimension, up to a logarithmic factor, when the right-hand side is unknown; for well-conditioned matrices of full rank, this bound is quadratic in $\rho$ up to a logarithmic factor. Mande and Shao also use communication complexity to prove lower bounds quadratic in the Frobenius norm for several sampling problems in linear algebra \cite{MandeShao2025}. For linear systems, their hard matrices have condition number one, so their bound is again quadratic in $\rho$. They note that, except for a scalar estimation task from supervised clustering, their lower bounds do not match the known upper bounds in the dependence on the Frobenius norm, and they pose the improvement of either the lower bounds or the dequantized algorithms as an open problem. Gr{\o}nlund and Larsen prove an exponential separation in the dimension for sparse linear systems whose condition number grows logarithmically with the dimension \cite{GronlundLarsen2024}, but their classical lower bound grows more slowly than $N$ for $N\times N$ matrices with $\rho^2=\Theta(N)$, so it is below quadratic in $\rho$. Finally, oracle problems such as Fourier sampling and Forrelation have large separations between quantum and classical query complexity \cite{AaronsonChen2017,AaronsonAmbainis2015}, but these problems are not linear-algebra tasks parameterized by $\rho$ and the condition number.

One cannot prove a query lower bound larger than $\widetilde O(\rho^2)$ with communication arguments in which a constant number of players hold fixed subsets of the matrix entries and the right-hand side is known to all players. Indeed, in Appendix~\ref{app:communication} we show that, when $k$ players hold the entries of a well-conditioned $N\times N$ matrix according to a partition known to all players, a coordinator, a party that holds no entries of the matrix and exchanges messages with the players, can sample from the solution with $O(kN(B+\log N))$ bits of communication, where the entries of the matrix and of the right-hand side are integer multiples of $2^{-B}$. For constant $k$, full-rank matrices, and $B=O(\log N)$, this is $\widetilde O(\rho^2)$ bits, because $N\le\kappa^2\rho^2$.

\subsection{Our results}\label{sec:results}

We answer Question~\ref{q:main}: $\widetilde\Omega(\rho^4)$ queries are necessary. For linear systems and several related tasks, we construct explicit families of dense, well-conditioned matrices for which every classical algorithm with sampling and query access requires $\Omega(\rho^4/\log\rho)$ queries to sample from the output within any sufficiently small constant error, while $\widetilde\Theta(\rho)$ quantum queries suffice and are necessary. An $N\times N$ matrix with condition number $\kappa$ satisfies $N\le\kappa^2\rho^2$, so our matrices have dimension $O(\rho^2)$, and reading all of their entries takes $O(\rho^4)$ queries. At fixed condition number and accuracy, the dequantized algorithms cited in Table~\ref{tab:results} also use $\widetilde O(\rho^4)$ queries, without a bound on the dimension, for linear systems on every invertible matrix and for the other tasks on our families. Hence, at constant condition number and for sufficiently small constant errors, the worst-case query complexity of sampling from the solution of a linear system is $\widetilde\Theta(\rho^4)$ for classical algorithms with sampling and query access and $\widetilde\Theta(\rho)$ for quantum algorithms with QRAM (Section~\ref{sec:larger-separations}), and the same quartic gap holds for the other tasks on our families. In our families, $\Theta(\rho^4)$ of the entries, a constant fraction, are unknown, that is, not the same on every input of the family. At constant condition number, every family with a classical lower bound of order $\rho^4/\log\rho$ must have $\Omega(\rho^4/\log\rho)$ unknown entries, a fraction $\Omega(1/\log\rho)$ of all entries, because reading only the unknown entries suffices to solve the task (see the paragraph \emph{Constraints on hard families} in Section~\ref{sec:overview}).

The tasks covered are linear systems, positive-definite linear systems, least squares, ridge regression, matrix powers, square roots and inverse square roots, imaginary-time and real-time Hamiltonian evolution, and sampling from the top principal component. Table~\ref{tab:results} summarizes the results.

\paragraph{Input and output models.}
We compare quantum algorithms with QRAM and classical algorithms with the sampling and query access typically considered in the dequantization literature; precise definitions are given in Section~\ref{sec:input-models}. A classical algorithm has sampling and query access to the input matrix $A$, as described in Section~\ref{sec:motivation}, and a quantum algorithm has coherent versions of these operations: reversible entry queries and unitaries that prepare normalized rows, normalized columns, and the vector of row norms. For both kinds of algorithms, the right-hand side or the initial vector of each task is fixed and known in advance. The two kinds of algorithms differ in their outputs: a dequantized algorithm returns sampling and query access to an approximate solution, whereas a quantum algorithm returns a quantum state close to the normalized solution. From either output, however, one can draw an approximate sample from the length-squared distribution $p_x(j)=|x_j|^2/\|x\|^2$ of the solution $x$. We therefore compare the two kinds of algorithms on the following task: output one sample whose distribution is within total variation distance $\varepsilon$ of $p_x$ on every input of the family. In particular, a lower bound for this task bounds the cost of computing the output of a dequantized algorithm together with one sample from it (Lemma~\ref{lem:solution-access}). We count queries in the worst case. The data structures that implement the input operations are part of the input, so the cost of building them is not counted for either kind of algorithm.

\paragraph{Ranges of parameters.}
The bound on the condition number and the range of errors in each of our results are determined by a parameter of the matrix family. Each of our matrix families is indexed by an unknown matrix $W\in\{\pm1\}^{n\times n}$ of signs and has a coupling parameter, which sets the size of the unknown entries. The linear systems of Theorem~\ref{thm:intro-linear} use a parameter $\lambda\in(0,1]$. Positive-definite systems and matrix functions such as square roots require symmetric positive-definite matrices, so all other tasks use matrices built from $tT_W$ with a parameter $t\in(0,1/64]$, where $T_W$ is a symmetric matrix whose nonzero blocks are $W/\sqrt n$, the transpose of $W/\sqrt n$, and a normalized Walsh--Hadamard matrix (Section~\ref{sec:pd-family}). For example, the positive-definite linear systems use the matrices $P_W=I-tT_W$, whose unknown entries equal $\pm t/\sqrt n$. A smaller parameter brings the condition number closer to one (to two for principal components), but it also weakens the dependence of the output distribution on the unknown signs. Our classical lower bounds therefore hold for all errors below a constant multiple of a power of the parameter: $\lambda^2$ for the linear systems of Theorem~\ref{thm:intro-linear}, $t^2$ for Hamiltonian evolution, and $t^4$ for the remaining tasks (Section~\ref{sec:error-ranges}). The parameter can be any fixed rational number in its range. It is required to be rational so that every entry has an exact finite encoding, and the upper bound $1/64$ on $t$ is used in the error estimates of Section~\ref{sec:positive-definite}. For linear systems, explicit constants for one value of $\lambda$ are given after Theorem~\ref{thm:intro-linear}, and Table~\ref{tab:results} states all results in asymptotic notation.

\paragraph{Main results.}
Our first result concerns linear systems.

\begin{theorem}[Linear systems; informal version of Theorem~\ref{thm:linear-systems}]\label{thm:intro-linear}
For every rational constant $\lambda\in(0,1]$ and every power of four $n=4^\ell$ with $\ell$ sufficiently large, there is an explicit family of matrices $A$ with condition number at most $1+152\lambda$ and $\rho=\Theta(\sqrt n)$, together with a right-hand side $b$, such that the following holds for every constant error $\varepsilon\in(0,\lambda^2/12]$.
\begin{enumerate}
\item Every classical algorithm that samples from $p_{A^{-1}b}$ within total variation distance $\varepsilon$ on every matrix in the family makes $\Omega(\rho^4/\log\rho)$ queries on some matrix in the family. Reading all entries, with $O(\rho^4)$ queries, suffices.
\item A quantum algorithm solves the same task with $O(\rho)$ queries, and every quantum algorithm requires $\Omega(\rho)$ queries.
\end{enumerate}
\end{theorem}

We restrict $n$ to powers of four so that the normalized Walsh--Hadamard matrix of order $n$ is defined and its entries $\pm1/\sqrt n$, like those of $W/\sqrt n$, are dyadic rationals with an exact finite encoding (Section~\ref{sec:hidden-signs}).

The bounds $1+152\lambda$ and $\lambda^2/12$ in Theorem~\ref{thm:intro-linear} are simplified, and Theorem~\ref{thm:linear-systems} states sharper ones. For example, for $\lambda=1/4$, the condition number is at most $10.5$, and the bounds of Theorem~\ref{thm:linear-systems} hold for every error below $0.0116$ (Table~\ref{tab:linear-examples}).

A separation of the same order holds for the other tasks in Table~\ref{tab:results}, for errors below the smaller thresholds given in the table.

\begin{theorem}[Other tasks; informal]\label{thm:intro-other}
For each task in Table~\ref{tab:results} other than linear systems and Fourier sampling on a set of characters, and for every fixed rational value of the parameter $t$ in the range given in the corresponding section, there is an explicit family of matrices with constant condition number and $\rho=\Theta(\sqrt n)$, together with a right-hand side or initial vector, such that sampling from the output within any fixed positive error in the range stated in Table~\ref{tab:results} requires $\Omega(\rho^4/\log\rho)$ classical queries, while $\widetilde\Theta(\rho)$ quantum queries suffice and are necessary.
\end{theorem}

\begin{table}[htbp]
\centering
\small
\renewcommand{\arraystretch}{1.25}
\setlength{\tabcolsep}{5pt}
\begin{tabular}{@{}>{\raggedright\arraybackslash}p{4.2cm}cccccc@{}}
\toprule
& & & \multicolumn{2}{c}{Classical queries} & \multicolumn{2}{c}{Quantum queries} \\
\cmidrule(lr){4-5}\cmidrule(l){6-7}
Task & \begin{tabular}[b]{@{}c@{}}Condition\\number\end{tabular} & Error & Lower & Upper & Lower & Upper \\
\midrule
Linear systems (Section~\ref{sec:linear-systems}) & $1+O(\lambda)$ & $O(\lambda^2)$ & \ours{$\Omega(\rho^4/\log\rho)$} & \withcite{$\widetilde O(\rho^4)$}{\cite{ShaoMontanaro2022,BakshiTang2024}\\\footnotesize\cite{ChenEtAl2025}} & $\Omega(\rho)$ & \withcite{$O(\rho)$}{\cite{CostaEtAl2022}} \\
Positive-definite systems; least squares (Section~\ref{sec:pd-systems}) & $1+O(t)$ & $O(t^4)$ & \ours{$\Omega(\rho^4/\log\rho)$} & \withcite{$\widetilde O(\rho^4)$}{\cite{BakshiTang2024}} & $\Omega(\rho)$ & \withcite{$\widetilde O(\rho)$}{\cite{ChakrabortyEtAl2019,GilyenEtAl2019}} \\
Ridge regression; $A^p$; $A^{1/2}$; $A^{-1/2}$; $e^{-\tau A}$; $e^{-i\tau A}$ (Section~\ref{sec:functions}) & $1+O(t)$ & $O(c_gt^4)$ & \ours{$\Omega(\rho^4/\log\rho)$} & \withcite{$\widetilde O(\rho^4)$}{\cite{BakshiTang2024}} & $\Omega(\rho)$ & \withcite{$\widetilde O(\rho)$}{\cite{ChakrabortyEtAl2019,GilyenEtAl2019}} \\
Hamiltonian evolution $e^{-iA}$ (Section~\ref{sec:hamiltonian}) & $1+O(t)$ & $O(t^2)$ & \ours{$\Omega(\rho^4/\log\rho)$} & \withcite{$\widetilde O(\rho^4)$}{\cite{BakshiTang2024}} & $\Omega(\rho)$ & \withcite{$\widetilde O(\rho)$}{\cite{ChakrabortyEtAl2019,GilyenEtAl2019}} \\
Top principal component; a row of the rank-one truncation (Section~\ref{sec:pca}) & $O(1)$ & $O(t^4)$ & \ours{$\Omega(\rho^4/\log\rho)$} & \withcite{$\widetilde O(\rho^4)$}{\cite{BakshiTang2024}} & $\Omega(\rho)$ & \withcite{$\widetilde O(\rho)$}{\cite{ChakrabortyEtAl2019,GilyenEtAl2019}} \\
Fourier sampling on $K$ of $M$ characters (Section~\ref{sec:fourier}) & -- & $O(1)$ & \ours{$\Omega(M/\log K)$} & $O(M)$ & -- & \withcite{$O(\sqrt{M/K})$}{\cite{BrassardEtAl2002}} \\
\bottomrule
\end{tabular}
\caption{Summary of the results. The classical lower bounds, shown in red, are new (in the last row, only for $K<M$). In each matrix family, $\rho^2=\Theta(n)$, and $\lambda\in(0,1]$ or $t\in(0,1/64]$ is a fixed rational parameter. Each matrix function $g$ is fixed, with $g(1)g'(1)g''(1)\ne0$ (for example, $p\ge2$ and $\tau\ne0$), the parameter $t$ must also satisfy the conditions of Theorem~\ref{thm:functions}, and $c_g=|g''(1)/(2g(1))|^2$. An entry $O(\cdot)$ in the error column means that the lower bounds hold for every fixed positive error up to an absolute constant times the given quantity. The classical upper bounds for the matrix families are those of the cited dequantized algorithms, which hold without a bound on the dimension; for our families, reading all entries also uses $O(\rho^4)$ queries. The quantum lower bounds follow from the lower bound for unstructured search \cite{BennettEtAl1997} by the reduction in Lemma~\ref{lem:search}. The quantum upper bounds for the matrix families come from the cited algorithms, applied to the standard block encoding of $A/\|A\|_F$, whose singular values are those of $A/\|A\|$ divided by $\rho$. For Fourier sampling on a set of characters, the classical upper bound comes from reading all $M$ values, and the quantum upper bound is proved in Theorem~\ref{thm:fourier-quantum} by amplitude amplification. For $K=M$, the classical lower bounds $\Omega(M/\log M)$ \cite{AaronsonAmbainis2015} and, at sufficiently small constant errors, $\Omega(M)$ \cite{AaronsonChen2017,Montanaro2019} were known for Fourier sampling (Section~\ref{sec:fourier}). Fourier sampling on a set of characters has no condition number, and we do not prove a quantum lower bound for it. The notation $\widetilde O$ hides factors polylogarithmic in $\rho$.}\label{tab:results}
\end{table}

\paragraph{The framework.}
All of the classical lower bounds follow from the same framework, developed in Section~\ref{sec:framework}. The framework is based on a lower bound for sampling from distributions that depend on a fixed set of $K$ Fourier coefficients of an unknown Boolean function on $M$ points. The simplest such task is the following sampling problem: output either one of the $K$ characters, with probability proportional to its squared Fourier coefficient, or an additional symbol $\bot$. For our linear-algebra tasks, the output vector is close, on a set of coordinates, to an affine function of $K$ Fourier coefficients, and the transfer theorem (Theorem~\ref{thm:transfer}) shows that a lower bound of the same order as in Theorem~\ref{thm:intro-fourier} holds for these tasks, for errors smaller than a threshold that depends on the task (see the paragraph \emph{The general framework} in Section~\ref{sec:overview}). The following theorem states the query complexity of the sampling problem.

\begin{theorem}[Fourier sampling on a set of characters; informal version of Theorems~\ref{thm:fourier-classical} and~\ref{thm:fourier-quantum}]\label{thm:intro-fourier}
Let $f:\{0,1\}^m\to\{-1,1\}$ be an unknown Boolean function on $M=2^m$ points, accessed by queries, with Fourier coefficients $\widehat f(s)$, and let $S$ be a fixed set of $K\ge128$ Fourier characters. Put $D_f=1+(M/K)\sum_{s\in S}\widehat f(s)^2$, and consider the task of outputting an additional symbol $\bot$ with probability $1/D_f$ and otherwise an element $s\in S$ with probability proportional to $\widehat f(s)^2$. Every classical algorithm that solves this task within a total variation distance $\varepsilon<31/1536$, for example $\varepsilon=1/100$, makes more than $\lfloor M/(32\log(512K))\rfloor$ queries on some input, while a quantum algorithm solves it within any constant total variation distance with $O(\sqrt{M/K})$ queries.
\end{theorem}

Because of the symbol $\bot$, the task is well defined for every $f$, including functions whose Fourier coefficients on $S$ all vanish. For uniformly random $f$, each $\widehat f(s)^2$ has mean $1/M$, so $D_f$ has mean $2$, and the variance of $D_f$ is at most $2/K$ (Lemma~\ref{lem:normalization}). Hence, for most $f$, the symbol $\bot$ is output with probability close to $1/2$. The assumption $K\ge128$ ensures that the variance of $D_f$ is at most $1/64$, which the proof of Theorem~\ref{thm:fourier-classical} uses to obtain the constant $31/1536$.

The size of the separation in Theorem~\ref{thm:intro-fourier} depends on the number $K$ of characters. If $K=M^\theta$ with $0<\theta<1$, the classical lower bound is the $(2/(1-\theta))$-th power of the quantum upper bound, up to a logarithmic factor. Our lower bounds instantiate the framework with $M=n^2$ and $K=n$: the matrices contain a dense $n\times n$ block of unknown signs, and the output of each task depends on $n$ of the Fourier coefficients of these signs. Here $\theta=1/2$, and since $\rho^2=\Theta(n)$, the two bounds become $\Omega(\rho^4/\log\rho)$ and $O(\rho)$.

If $K=M$, then $D_f=2$ by Parseval's identity, so the task is to output $\bot$ with probability $1/2$ and otherwise a character $s$ with probability $\widehat f(s)^2$, as in Fourier sampling. Classical lower bounds for Fourier sampling were known and also hold for this task \cite{AaronsonAmbainis2015,AaronsonChen2017,Montanaro2019} (Section~\ref{sec:fourier}), and Theorem~\ref{thm:intro-fourier} extends them to the sets of $K<M$ characters that the matrix families of Table~\ref{tab:results} require. If the dense block of unknown signs is replaced by a diagonal block of $N$ unknown signs, the solution of the resulting linear system contains all $N$ Fourier coefficients of these signs, so $K=M=N$. One quantum query to the Boolean function given by these signs then suffices, while $\Omega(N)$ classical queries are necessary at sufficiently small constant errors, by the lower bounds for Fourier sampling of Aaronson and Chen \cite{AaronsonChen2017} and of Montanaro \cite{Montanaro2019} (Appendix~\ref{app:diagonal}). For that family, the classical query complexity is not bounded by any power of the quantum query complexity, but it is only $\Theta(\rho^2)$ (Section~\ref{sec:larger-separations}).

\paragraph{Scope.}
Our results concern query complexity in the worst case over explicit families of inputs, at fixed condition number and fixed positive error; the error is positive because the quantum algorithms for the matrix families of Table~\ref{tab:results} sample only approximately. The results also concern only sampling from an output vector. We do not prove lower bounds for estimating scalar quantities, such as eigenvalues, traces, or quadratic forms. Such quantities can be easy to estimate on our families: on the family for principal components, for example, the top eigenvalue can be estimated to error $1/100$ without any query (Section~\ref{sec:not-covered}). Finally, the factor $1/\log\rho$ in the classical lower bounds comes from a union bound over the output coordinates (Section~\ref{sec:open}).

\subsection{Proof overview}\label{sec:overview}

We outline the proofs for linear systems first and then their extension to the other tasks. For linear systems, we construct a family of well-conditioned matrices indexed by an unknown matrix $W$ of signs. For this family, every classical algorithm that samples from the solution within a sufficiently small constant total variation distance requires $\Omega(\rho^4/\log\rho)$ queries, a quantum algorithm uses $O(\rho)$ queries, and every quantum algorithm requires $\Omega(\rho)$ queries. The outline has five steps. First, we derive two constraints on the construction of a family with a fourth-power classical lower bound. Second, we describe a family of linear systems that satisfies these constraints and whose solution contains Fourier coefficients of the unknown signs. Third, we prove the classical lower bound by comparing the output of a classical algorithm with exact samples through a bounded function of the signs and of the output, which we call a test function. Fourth, we explain why the quantum upper bound $O(\rho)$ holds when the condition number is constant, and we derive the quantum lower bound from unstructured search. Fifth, we describe the framework of Section~\ref{sec:framework}, which states these arguments for general sets of Fourier coefficients, and its application to the other tasks.

\paragraph{Constraints on hard families.}
The first constraint concerns the number of unknown entries. A lower bound is proved for a family of inputs; entries that take the same value on every input of the family are known to the algorithm, and only the remaining entries, which we call unknown, carry information about the input. For an invertible $N\times N$ matrix $A$ with condition number $\kappa$, each of the $N$ singular values is at least $\|A\|/\kappa$, so $\rho^2=\|A\|_F^2/\|A\|^2\ge N/\kappa^2$. Reading all $N^2\le\kappa^4\rho^4$ entries therefore solves any task with $O(\rho^4)$ queries when the condition number is constant. Reading only the unknown entries also suffices, so a classical lower bound of order $\rho^4/\log\rho$ requires $\Omega(\rho^4/\log\rho)$ unknown entries, and hence dimension $\Omega(\rho^2/\sqrt{\log\rho})$. At constant condition number, the matrices of a hard family must therefore be dense: their number of unknown entries must be of the largest possible order $\rho^4$, up to a logarithmic factor.

The second constraint concerns the sampling operations. A norm query or a sample from a length-squared distribution depends on the magnitudes of many entries at once, so a single such query could reveal information about many unknown entries. We therefore fix the magnitude of every entry in advance and let the input determine only the signs of the unknown entries. Every norm and every sampling distribution is then the same for all inputs of the family, and each classical query reveals at most one unknown sign (Lemma~\ref{lem:access}).

\paragraph{The family of linear systems.}
Our linear systems use the matrices
\[
A_W=\begin{pmatrix}I&0&0\\-\lambda R&I&0\\0&-\lambda H&I\end{pmatrix},\qquad R=\frac W{\sqrt n},
\]
where $W\in\{\pm1\}^{n\times n}$ is an unknown matrix of signs, $H$ is the normalized Walsh--Hadamard matrix, with entries $\pm1/\sqrt n$, and $\lambda\in(0,1]$ is the parameter (Figure~\ref{fig:construction}). The family of Section~\ref{sec:nonsymmetric-family} also rescales $A_W$ and adds an identity block; these changes only normalize the operator norm and do not affect the argument. We consider only matrices $W$ with $\|R\|\le8$; this condition excludes an exponentially small fraction of all $W\in\{\pm1\}^{n\times n}$. Under this condition, $\|A_W\|$ and $\|A_W^{-1}\|$ are bounded by constants (Lemma~\ref{lem:nonsymmetric-parameters}), so the condition number of $A_W$ is bounded by a constant. The normalized Frobenius norm, in contrast, grows with $n$: $\rho^2=\Theta(n)$. The family satisfies both constraints: its unknown entries, the $n^2$ entries $-\lambda W_{ab}/\sqrt n$ of the block $-\lambda R$, form a constant fraction of all entries, the dimension $3n$ is of order $\rho^2$, and every unknown entry has the known magnitude $\lambda/\sqrt n$. For the right-hand side $(u,u,u)$, where $u$ is the uniform unit vector, forward substitution shows that the solution $x=A_W^{-1}(u,u,u)$ has the blocks
\[
x_1=u,\qquad x_2=u+\lambda v,\qquad x_3=u+\lambda e_0+\lambda^2Hv,
\]
where $v=Ru$ is the vector of normalized row sums of $W$, and $e_0$ is the standard basis vector with a one in coordinate $0$.

The construction is designed so that the third block of the solution depends on all $n^2$ unknown signs and typically carries a constant fraction of the squared norm of the solution. The second block does not have this property: each of its coordinates, $n^{-1/2}+\lambda v_a$, depends only on the $n$ signs in row $a$, so a classical algorithm can compute the squared magnitude of any coordinate of the second block exactly with $n=\Theta(\rho^2)$ queries. For random signs, with high probability, each of these squared magnitudes is $O((\log n)/n)$ and their sum is at least $1/2$, so rejection sampling from the second block uses only $\widetilde O(\rho^2)$ queries, and one cannot prove a fourth-power lower bound for random signs from the second block alone. In the third block, by contrast, every coordinate depends on all $n^2$ signs. Its coordinates are $\lambda^2(Hv)_j$ plus known terms, and
\[
Z_j=\sqrt n\,(Hv)_j=\frac1n\sum_{a,b}h_{ja}W_{ab},\qquad h_{ja}=\sqrt n\,H_{ja}=\pm1,
\]
is a sum of all $n^2$ signs with weights of equal magnitude. In the language of Fourier analysis, $Z_j$ is a normalized Fourier coefficient of the Boolean function $(a,b)\mapsto W_{ab}$ on $M=n^2$ points: it is the sum of $W_{ab}$ times the character $(a,b)\mapsto h_{ja}$, divided by $\sqrt M$. These $K=n$ characters do not depend on the column index $b$. Since $H^2=I$, the second block is also determined by these coefficients, through $v=H(Hv)$, so the solution depends on the signs only through $K=n$ of the $M=n^2$ Fourier coefficients.

\begin{figure}[htbp]
\centering
\begin{tikzpicture}[x=1cm,y=1cm,font=\small]
\draw (0,0) rectangle (3,3);
\draw (1,0) -- (1,3);
\draw (2,0) -- (2,3);
\draw (0,1) -- (3,1);
\draw (0,2) -- (3,2);
\node at (0.5,2.5) {$I$};
\node at (1.5,1.5) {$I$};
\node at (2.5,0.5) {$I$};
\node at (0.5,1.5) {$-\lambda R$};
\node at (1.5,0.5) {$-\lambda H$};
\node at (1.5,-0.45) {$A_W$};
\draw (3.7,0) rectangle (6.5,3);
\draw (3.7,1) -- (6.5,1);
\draw (3.7,2) -- (6.5,2);
\node at (5.1,2.5) {$u$};
\node at (5.1,1.5) {$u+\lambda v$};
\node at (5.1,0.5) {$u+\lambda e_0+\lambda^2Hv$};
\node at (5.1,-0.45) {$x=A_W^{-1}(u,u,u)$};
\node[anchor=west,text width=6cm] at (6.9,2.5) {does not depend on $W$};
\node[anchor=west,text width=6cm] at (6.9,1.5) {$v=Ru$: each $v_a$ depends on the $n$ signs in row $a$ of $W$};
\node[anchor=west,text width=6cm] at (6.9,0.5) {$Hv$: each coordinate $Z_j/\sqrt n$ depends on all $n^2$ signs};
\end{tikzpicture}
\caption{The matrix $A_W$ and the solution $x$ of $A_Wx=(u,u,u)$. The second block of $x$ contains the normalized row sums $v=Ru$ of the unknown signs, and the third block contains their Walsh--Hadamard transform $Hv$.}\label{fig:construction}
\end{figure}

\paragraph{Test functions and the classical lower bound.}
To prove the classical lower bound, we run a classical algorithm on independent and uniformly random signs. An algorithm with small error on every input of the family also has small error on random inputs, because random signs violate the condition $\|R\|\le8$ only with exponentially small probability, so it suffices to show that no algorithm with few queries has small error on random inputs. For random signs, each $Z_j$ is approximately a standard Gaussian variable. Coordinate $j\ne0$ of the third block has squared magnitude $(1+\lambda^2Z_j)^2/n$, so an exact sampler returns third-block coordinates with large positive $Z_j$ more often than a uniform choice would.

To show that a classical algorithm with few queries cannot reproduce this dependence on $Z_j$, we use a test function: a bounded function $F(W,j)$ of the signs $W$ and of an outcome $j$, here a coordinate of the solution. We compare the expectation of $F(W,J)$, where $J$ is the random output of a sampler, under exact sampling and at the output of the algorithm. If $F$ takes values in an interval of length two, then the expectation of $F(W,J)$ under a sampler with total variation error $\varepsilon$ differs from its expectation under exact sampling by at most $2\varepsilon$. Hence, if the expectation under exact sampling and the expectation at the output of every algorithm with $o(n^2/\log n)$ queries differ by at least $c$, then every such algorithm has total variation error at least $c/2$.

We use the sine test, which equals $\sin Z_j$ at each coordinate $j$ of the third block and zero at the coordinates of the other blocks. The function that equals $Z_j$ at these coordinates and zero elsewhere has a positive expectation under exact sampling, because coordinates with large positive $Z_j$ are sampled more often, but its values are not bounded independently of $n$. In the sine test, $Z_j$ is replaced by $\sin Z_j$, which is bounded by one and, like $Z_j$, changes sign when all signs are flipped.

Under exact sampling, the expectation of the sine test is approximately $2\lambda^2e^{-1/2}/d$, where $d$ is the expected squared norm of the solution $x$. Indeed, coordinate $j\ne0$ of the third block is sampled with probability $(1+\lambda^2Z_j)^2/(n\|x\|^2)$, and $\|x\|^2$ is close to $d$ for most $W$ (Lemma~\ref{lem:nonsymmetric-solution}), so this coordinate contributes approximately $\E[(1+\lambda^2Z_j)^2\sin Z_j]/(nd)$ to the expectation. In the expansion $(1+\lambda^2Z_j)^2\sin Z_j=(1+2\lambda^2Z_j+\lambda^4Z_j^2)\sin Z_j$, the terms $\sin Z_j$ and $\lambda^4Z_j^2\sin Z_j$ have expectation zero, because they are odd functions of $Z_j$, whose distribution is symmetric about zero. The remaining term has expectation $2\lambda^2\E[Z_j\sin Z_j]$. Since $Z_j$ is approximately a standard Gaussian variable $G$, the expectation $\E[Z_j\sin Z_j]$ is close to $\E[G\sin G]=e^{-1/2}$ (Lemma~\ref{lem:moments}). The $n-1$ coordinates $j\ne0$ therefore contribute approximately $2\lambda^2e^{-1/2}/d$ in total. Coordinate $0$ of the third block is sampled with constant probability, but it contributes only $O(1/\sqrt n)$, because its squared magnitude is $\lambda^2$ plus terms of order $1/\sqrt n$, and $\E\sin Z_0=0$. We call the expectation of the sine test under exact sampling, computed with $\|x\|^2$ replaced by $d$, the bias of the sine test.

For every classical algorithm with $q=o(n^2/\log n)$ queries, in contrast, the expectation of the sine test at its output is $o(1)$. At any time, let $\mu_j$ be the sum of the terms $h_{ja}W_{ab}/n$ of $Z_j$ whose signs $W_{ab}$ have been queried; it is determined by the results of the previous queries. Conditioned on the previous queries, the unqueried signs are still independent and uniformly random, so the sum $Z_j-\mu_j$ of the other terms has a distribution that is symmetric about zero. Since $\sin Z_j=\sin\mu_j\cos(Z_j-\mu_j)+\cos\mu_j\sin(Z_j-\mu_j)$ and the second term has conditional expectation zero, the conditional expectation of $\sin Z_j$ is $\sin\mu_j$ times the conditional expectation of $\cos(Z_j-\mu_j)$, so its absolute value is at most $|\mu_j|$. A query of a new sign changes each $\mu_j$ by $\pm1/n$ with conditional mean zero, even when the queried position depends on the results of the previous queries, so for each $j$ the values of $\mu_j$ after successive queries form a martingale with increments at most $1/n$ in absolute value. By the Azuma--Hoeffding inequality and a union bound over the $n$ coordinates, with high probability every $|\mu_j|$ is $O(\sqrt{q\log n}/n)$, which is $o(1)$. This bound holds for all coordinates at once, so it also holds at the output coordinate, which may depend on the previous queries.

The expectations of the sine test under exact sampling and at the output of every algorithm with $o(n^2/\log n)$ queries therefore differ by approximately the bias. Since the sine test takes values in $[-1,1]$, every such algorithm has total variation error at least about half of the bias, so sampling with any smaller error requires $\Omega(n^2/\log n)=\Omega(\rho^4/\log\rho)$ queries.

\paragraph{The quantum bounds.}
The quantum upper bound $O(\rho)$ holds because the condition number is constant. Quantum linear-system algorithms for dense matrices use the standard block encoding of $A/\|A\|_F$ \cite{ChakrabortyEtAl2019,GilyenEtAl2019}, which costs $O(1)$ queries per use and whose smallest singular value is $1/(\kappa\rho)$. The algorithm of Costa et al.\ uses this block encoding $O(\kappa\rho)$ times at constant error \cite{CostaEtAl2022}, so it makes $O(\rho)$ queries for constant $\kappa$.

The quantum upper bound of Theorem~\ref{thm:linear-systems} is proved with a simpler algorithm for our family, which also makes $O(\rho)$ queries. In this algorithm, one query applies the $n^2$ signs as phases to a uniform superposition over positions $(a,b)$. A Walsh--Hadamard transform on the column register, followed by a projection of that register onto the basis state $|0\cdots0\rangle$, leaves the vector $n^{-1/2}\sum_av_a|a\rangle$, whose squared norm $\|v\|^2/n$ is $O(1/n)$. Applying $H$ to this vector produces $n^{-1/2}Hv$. The solution $x$ is the sum of $(0,\lambda v,\lambda^2Hv)$ and of the vector $(u,u,u+\lambda e_0)$, which does not depend on $W$. A linear combination of the preparations of $v$, of $Hv$, and of $(u,u,u+\lambda e_0)$, with a register that selects the block, therefore prepares a state whose projection onto a fixed subspace is proportional to $x$ and has squared norm of order $1/n$. Amplitude amplification raises the success probability of this projection to a constant with $O(\sqrt n)=O(\rho)$ queries, and measuring the normalized solution then samples from the output distribution.

The quantum lower bound comes from unstructured search. If one row of an otherwise fixed matrix $R$ is replaced by $u^{\mathsf T}$, the solution has a coordinate of constant probability at the position of that row. A sampler therefore finds the modified row, and finding it is an instance of unstructured search, which requires $\Omega(\sqrt n)$ queries \cite{BennettEtAl1997}.

\paragraph{The general framework.}
We extend the lower-bound argument for the third block of our linear systems to every task whose output vector is close, on a set of coordinates, to an affine function of Fourier coefficients of a Boolean function. Let $f$ be a Boolean function on $\{0,1\}^m$, with $M=2^m$ points. For $s,\sigma\in\{0,1\}^m$, let $\chi_s(\sigma)=(-1)^{s\cdot\sigma}$, and let $Z_s=M^{-1/2}\sum_\sigma\chi_s(\sigma)f(\sigma)=\sqrt M\,\widehat f(s)$ be the normalized Fourier coefficient of $f$ at the character $\chi_s$. For our linear systems, $f(a,b)=W_{ab}$, and $Z_j$ is the coefficient at the character $\chi_s$ with $s=(j,0)$. The arguments are developed at this level of generality in Section~\ref{sec:framework}, for a fixed set $S$ of $K$ characters, and the sampling problem of Theorem~\ref{thm:intro-fourier} is the simplest example. The transfer theorem (Theorem~\ref{thm:transfer}) and Lemma~\ref{lem:two-conditions} then show that a task requires $\Omega(M/\log K)$ classical queries, for every total variation error below about half of the bias of a test function, whenever its output vector satisfies the following two conditions up to approximation errors that are small compared with the bias: on the coordinates indexed by $S$, it is close to $(\gamma_s+\beta Z_s)/\sqrt K$, an affine function of the Fourier coefficients, with $\gamma$ and $\beta$ independent of the input, and its squared norm is close to a fixed value on average. For our linear systems, $S$ indexes the third block, $K=n$, $\gamma_s=1$ for all but one element of $S$, and $\beta=\lambda^2$, so the bias of the sine test is of order $\beta$. When $\gamma=0$, however, the bias of the sine test is zero, because the weights $|\beta|^2Z_s^2$ do not change when all signs are flipped while the sine test changes sign; the lower bounds then use the cosine test, which equals $\cos Z_s$ minus its mean at each $s\in S$ and has a bias of order $|\beta|^2$ (Section~\ref{sec:bias}).

\paragraph{Other tasks.}
For positive-definite systems, and for matrix functions such as square roots, the matrix must be symmetric and positive definite. For all tasks other than linear systems, we therefore use matrices built from the symmetric matrix $T_W$, such as $P_W$:
\[
T_W=\begin{pmatrix}0&R^{\mathsf T}&0\\R&0&H\\0&H&0\end{pmatrix},\qquad P_W=I-tT_W,
\]
where $t\in(0,1/64]$ is the parameter; since $\|T_W\|\le9$ under the condition $\|R\|\le8$, the matrix $P_W$ is positive definite. With the right-hand side $b=(u,0,0)$, the third block of $P_W^{-1}b$ equals $t^2Hv$ up to a relative error of order $t^2$. This third block has no component that is independent of the input, so $\gamma=0$, and we use the cosine test, whose bias is of order $t^4$. More generally, for an analytic function $g$, the expansion of $g(P_W)b$ in powers of $tT_W$ shows that its third block equals $\frac12g''(1)t^2Hv$ up to terms of order $t^4$, so the classical lower bound holds for every function with $g(1)g''(1)\ne0$, for $t$ small enough. Examples include the functions for ridge regression, matrix powers, square roots, inverse square roots, and imaginary-time and real-time evolution. For Hamiltonian evolution, we choose the initial vector $(u,0,u)/\sqrt2$ instead of $b$, so that the third block of the output contains a component that does not depend on the input, and $\gamma\ne0$. The bias of the sine test is then of order $t^2$, and the lower bound holds for all errors below a constant multiple of $t^2$. For principal components, a known rank-one term creates a constant spectral gap, and the third block of the top eigenvector is again close to a multiple of $t^2Hv$.

\subsection{Related work}\label{sec:related-work}

Several works extend, analyze, or evaluate the quantum and dequantized algorithms compared in this paper. Le Gall shows that dequantized algorithms for low-rank and sparse QSVT remain efficient when the input sampling distributions are only close to the exact ones in total variation distance \cite{LeGall2025}. Arrazola et al.\ evaluate the practical performance of quantum-inspired algorithms \cite{ArrazolaEtAl2020}. Cha and Lee analyze sampling and inner-product estimation under assumptions on the input entries \cite{ChaLee2025}, and Isogai et al.\ give a classical algorithm for diagonally weighted matrix functions and use it to dequantize the quantum sampler for learning with optimized random features \cite{IsogaiEtAl2026}. On the quantum side, Costa et al.\ give a linear-system solver with optimal dependence on the condition number \cite{CostaEtAl2022}, and Low and Su give a linear-system solver with an optimal number of calls to the state-preparation oracle \cite{LowSu2026}. Another line of work reduces the dependence of quantum linear-system solvers on the condition number \cite{KimEtAl2026,DalzellEtAl2026}. For our families, whose condition number is a constant, the quantum query complexity is $\widetilde\Theta(\rho)$ (Table~\ref{tab:results}); in particular, every quantum algorithm in the input model of Section~\ref{sec:input-models} requires $\Omega(\rho)$ queries, whatever its dependence on the condition number.

Other works study the estimation of scalar quantities instead of sampling. Montanaro and Shao prove classical lower bounds exponential in the approximate degree for estimating entries of functions of sparse Hermitian matrices \cite{MontanaroShao2024Functions}, and Gharibian and Le Gall give classical algorithms for related scalar quantities of sparse matrices \cite{GharibianLeGall2023}. Edenhofer, Hasegawa, and Le Gall dequantize quantum algorithms for spectral sums of sparse matrices, such as the log-determinant, and prove that estimating normalized traces of powers and inverses of log-local Hamiltonians to inverse-polynomial accuracy is complete for DQC1, the class of problems solved efficiently in the one-clean-qubit model \cite{EdenhoferEtAl2025}. Andoni et al.\ prove lower bounds for estimating individual coordinates of the solutions of linear systems \cite{AndoniEtAl2019}. Mande and Shao also prove a classical lower bound of fourth order in $\|A\|_F\|b\|$ for a scalar estimation task from supervised clustering, for which a quantum algorithm whose cost is of second order in $\|A\|_F\|b\|$ is known, so the separation is quadratic \cite{MandeShao2025}. In that task, $\|A\|_F^2=2$ and $\|A\|\ge1$, so $\rho\le\sqrt2$, and the bound grows with the norm of the vector $b$.

Quantum and classical algorithms have also been compared in other input models. Cotler, Huang, and McClean show that, for some learning tasks, classical algorithms with sampling and query access to an input vector are exponentially faster than quantum algorithms that receive copies of the quantum state proportional to this vector, so sampling and query access can be more powerful than quantum state inputs \cite{CotlerEtAl2021}. In our comparison, the quantum algorithms instead have coherent access to the same input as the classical algorithms (Section~\ref{sec:input-models}). Zhao et al.\ prove that a quantum computer of polylogarithmic size can perform classification and dimension reduction on large classical data sets by processing the data samples as they arrive, whereas every classical machine with the same prediction performance requires exponentially more memory; their quantum algorithms access the data through quantum oracle sketching \cite{ZhaoEtAl2026}. The resource in their separation is memory in a streaming setting, whereas our results concern the number of queries to an input that can be accessed repeatedly.

The sampling problem of Theorem~\ref{thm:intro-fourier} generalizes Fourier sampling: for $K=M$, the task is to output $\bot$ with probability $1/2$ and otherwise a character $s$ with probability $\widehat f(s)^2$, as in Fourier sampling. Aaronson and Ambainis prove that Fourier sampling requires $\Omega(M/\log M)$ classical queries \cite{AaronsonAmbainis2015}. Aaronson and Chen \cite{AaronsonChen2017} and, independently, Montanaro \cite{Montanaro2019} prove that Fourier sampling requires $\Omega(M)$ classical queries at sufficiently small constant errors, and Montanaro also proves that distributed Fourier sampling requires classical communication linear in the size of the truth table. Forrelation is a decision problem that is also defined through the Walsh--Hadamard transform. Bansal and Sinha prove that $k$-fold Forrelation, which a quantum algorithm solves with $\lceil k/2\rceil$ queries, requires a number of randomized classical queries that is, up to logarithmic factors, at least the $(1-1/k)$-th power of the input length \cite{BansalSinha2021}. Sherstov, Storozhenko, and Wu prove the same separation for $k$-fold rorrelation, in which a random orthogonal matrix replaces the Walsh--Hadamard transform \cite{SherstovEtAl2021}. These separations are optimal, up to logarithmic factors, for partial functions computed with a constant number of quantum queries, because Aaronson and Ambainis show that every partial function computed with $q$ quantum queries can be computed with $O(N^{1-1/(2q)})$ classical queries, where $N$ is the input length \cite{AaronsonAmbainis2015}.

Quartic quantum speedups over specific classical algorithms are also known for planted inference problems. Hastings gives a quantum algorithm for tensor principal component analysis with a quartic speedup over the fastest classical spectral algorithm \cite{Hastings2020}. Schmidhuber, O'Donnell, Kothari, and Babbush give a quantum algorithm for the planted noisy $k$XOR problem with a nearly quartic speedup over the best known classical algorithm, generalizing the result of Hastings with a framework based on the Kikuchi method \cite{SchmidhuberEtAl2025}.

\subsection{Organization}\label{sec:organization}

The input and output models and standard tools are given in Section~\ref{sec:preliminaries}. In Section~\ref{sec:framework} we develop the framework: the test functions and their biases, the limits of classical algorithms with few queries, the transfer theorem, and Fourier sampling on a set of characters. The framework is applied to linear systems in Section~\ref{sec:linear-systems}, and to positive-definite systems, least squares, matrix functions, Hamiltonian evolution, and principal components in Section~\ref{sec:positive-definite}. The worst-case separation as a function of $\rho$, the ranges of errors, settings that are not covered by our framework, and open problems are discussed in Section~\ref{sec:discussion}. In the appendices, we give explicit dimension thresholds (Appendix~\ref{app:regimes}), prove the communication upper bounds (Appendix~\ref{app:communication}), describe a linear system whose solution contains all Fourier coefficients (Appendix~\ref{app:diagonal}), prove lower bounds for procedures whose output approximates the solution only with probability $2/3$ (Appendix~\ref{app:output-access}), and prove a separation in memory between quantum and classical streaming algorithms that sample from the solutions of the linear systems of Section~\ref{sec:linear-systems}, in a streaming model of Zhao et al.\ \cite{ZhaoEtAl2026} (Appendix~\ref{app:streaming}).

\section{Preliminaries}\label{sec:preliminaries}

This section collects the definitions and standard tools used in later sections. After fixing notation in Section~\ref{sec:notation}, we define the input and output models in Section~\ref{sec:input-models}. In Section~\ref{sec:accuracy}, we recall the standard argument by which lower bounds for sampling also hold for algorithms that return sampling and query access to an approximate solution (Lemma~\ref{lem:solution-access}). The unknown sign matrices used in all our matrix families, and the promise on their norm, are introduced in Section~\ref{sec:hidden-signs}, where Lemma~\ref{lem:spectral-tail} shows that uniformly random signs satisfy the promise except with exponentially small probability. In Section~\ref{sec:known-magnitudes}, we define queries to Boolean functions and show that, when the unknown entries have known magnitudes, the input operations reveal nothing beyond such queries (Lemma~\ref{lem:access}). Finally, the quantum tools are stated in Section~\ref{sec:quantum-tools}: amplitude amplification with an unknown success probability (Lemma~\ref{lem:amplification}), the quantum algorithm for the positive-definite families (Lemma~\ref{lem:qsvt}), and the reduction from unstructured search on which all our quantum lower bounds are based (Lemma~\ref{lem:search}), together with the sign matrices used in this reduction (Lemma~\ref{lem:search-instances}).

\subsection{Notation}\label{sec:notation}

For a matrix $A$, $\|A\|$ denotes the operator norm and $\|A\|_F$ the Frobenius norm. For an invertible matrix, $\kappa(A)=\|A\|\|A^{-1}\|$ is the condition number, and for a nonzero matrix, $\rho(A)=\|A\|_F/\|A\|$ is the normalized Frobenius norm. For a vector, $\|\cdot\|$ is the Euclidean norm. For a nonzero vector $x$ indexed by a finite set, $p_x(j)=|x_j|^2/\|x\|^2$ is its length-squared distribution, and $\TV(p,q)=\frac12\sum_j|p(j)-q(j)|$ is the total variation distance between two distributions. Throughout, $\log$ is the natural logarithm, and $\mathbf 1[\cdot]$ is the indicator of an event. The notation $\widetilde O$, $\widetilde\Omega$, and $\widetilde\Theta$ hides factors polylogarithmic in $\rho$ when the condition number and the error are fixed. In all our families the dimension is $\Theta(\rho^2)$, so these factors are also polylogarithmic in the dimension.

\subsection{Input and output models}\label{sec:input-models}

We define the classical input model, the quantum input model, and the sampling task in turn. The classical input model is sampling and query access to rows and columns.

\begin{definition}[Sampling and query access]\label{def:sq}
For a nonzero matrix $A\in\mathbb R^{N_1\times N_2}$, the classical input operations $\SQ(A)$ are the following.
\begin{enumerate}
\item Entry queries return $A_{ij}$ for given $(i,j)$.
\item Norm queries return the norm of a given row or column, or $\|A\|_F$.
\item Row and column sampling return a row index $i$ with probability $\|A_{i,*}\|^2/\|A\|_F^2$, or a column index $j$ with probability $\|A_{*,j}\|^2/\|A\|_F^2$.
\item For a given nonzero row $i$, conditional sampling returns each index $j$ with probability $A_{ij}^2/\|A_{i,*}\|^2$, which is proportional to $A_{ij}^2$. For a given nonzero column $j$, it returns each index $i$ with probability $A_{ij}^2/\|A_{*,j}\|^2$. For a zero row or column, it returns a fixed index.
\end{enumerate}
Each operation counts as one query. The right-hand side or initial vector of a task is fixed and known in advance, so access to it requires no query.
\end{definition}

Definition~\ref{def:sq} includes column norms, column sampling, and conditional sampling in a column. Sampling and query access to $A$ alone, as defined by Chia et al.\ \cite{ChiaEtAl2022}, does not provide these operations, so our classical lower bounds also hold in that model. The quantum input model consists of the coherent versions of the operations in Definition~\ref{def:sq}.

\begin{definition}[Coherent access]\label{def:coherent}
For a nonzero matrix $A\in\mathbb R^{N_1\times N_2}$ whose entries have a fixed binary encoding, the quantum input operations are the following unitaries.
\begin{enumerate}
\item A reversible entry query maps $|i,j,z\rangle$ to $|i,j,z\oplus A_{ij}\rangle$.
\item Row preparation maps $|i\rangle|0\rangle$ to $|i\rangle\sum_j(A_{ij}/\|A_{i,*}\|)|j\rangle$ for a nonzero row $i$, and column preparation maps $|j\rangle|0\rangle$ to $|j\rangle\sum_i(A_{ij}/\|A_{*,j}\|)|i\rangle$ for a nonzero column $j$. For a zero row or column, the corresponding preparation maps $|i\rangle|0\rangle$ or $|j\rangle|0\rangle$ to itself.
\item Norm preparation maps $|0\rangle$ to $\sum_i(\|A_{i,*}\|/\|A\|_F)|i\rangle$, or to the analogous vector of column norms.
\end{enumerate}
Each of these unitaries, its inverse, and its controlled version counts as one query. Row, column, and norm preparation are specified only on the states listed above. Their action on other states is arbitrary, and an algorithm must succeed for every choice of this action. A fixed right-hand side or initial vector is prepared by a circuit that does not depend on $A$.
\end{definition}

In both models, the operations are supplied with the input, and no other information about $A$ is available. The classical data structures used by dequantized algorithms \cite{Tang2019,ChiaEtAl2022} and the QRAM data structures used in quantum linear algebra \cite{KerenidisPrakash2017,ChakrabortyEtAl2019} store a matrix by rows, and they provide all of these operations when they store both $A$ and $A^{\mathsf T}$. In the notation of Chia et al.\ \cite{ChiaEtAl2022}, the operations of Definition~\ref{def:sq} are sampling and query access to both $A$ and $A^{\mathsf T}$.

The output model is sampling from a target vector.

\begin{definition}[Sampling task]\label{def:sampling-task}
Fix a family of inputs and, for each input, a nonzero target vector $x$. An algorithm solves the sampling task with error $\varepsilon$ if, on every input of the family, the distribution of its output is within total variation distance $\varepsilon$ of $p_x$. Its query complexity is the largest number of queries it makes on an input of the family, over all runs. For randomized classical algorithms, its expected query complexity is the largest expected number of queries on an input of the family.
\end{definition}

\subsection{From approximate solutions to samples}\label{sec:accuracy}

Dequantized algorithms return a classical description of an approximate solution that supports sampling and query access, rather than a single sample. We apply our lower bounds to such algorithms through a standard argument: constructing the description and then drawing one sample from it is a sampling algorithm. The following lemma bounds the error of this sampling algorithm in terms of the relative error of the approximate vector and the probabilities of failure.

\begin{lemma}\label{lem:output-error}
Let $x\ne0$ and $0\le\eta<1$. If $\|w-x\|\le\eta\|x\|$, then $w\ne0$ and $\TV(p_w,p_x)\le2\eta$. More generally, suppose a randomized procedure produces such a vector $w$ with probability at least $1-\delta$, and that, on that event, a sampling routine has total variation error at most $\zeta$ relative to $p_w$. If the procedure outputs an arbitrary index on all other outcomes, the distribution of its output is within total variation distance $2\eta+\delta+\zeta$ of $p_x$.
\end{lemma}

\begin{proof}
Put $x'=x/\|x\|$ and $w'=w/\|w\|$. The reverse triangle inequality and normalization imply $\|x'-w'\|\le2\eta$. By the Cauchy--Schwarz inequality,
\[
\frac12\sum_j\bigl||x'_j|^2-|w'_j|^2\bigr|\le\frac12\Bigl(\sum_j\bigl||x'_j|-|w'_j|\bigr|^2\Bigr)^{1/2}\Bigl(\sum_j(|x'_j|+|w'_j|)^2\Bigr)^{1/2}\le\|x'-w'\|.
\]
For the randomized statement, condition on the event that the vector approximation succeeds and use the convexity of the total variation distance. The complementary event contributes at most its probability $\delta$.
\end{proof}

The next lemma states the standard argument. The sampling routine may fail with small probability, and its cost may be bounded only in expectation, so the lemma concerns the expected number of queries.

\begin{lemma}\label{lem:solution-access}
Fix a family of inputs, with a nonzero target vector $x$ for each input, and an error $\varepsilon\in(0,1)$. Suppose that every classical algorithm that samples from $p_x$ within total variation distance $\varepsilon$ on every input of the family makes more than $Q$ queries in expectation on some input. Suppose that a classical procedure uses the input operations of Definition~\ref{def:sq} and returns a description of a vector $w$ such that, on every input of the family, with probability at least $1-\delta$, $\|w-x\|\le\eta\|x\|$ and the description supports a sampling routine whose output distribution is within total variation distance $\zeta$ of $p_w$, where $2\eta+\delta+\zeta\le\varepsilon$. Then constructing the description and drawing one sample from it use more than $Q$ queries in expectation on some input.
\end{lemma}

\begin{proof}
Run the procedure, and then run the sampling routine once; if the routine fails, output an arbitrary index. This is a sampling algorithm, and by Lemma~\ref{lem:output-error} its output distribution is within total variation distance $2\eta+\delta+\zeta\le\varepsilon$ of $p_x$ on every input of the family. It therefore makes more than $Q$ queries in expectation on some input.
\end{proof}

Every classical lower bound in Sections~\ref{sec:linear-systems} and~\ref{sec:positive-definite} holds for the expected number of queries, so Lemma~\ref{lem:solution-access} applies to each of them. The sampling algorithm in the proof of Lemma~\ref{lem:solution-access} draws a single sample and makes no norm queries to the returned vector. Procedures whose output approximates the solution only with probability $2/3$ require an additional step that amplifies the success probability; they are treated in Appendix~\ref{app:output-access}, under the assumption of exact access to the returned vector.

\subsection{Unknown sign matrices}\label{sec:hidden-signs}

All our matrix families are built from the following objects. Let $n=2^k$ be a power of four, so that $k=\log_2n$ is even, and index the rows and columns of $n\times n$ matrices by $\{0,1\}^k$. For $j,a\in\{0,1\}^k$, put $h_{ja}=(-1)^{j\cdot a}$, where $j\cdot a$ is the inner product modulo two. The normalized Walsh--Hadamard matrix $H$ has entries $h_{ja}/\sqrt n$; it is real, symmetric, and orthogonal. Let $u=n^{-1/2}(1,\ldots,1)^{\mathsf T}$ be the uniform unit vector, and let $e_0$ be the standard basis vector indexed by $0^k$, so that $He_0=u$ and $Hu=e_0$. For an unknown sign matrix $W\in\{\pm1\}^{n\times n}$, put
\begin{equation}\label{eq:hidden-notation}
R=\frac W{\sqrt n},\qquad v=Ru,\qquad y=Hv,\qquad Z_j=\sqrt n\,y_j=\frac1n\sum_{a,b}h_{ja}W_{ab},\qquad \nu=\|v\|^2.
\end{equation}
Every entry of $R$ and $H$ has absolute value $1/\sqrt n$, so $\|R\|_F^2=\|H\|_F^2=n$. We take $n$ to be a power of four, and not only a power of two, so that $\sqrt n=2^{k/2}$ is a power of two. The entries of $R$ and $H$ are then dyadic rationals with $O(\log n)$ bits, which the reversible entry queries of Definition~\ref{def:coherent} can write exactly into a register. Since the parameters of all our families are fixed rational numbers, every entry of our matrices is then a rational number whose numerator and denominator have $O(\log n)$ bits. Our constructions impose the promise
\begin{equation}\label{eq:promise}
\|R\|\le8,
\end{equation}
and we write $\mathcal G$ for the set of sign matrices that satisfy it. Under the promise, the condition numbers of all our matrices are bounded by constants (Lemmas~\ref{lem:nonsymmetric-parameters}, \ref{lem:pd-parameters}, and~\ref{lem:pca-spectrum}). The constant $8$ is not optimized; it is chosen to keep the proof of Lemma~\ref{lem:spectral-tail} short.

The classical lower bounds are proved for uniformly random signs, for which the normalized row sums $v_a=n^{-1}\sum_bW_{ab}$ are independent. A sum of $n$ independent uniform signs has second moment $n$ and fourth moment $3n^2-2n$, so
\begin{equation}\label{eq:nu-moments}
\E\nu=1,\qquad \E\nu^2=1+\frac{2(n-1)}{n^2}\le3.
\end{equation}
These moments enter the error bounds in the applications of the transfer theorem (Theorem~\ref{thm:transfer}) in Sections~\ref{sec:linear-systems} and~\ref{sec:positive-definite}. Random signs satisfy the promise~\eqref{eq:promise} except with exponentially small probability.

\begin{lemma}\label{lem:spectral-tail}
If the entries of $W$ are independent uniform signs, then
\[
\Prb(W\notin\mathcal G)\le\delta_n:=2\exp\bigl(-(8-2\log9)n\bigr).
\]
\end{lemma}

\begin{proof}
For real unit vectors $r,s$, the random variable $r^{\mathsf T}Rs$ is a sum of independent signs whose squared coefficients sum to $1/n$. Hoeffding's inequality bounds $\Prb(|r^{\mathsf T}Rs|>4)$ by $2e^{-8n}$. Choose a $1/4$-net of the unit sphere with at most $9^n$ points. If $\|R\|>8$, approximating a pair of maximizing unit vectors by net points changes the bilinear form by at most $\|R\|/2$, so some pair of net points has bilinear form larger than $4$. A union bound over at most $9^{2n}$ pairs completes the proof.
\end{proof}

\subsection{Boolean functions and matrices with known magnitudes}\label{sec:known-magnitudes}

In our lower bounds, the unknown input is a Boolean function, that is, a map $f:\{0,1\}^m\to\{-1,1\}$; as usual in Fourier analysis, we write its values as $\pm1$ rather than as bits. For our matrix families, $f(a,b)=W_{ab}$. A classical query returns $f(\sigma)$ for a chosen point $\sigma$, and a quantum query applies the oracle $O_f|\sigma,c\rangle=|\sigma,c\oplus(1-f(\sigma))/2\rangle$. A controlled phase query $|b\rangle|\sigma\rangle\mapsto f(\sigma)^b|b\rangle|\sigma\rangle$ uses one call: prepare an ancilla in $|-\rangle$ when $b=1$ and in $|+\rangle$ when $b=0$, apply $O_f$ to $|\sigma\rangle$ and the ancilla, and undo the preparation of the ancilla.

The lower bounds of Section~\ref{sec:framework} concern algorithms that query a Boolean function directly. The next lemma shows that the input models of Section~\ref{sec:input-models} provide no additional information when every entry that depends on $f$ is an unknown sign times a known nonzero number.

\begin{lemma}\label{lem:access}
Let $\{A(f)\}$ be a family of real $N_1\times N_2$ matrices indexed by Boolean functions $f$ on $\{0,1\}^m$. Suppose every entry $A_{ij}(f)$ either does not depend on $f$ or equals $\mu_{ij}f(\sigma_{ij})$ for a fixed nonzero real number $\mu_{ij}$ and a fixed point $\sigma_{ij}$, and that the map $(i,j)\mapsto\sigma_{ij}$ is computed by a fixed reversible circuit.
\begin{enumerate}
\item Every entry magnitude, every row and column norm, the Frobenius norm, and every sampling distribution in Definition~\ref{def:sq} is independent of $f$. Every classical input operation can therefore be simulated with at most one query to $f$, and a lower bound for classical algorithms that query $f$ directly is also a lower bound in the model of Definition~\ref{def:sq}.
\item Every coherent input operation, its inverse, and its controlled version can be implemented with $O(1)$ calls to $O_f$ and fixed circuits. Conversely, if every point appears as some $\sigma_{ij}$, one call to $O_f$ can be implemented with two reversible entry queries. Quantum query complexities in the two models therefore agree up to constant factors.
\end{enumerate}
\end{lemma}

\begin{proof}
The norms and sampling distributions depend only on the magnitudes $|A_{ij}(f)|$, which are fixed. A classical algorithm can therefore carry out norm queries and all sampling operations without queries, and an entry query reads at most one value $f(\sigma_{ij})$.

A reversible entry query computes $\sigma_{ij}$, calls $O_f$, writes $\mu_{ij}f(\sigma_{ij})$ or the fixed entry into the output register, and uncomputes the call with a second call to $O_f$. Row preparation first prepares the state $\sum_j(|A_{ij}|/\|A_{i,*}\|)|j\rangle$, which does not depend on $f$. It then applies the known signs, which are the signs of the fixed entries and of the numbers $\mu_{ij}$, by a fixed phase circuit, and it applies the unknown signs $f(\sigma_{ij})$ by one controlled phase query at the point $\sigma_{ij}$. Column preparation is similar, and the norm vectors do not depend on $f$. Inverses and controlled versions use the same number of calls. Conversely, to implement one call to $O_f$ at a point $\sigma$, use a fixed position $(i,j)$ with $\sigma_{ij}=\sigma$. A reversible entry query writes $\mu_{ij}f(\sigma)$ into an ancilla register, a fixed circuit adds the bit $(1-f(\sigma))/2$ to the target qubit by comparing the sign of this entry with the known sign of $\mu_{ij}$, and a second entry query uncomputes the ancilla register.
\end{proof}

The quantum upper bounds also bound the number of gates. In our families, the positions of each row and column split, independently of $W$, into a constant number of sets, and all entries in one set have the same known magnitude. Each set is the intersection of the row or column with one $n\times n$ block, such an intersection with one position removed, or a single position. The states prepared in the proof of Lemma~\ref{lem:access} are therefore combinations of a constant number of uniform superpositions over such sets, and each input operation uses $O(\log n)$ gates apart from the synthesis of fixed rotations.

\subsection{Quantum algorithms and quantum lower bounds}\label{sec:quantum-tools}

Our quantum algorithms for Fourier sampling on a set of characters and for the linear systems of Section~\ref{sec:linear-systems} prepare a state whose projection onto a fixed subspace is proportional to the target state. The squared norm of this projection depends on the input and is unknown, but it has a known lower bound. In this situation, we use the following randomized form of amplitude amplification \cite{BrassardEtAl2002}.

\begin{lemma}\label{lem:amplification}
Let $V$ be a quantum circuit that makes $r$ queries, and let $\Pi$ be an orthogonal projection that does not depend on the input. Suppose that $p=\|\Pi V|0\rangle\|^2$ satisfies $p\ge p_0$ for a known number $p_0\in(0,1]$. For every $\varepsilon\in(0,1)$, there is a quantum algorithm that makes $O(r(1+\log(1/\varepsilon))/\sqrt{p_0})$ queries and, with probability at least $1-\varepsilon$, outputs the state $\Pi V|0\rangle/\sqrt p$; otherwise, it reports failure.
\end{lemma}

\begin{proof}
Write $p=\sin^2\phi$ with $\phi\in(0,\pi/2]$, let $G=-V(I-2|0\rangle\langle0|)V^\dagger(I-2\Pi)$, and put $L=\lceil2/\sqrt{p_0}\rceil$. The operator $G$ makes $2r$ queries, and $\Pi G^jV|0\rangle$ is a real multiple of $\Pi V|0\rangle$ with squared norm $\sin^2((2j+1)\phi)$ \cite{BrassardEtAl2002}. A trial chooses with equal probability between measuring $\Pi$ on $V|0\rangle$ and measuring $\Pi$ on $G^jV|0\rangle$ for $j$ uniform in $\{0,\ldots,L-1\}$. If $p>3/4$, the first choice succeeds with probability greater than $3/4$. If $p\le3/4$, then $\phi\le\pi/3$, so $\sin2\phi=2\sin\phi\cos\phi\ge\sin\phi\ge\sqrt{p_0}$, and the second choice succeeds with probability
\[
\frac1L\sum_{j=0}^{L-1}\sin^2((2j+1)\phi)=\frac12-\frac{\sin(4L\phi)}{4L\sin2\phi}\ge\frac12-\frac1{4L\sqrt{p_0}}\ge\frac38.
\]
Every trial therefore succeeds with probability at least $3/16$ and makes at most $2rL$ queries. On success, the state after the measurement is $\Pi V|0\rangle/\sqrt p$, up to a global sign. Run up to $\lceil\log(1/\varepsilon)/\log(16/13)\rceil$ trials, and report failure if all of them fail. The failure probability is at most $\varepsilon$, and the number of queries is $O(rL(1+\log(1/\varepsilon)))$.
\end{proof}

Our quantum algorithms for positive-definite matrices combine block encodings and quantum singular value transformation in the standard way.

\begin{lemma}\label{lem:qsvt}
Let $\{A(f)\}$ be a family of real symmetric matrices of dimension $O(n)$ that satisfies the hypotheses of Lemma~\ref{lem:access}, with $\|A(f)\|_F$ known and of order $\Theta(\sqrt n)$, and with all eigenvalues in a fixed interval $[\lambda_-,\lambda_+]\subset(0,\infty)$. Suppose the coherent input operations use $O(\log n)$ gates apart from fixed rotations. Let $b$ be a fixed unit vector prepared with $O(\log n)$ gates, let $g$ be analytic on an open set that contains $[\lambda_-,\lambda_+]$, and suppose $\|g(A(f))b\|\ge c_0$ for a constant $c_0>0$. Then, for every constant $\varepsilon>0$, a quantum algorithm samples from $p_{g(A(f))b}$ within total variation distance $\varepsilon$ with $\widetilde O(\sqrt n)$ queries and gates. The same holds for the orthogonal projector $\Pi_\theta(A)$ onto the span of the eigenvectors with eigenvalues above a threshold $\theta$, if every eigenvalue is at a constant distance from $\theta$ and $\|\Pi_\theta(A)b\|$ is at least a positive constant.
\end{lemma}

\begin{proof}
The isometries $|i\rangle\mapsto|i\rangle\sum_j(A_{ij}/\|A_{i,*}\|)|j\rangle$ and $|j\rangle\mapsto\sum_i(\|A_{i,*}\|/\|A\|_F)|i\rangle|j\rangle$, built from row and norm preparation, have overlap matrix $A/\|A\|_F$. Their unitary extensions therefore form a block encoding of $A/\|A\|_F$ \cite{ChakrabortyEtAl2019,GilyenEtAl2019}. Uniform singular value amplification turns it into a block encoding of $A/(2\lambda_+)$, with operator-norm error $\varepsilon'$, using $O((\|A\|_F/\lambda_+)\log(\|A\|_F/\varepsilon'))=\widetilde O(\sqrt n)$ uses of the block encoding \cite{GilyenEtAl2019}. Since $g$ is analytic on a neighborhood of the spectrum, there is a polynomial of degree $O(\log(1/\varepsilon'))$, with constants that depend on $g$ and on the interval, that approximates a bounded multiple of $x\mapsto g(2\lambda_+x)$ on $[\lambda_-/(2\lambda_+),1/2]$ and is bounded by one on $[-1,1]$. Quantum singular value transformation implements this polynomial of the block encoding \cite{GilyenEtAl2019}. If the polynomial has no definite parity, its even and odd parts are implemented separately and combined by a linear combination of unitaries, at the cost of a constant factor. For complex-valued $g$, the real and imaginary parts are implemented separately and combined by a linear combination of unitaries. Applying the result to $b$ and postselecting succeeds with probability at least a constant multiple of $c_0^2$, so amplitude amplification (Lemma~\ref{lem:amplification}) adds only a constant factor. For the projector, a polynomial approximation of the step function at $\theta$ with a constant gap has degree $O(\log(1/\varepsilon'))$. Choosing $\varepsilon'$ as a sufficiently small constant that depends on $\varepsilon$, $c_0$, $g$, and the interval, and synthesizing the fixed rotations to inverse-polynomial precision \cite{RossSelinger2016}, keeps the total variation error below $\varepsilon$ and adds only logarithmic factors.
\end{proof}

All our quantum lower bounds follow from the lower bound for unstructured search by the following reduction, which we state once for all families.

\begin{lemma}\label{lem:search}
Fix a family of inputs and a sampling task. Suppose that, for every sufficiently large $n$, the family contains inputs $A^{(\emptyset)}$ and $A^{(a)}$ for $a$ in a set of size $n-1$, and that there are an injective map $a\mapsto j(a)$ into the set of outcomes and a constant $p_*>0$ with the following properties. Every coherent input operation on these inputs can be implemented with $O(1)$ queries to the bit oracle of a string $s\in\{0,1\}^{n-1}$, where $s=0$ for $A^{(\emptyset)}$ and $s$ is the indicator vector of $a$ for $A^{(a)}$. For every $a$, the target distribution of $A^{(a)}$ assigns probability at least $p_*$ to $j(a)$. Then every quantum algorithm that solves the sampling task with a constant error $\varepsilon<p_*$ makes $\Omega(\sqrt n)$ queries on some of these inputs.
\end{lemma}

\begin{proof}
Run the sampler $T=\lceil2/(p_*-\varepsilon)\rceil$ times on the unknown input. Whenever the output equals $j(a')$ for some $a'$, query $s_{a'}$, and report $a'$ if $s_{a'}=1$. If no index is reported, report that no position is marked. On $A^{(\emptyset)}$, no index is ever reported. On $A^{(a)}$, each run outputs $j(a)$ with probability at least $p_*-\varepsilon$, so the probability that $a$ is never found is at most $(1-(p_*-\varepsilon))^T\le e^{-2}<1/3$. This procedure decides with bounded error whether $s$ has a marked position. It uses $T$ times the query complexity of the sampler, up to a constant factor, plus $T$ queries. Deciding whether a string of length $n-1$ with at most one marked position has a marked position requires $\Omega(\sqrt n)$ quantum queries \cite{BennettEtAl1997}, and $T$ depends only on $p_*-\varepsilon$.
\end{proof}

In all our quantum lower bounds, we apply Lemma~\ref{lem:search} to the following sign matrices: the matrix with $R=H$, and the matrices obtained from it by replacing one row of $R$, other than row $0^k$, by $u^{\mathsf T}$.

\begin{lemma}\label{lem:search-instances}
Let $W^{(\emptyset)}=\sqrt n\,H$, so that $R=H$. For $a\in\{0,1\}^k\setminus\{0^k\}$, let $W^{(a)}$ be the sign matrix for which $R=W^{(a)}/\sqrt n$ equals $H$ except that row $a$ is replaced by $u^{\mathsf T}$.
\begin{enumerate}
\item These matrices satisfy the promise~\eqref{eq:promise}. For $W^{(\emptyset)}$, $v=e_0$. For $W^{(a)}$, $\|R\|=\sqrt2$, $v=e_0+e_a$, and $RR^{\mathsf T}v=2v$.
\item Index the strings $s\in\{0,1\}^{n-1}$ by $\{0,1\}^k\setminus\{0^k\}$, and put $W(s)=W^{(\emptyset)}$ for $s=0$ and $W(s)=W^{(a)}$ when $s$ is the indicator vector of $a$. For every family of matrices that satisfies the hypotheses of Lemma~\ref{lem:access} with $f(a,b)=W_{ab}$, every coherent input operation on the inputs with $W=W(s)$ can be implemented, as in the proof of Lemma~\ref{lem:access}, with $O(1)$ queries to the bit oracle of $s$.
\end{enumerate}
\end{lemma}

\begin{proof}
The rows of $H$ are orthonormal, and row $0^k$ of $H$ equals $u^{\mathsf T}$. The matrix $R$ of $W^{(a)}$ therefore has two rows equal to $u^{\mathsf T}$, in positions $0^k$ and $a$, and its other rows are orthonormal and orthogonal to $u$. Hence $RR^{\mathsf T}$ has the eigenvalues $2$, $0$, and $1$, so $\|R\|=\sqrt2\le8$. Moreover, $v=Ru=e_0+e_a$ and $R^{\mathsf T}v=2u$, so $RR^{\mathsf T}v=2v$. For $W^{(\emptyset)}$, $\|R\|=\|H\|=1$ and $v=Hu=e_0$. For the second claim, the entry $W(s)_{ab}$ equals $1$ if $a\ne0^k$ and $s_a=1$, and it equals $h_{ab}$ otherwise. One call to $O_f$ with $f=W(s)$ can therefore be implemented with two queries to $s$, one to compute $s_a$ and one to uncompute it, and Lemma~\ref{lem:access} implies the claim.
\end{proof}

\section{Fourier sampling on a set of characters and the transfer theorem}\label{sec:framework}

In this section we prove the lower bound from which all our classical lower bounds follow, and we apply it to Fourier sampling on a set of characters. The lower bound concerns an unknown Boolean function on $M$ points and a fixed set $S$ of $K$ Fourier characters. Let $Z_s$ be the normalized Fourier coefficient of the Boolean function at a character $s$. Suppose that each Boolean function determines a target vector whose coordinates indexed by $S$ are close to $(\gamma_s+\beta Z_s)/\sqrt K$, where $\gamma_s$ and $\beta$ do not depend on the Boolean function, and whose squared norm is close to a fixed value. Suppose further that the bias of a test function, defined below, exceeds the errors of these two approximations by a margin that does not depend on $M$. Then every classical algorithm that samples from the length-squared distribution of the target vector with an error below half of this margin requires $\Omega(M/\log(2K))$ queries. For Fourier sampling on a set of characters, a quantum algorithm uses only $O(\sqrt{M/K})$ queries.

The proof compares the expectation of a test function, which is a bounded function of the unknown Boolean function and of the output, under exact sampling and under a classical algorithm with few queries. In Section~\ref{sec:test-functions}, we define the Fourier coefficients of a Boolean function and the two test functions that we use, the sine test and the cosine test. In Section~\ref{sec:bias}, we compute the bias of each test function, which approximates its expectation under exact sampling, averaged over uniformly random signs, when the output distribution weights each character $s$ in $S$ by $|\gamma_s+\beta Z_s|^2$. In Section~\ref{sec:transcripts}, we show that a classical algorithm with few queries cannot reproduce this bias. Conditioned on its previous queries, the unqueried signs are still independent and uniformly random, and, with high probability, the queried signs contribute little to every $Z_s$. Both test functions therefore have conditional expectation close to zero at every possible output. In Section~\ref{sec:transfer}, the two facts are combined into the transfer theorem, and we show that the two conditions on the target vector imply its hypothesis. Finally, in Section~\ref{sec:fourier}, the transfer theorem is applied to Fourier sampling on a set of characters, and the quantum algorithm for this problem is given.

\subsection{Fourier coefficients and test functions}\label{sec:test-functions}

We first define the Fourier coefficients of a Boolean function and the two test functions used in all our lower bounds. Fix $m\ge2$, put $M=2^m$, and let $f$ be a Boolean function on $\{0,1\}^m$, accessed by the queries of Section~\ref{sec:known-magnitudes}. For $s,\sigma\in\{0,1\}^m$, let $\chi_s(\sigma)=(-1)^{s\cdot\sigma}$. The Fourier coefficients and the normalized Fourier coefficients of $f$ are
\begin{equation}\label{eq:fourier-coefficients}
\widehat f(s)=\frac1M\sum_{\sigma\in\{0,1\}^m}\chi_s(\sigma)f(\sigma),\qquad Z_s(f)=\sqrt M\,\widehat f(s)=\frac1{\sqrt M}\sum_{\sigma\in\{0,1\}^m}\chi_s(\sigma)f(\sigma).
\end{equation}
By Parseval's identity, $\sum_s\widehat f(s)^2=1$, so $\sum_sZ_s(f)^2=M$. The test functions below depend on a fixed set $S$ of $K$ characters.

In the matrix families of Sections~\ref{sec:linear-systems} and~\ref{sec:positive-definite}, $m=2k$, $M=n^2$, and $f(a,b)=W_{ab}$. For $j\in\{0,1\}^k$, the character $\chi_{(j,0^k)}(a,b)=h_{ja}$ does not depend on $b$, and $Z_{(j,0^k)}(f)$ equals $Z_j$ in~\eqref{eq:hidden-notation}. We write
\begin{equation}\label{eq:row-characters}
S_{\mathrm{row}}=\{(j,0^k):j\in\{0,1\}^k\}
\end{equation}
for this set of $K=n$ characters. In the lower bounds for these families, $S=S_{\mathrm{row}}$.

Let $\mathcal Y\supseteq S$ be a finite set of outcomes, which are the possible outputs of the samplers that we consider. In the matrix tasks, the outcomes are the coordinates of the output vector, and the character $(j,0^k)\in S_{\mathrm{row}}$ is identified with coordinate $j$ of the third block. For Fourier sampling on a set of characters (Definition~\ref{def:fourier-sampling}), the outcomes are the elements of $S$ and one additional outcome $\bot$. A test function is a bounded function $F(f,j)$ of a Boolean function $f$ and of an outcome $j\in\mathcal Y$. We write lowercase letters, such as $j$ and $s$, for fixed outcomes, and the capital letter $J$, possibly with a subscript, for the output of a sampler, which is a random element of $\mathcal Y$. For a Boolean function $f$, the sine test and the cosine test are
\begin{equation}\label{eq:tests}
F_{\sin}(f,s)=\sin Z_s(f),\qquad F_{\cos}(f,s)=\cos Z_s(f)-c_M\qquad(s\in S),\qquad\text{where }c_M=\cos(M^{-1/2})^M,
\end{equation}
and $F_{\sin}(f,j)=F_{\cos}(f,j)=0$ for $j\in\mathcal Y\setminus S$. Since $S$ is fixed in each application, we omit it from the notation of the test functions. For uniformly random $f$, Lemma~\ref{lem:moments} below shows that $\E\sin Z_s(f)=0$ and $\E\cos Z_s(f)=c_M$, so both test functions have expectation zero at every fixed outcome.

The proofs use three properties of the two test functions. First, each test function takes values in an interval of length two, so its expectation under a sampler with total variation error $\varepsilon$ differs from its expectation under exact sampling by at most $2\varepsilon$. We use this property to turn a difference between two expectations of a test function into a lower bound on the total variation error. Second, $e^{iZ_s(f)}$ is the product of the factors $e^{i\chi_s(\sigma)f(\sigma)/\sqrt M}$ over all points $\sigma$. When the values of $f$ at the unqueried points are independent and uniformly random, the conditional expectations of $\sin Z_s$ and $\cos Z_s$ therefore factor over these points and can be computed exactly. With this computation and the Azuma--Hoeffding inequality, we show in Lemma~\ref{lem:transcript} that the expectation of each test function at the output of a classical algorithm with few queries is close to zero, and hence such an algorithm cannot reproduce the bias. Third, replacing $f$ by $-f$ changes the sign of every $Z_s(f)$, so the sine test changes sign, while the cosine test does not change. This property determines which of the two test functions has a nonzero bias (Section~\ref{sec:bias}).

\subsection{The bias of a test function}\label{sec:bias}

In this subsection we compute the expectation of a test function under exact sampling, averaged over uniformly random Boolean functions. In all our applications, the target distribution assigns to each $s\in S$ a probability close to $|\gamma_s+\beta Z_s(f)|^2/(dK)$, where $\gamma\in\mathbb C^S$, $\beta\in\mathbb C$, and $d>0$ do not depend on $f$. Since a test function $F$ vanishes outside $S$, the expectation of $F$ under exact sampling is then close to
\begin{equation}\label{eq:Phi}
\Phi(f)=\frac1{dK}\sum_{s\in S}|\gamma_s+\beta Z_s(f)|^2F(f,s).
\end{equation}
We call the average $\Lambda_F=\E\Phi(f)$ over uniformly random $f$ the bias of $F$. The precise sense in which the expectation of $F$ under exact sampling is close to $\Phi(f)$ is the hypothesis of the transfer theorem (Theorem~\ref{thm:transfer}). The bias depends on $c_M$ and on two further moments of a normalized coefficient,
\begin{equation}\label{eq:moments}
b_M=\sqrt M\sin(M^{-1/2})\cos(M^{-1/2})^{M-1},\qquad a_M=\cos(M^{-1/2})^{M-2}\bigl(1-M\sin^2(M^{-1/2})\bigr).
\end{equation}

\begin{lemma}\label{lem:moments}
Let $f$ be uniformly random and $s\in\{0,1\}^m$. Then $Z_s=Z_s(f)$ has the distribution of $M^{-1/2}$ times a sum of $M$ independent uniform signs. In particular, $\E Z_s^2=1$ and $\E Z_s^4=3-2/M$. Moreover, $\E\sin Z_s=\E[Z_s\cos Z_s]=\E[Z_s^2\sin Z_s]=0$, and
\[
\E\cos Z_s=c_M,\qquad \E[Z_s^2\cos Z_s]=a_M,\qquad \E[Z_s\sin Z_s]=b_M.
\]
For $M\ge4$,
\begin{equation}\label{eq:moment-bounds}
\frac12\le c_M\le e^{-1/2},\qquad 0\le a_M\le\frac1{3M},\qquad \Bigl(1-\frac1{6M}\Bigr)e^{-1/2}\le b_M\le1,\qquad a_M-c_M\le-\frac5{12}.
\end{equation}
As $M\to\infty$, $c_M\to e^{-1/2}$, $a_M\to0$, and $b_M\to e^{-1/2}$.
\end{lemma}

\begin{proof}
For fixed $s$, the $M$ numbers $\chi_s(\sigma)f(\sigma)$ are independent uniform signs, which proves the first claim and the formulas for $\E Z_s^2$ and $\E Z_s^4$. Since $Z_s$ is symmetric, every odd function of $Z_s$ has expectation zero. The characteristic function is $\phi(\theta)=\E e^{i\theta Z_s}=\cos(\theta/\sqrt M)^M$, and its first two derivatives satisfy
\begin{gather*}
\E[Z_se^{i\theta Z_s}]=-i\phi'(\theta)=i\sqrt M\cos(\theta/\sqrt M)^{M-1}\sin(\theta/\sqrt M),\\
\E[Z_s^2e^{i\theta Z_s}]=-\phi''(\theta)=\cos(\theta/\sqrt M)^{M-2}\bigl(1-M\sin^2(\theta/\sqrt M)\bigr).
\end{gather*}
Taking $\theta=1$ and real or imaginary parts proves the three formulas.

For the bounds, put $x=M^{-1/2}\le1/2$. The inequalities $1-x^2/2\le\cos x\le e^{-x^2/2}$ and Bernoulli's inequality imply $1/2\le(1-1/(2M))^M\le c_M\le e^{-1/2}$. The inequalities $x-x^3/6\le\sin x\le x$ imply $x^2-x^4/3\le\sin^2x\le x^2$, hence $0\le1-M\sin^2x\le1/(3M)$ and $0\le a_M\le1/(3M)$. For $b_M$, $\sqrt M\sin x\ge1-1/(6M)$. Moreover, $\log\cos x\ge\log(1-x^2/2)\ge-x^2/(2-x^2)$, where the second inequality is $\log(1-y)\ge-y/(1-y)$ with $y=x^2/2$, so $\cos(x)^{M-1}\ge\exp(-(M-1)/(2M-1))\ge e^{-1/2}$. The upper bound $b_M\le1$ holds because $\sqrt M\sin x\le1$ and $0<\cos x\le1$. The limits follow from $\cos(x)^M=\exp(-1/2+O(1/M))$ and $\sqrt M\sin x=1+O(1/M)$. Finally, $a_M-c_M\le1/(3M)-1/2\le-5/12$ for $M\ge4$.
\end{proof}

\begin{lemma}\label{lem:bias}
Let $S$ be a set of $K$ characters, let $\gamma\in\mathbb C^S$, $\beta\in\mathbb C$, and $d>0$, and let $f$ be uniformly random. The biases of the sine test and the cosine test are
\begin{equation}\label{eq:transfer-bias}
\Lambda_{\sin}=\frac{2b_M}{dK}\sum_{s\in S}\operatorname{Re}(\overline{\gamma_s}\beta),\qquad \Lambda_{\cos}=\frac{|\beta|^2(a_M-c_M)}d.
\end{equation}
Moreover, every Boolean function $f$ satisfies $|\Phi(f)|\le\Phi_{\max}$, where $\Phi_{\max}=4(\|\gamma\|^2+|\beta|^2M)/(dK)$.
\end{lemma}

\begin{proof}
For each $s\in S$,
\[
|\gamma_s+\beta Z_s|^2=|\gamma_s|^2+2\operatorname{Re}(\overline{\gamma_s}\beta)Z_s+|\beta|^2Z_s^2.
\]
By Lemma~\ref{lem:moments}, $\E[|\gamma_s+\beta Z_s|^2\sin Z_s]=2\operatorname{Re}(\overline{\gamma_s}\beta)b_M$, because $\E\sin Z_s=\E[Z_s^2\sin Z_s]=0$, and $\E[|\gamma_s+\beta Z_s|^2(\cos Z_s-c_M)]=|\beta|^2(a_M-c_M)$, because $\E\cos Z_s=c_M$, $\E[Z_s\cos Z_s]=\E Z_s=0$, and $\E Z_s^2=1$. Summing over $s\in S$ proves~\eqref{eq:transfer-bias}. The bound on $\Phi$ follows from $|F|\le2$, $|\gamma_s+\beta Z_s|^2\le2|\gamma_s|^2+2|\beta|^2Z_s^2$, and $\sum_{s\in S}Z_s^2\le M$, which holds by Parseval's identity.
\end{proof}

The two biases depend differently on $\gamma$ and $\beta$. The bias of the sine test is linear in $\beta$, and it vanishes when $\gamma=0$: the weights $|\beta Z_s|^2$ do not change when $f$ is replaced by $-f$, while the sine test changes sign. The bias of the cosine test does not depend on $\gamma$, and by~\eqref{eq:moment-bounds} it is at most $-5|\beta|^2/(12d)$, so it is negative for every $\beta\ne0$. We use the sine test for the linear systems of Section~\ref{sec:linear-systems} and for Hamiltonian evolution, where $\beta$ is real and every $\gamma_s$ is positive, so that $|\Lambda_{\sin}|$ is of order $|\beta|$. For the other tasks, $\gamma=0$, and we use the cosine test, whose bias is of order $|\beta|^2$.

\subsection{Classical algorithms with few queries}\label{sec:transcripts}

We now show that a classical algorithm with few queries cannot reproduce the bias: at its output, the expectation of each test function is close to zero. For $\xi>0$ and $\delta\in(0,1)$, define the query threshold
\begin{equation}\label{eq:query-threshold}
q_0(\xi,\delta)=\left\lfloor\frac{\xi^2M}{2\log(2K/\delta)}\right\rfloor.
\end{equation}
The proof of the next lemma uses two facts. Conditioned on the previous queries, the values of $f$ at the unqueried points are still independent and uniformly random, so the conditional expectations of the test functions factor exactly over these points. With high probability, the queried values contribute little to every coefficient $Z_s$.

\begin{lemma}\label{lem:transcript}
Let $\xi>0$ and $\delta\in(0,1)$, and let $f$ be uniformly random. Let $\mathcal A$ be a randomized classical algorithm that queries values of $f$, makes at most $q\le q_0(\xi,\delta)$ queries on every input and every run, and outputs $J_{\mathcal A}\in\mathcal Y$. Then
\[
|\E F_{\sin}(f,J_{\mathcal A})|\le\xi+\delta,\qquad |\E F_{\cos}(f,J_{\mathcal A})|\le\frac{\xi^2}2+\frac q{2M}+2\delta.
\]
In particular, $|\E F(f,J_{\mathcal A})|\le\tau_F(\xi,\delta)$ for both test functions, where
\begin{equation}\label{eq:transcript-tolerance}
\tau_{\sin}(\xi,\delta)=\xi+\delta,\qquad \tau_{\cos}(\xi,\delta)=\xi^2+2\delta.
\end{equation}
\end{lemma}

\begin{proof}
Both bounds are averages over the internal randomness of $\mathcal A$, so it suffices to prove them for deterministic algorithms. Let the transcript $\mathcal T$ consist of the queried points and the values of $f$ at these points. The output $J_{\mathcal A}$ is a function of $\mathcal T$. Conditional on any transcript of positive probability, the values of $f$ at unqueried points are independent uniform signs, because each query point depends only on the values returned by earlier queries. Let $q'\le q$ be the number of distinct queried points, and put
\[
\mu_s(\mathcal T)=\frac1{\sqrt M}\sum_{\sigma\text{ queried}}\chi_s(\sigma)f(\sigma).
\]
The difference $Z_s-\mu_s(\mathcal T)$ is a sum of $M-q'$ independent terms $\pm M^{-1/2}$, each with characteristic function $\theta\mapsto\cos(\theta/\sqrt M)$. Hence, with $\vartheta=\cos(M^{-1/2})$,
\[
\E\bigl[e^{iZ_s}\bigm|\mathcal T\bigr]=e^{i\mu_s(\mathcal T)}\vartheta^{M-q'}.
\]
When $J_{\mathcal A}=s\in S$, this identity implies $|\E[F_{\sin}\mid\mathcal T]|=|\sin\mu_s|\vartheta^{M-q'}\le|\mu_s|$. Since $c_M=\vartheta^M$, it also implies
\[
\bigl|\E[F_{\cos}\mid\mathcal T]\bigr|=\bigl|\cos\mu_s\,\vartheta^{M-q'}-\vartheta^M\bigr|\le(1-\cos\mu_s)+(\vartheta^{M-q'}-\vartheta^M)\le\frac{\mu_s^2}2+\frac q{2M},
\]
because $\vartheta^{M-q'}-\vartheta^M\le1-\vartheta^{q'}\le q'(1-\vartheta)$ and $1-\vartheta\le1/(2M)$. When $J_{\mathcal A}\notin S$, both conditional expectations are zero. If $q=0$, then $\mu_s=0$ for every $s$, and the lemma follows.

Now let $q\ge1$. For each fixed $s$, the values of $\mu_s$ after successive queries form a martingale with increments of absolute value at most $M^{-1/2}$: a query at a new point adds $M^{-1/2}\chi_s(\sigma)f(\sigma)$, which has conditional mean zero, and a repeated query adds nothing. After $\mathcal A$ stops, extend the martingale by zero increments up to length $q$. By the Azuma--Hoeffding inequality, a union bound over $S$, and $q\le q_0(\xi,\delta)$,
\[
\Prb\Bigl(\max_{s\in S}|\mu_s(\mathcal T)|>\xi\Bigr)\le2K\exp\Bigl(-\frac{\xi^2M}{2q}\Bigr)\le\delta.
\]
On the complementary event, the bounds above hold with $|\mu_s|\le\xi$ for every possible output. On the event itself, $|F_{\sin}|\le1$ and $|F_{\cos}|\le2$. Taking expectations over $\mathcal T$ proves the two bounds of the lemma. Finally, $q\le q_0(\xi,\delta)$ implies $q/(2M)\le\xi^2/(4\log(2K/\delta))\le\xi^2/(4\log2)\le\xi^2/2$, so the bound for the cosine test is at most $\tau_{\cos}(\xi,\delta)$.
\end{proof}

\subsection{The transfer theorem}\label{sec:transfer}

The transfer theorem is obtained by combining two facts: the bias of a test function under exact sampling (Lemma~\ref{lem:bias}) and the bound on its expectation at the output of a classical algorithm with few queries (Lemma~\ref{lem:transcript}). The theorem concerns target vectors $x(f)$ indexed by the Boolean functions $f$ in a set $\mathcal G$, such as the set of sign matrices that satisfy the promise~\eqref{eq:promise}. Its hypothesis states that the expectation of a test function under exact sampling is within $\eta$ of $\Phi(f)$ on average over uniformly random $f$, where inputs outside $\mathcal G$ contribute zero, for parameters $\gamma$, $\beta$, and $d$ that do not depend on $f$. Its conclusion states that a classical algorithm with few queries cannot sample from $p_{x(f)}$ with an error below about half of the bias.

\begin{theorem}[Transfer theorem]\label{thm:transfer}
Let $S$ be a set of $K$ characters, let $\mathcal Y\supseteq S$ be a finite set of outcomes with $\mathcal Y\setminus S\ne\emptyset$, and let $\mathcal G$ be a set of Boolean functions with $\delta_{\mathcal G}=\Prb(f\notin\mathcal G)$ for uniformly random $f$. For each $f\in\mathcal G$, let $x(f)\in\mathbb C^{\mathcal Y}$ be a nonzero target vector. Let $F$ be the sine test or the cosine test. Suppose there are $\gamma\in\mathbb C^S$, $\beta\in\mathbb C$, $d>0$, and $\eta\ge0$, all independent of $f$, such that, with $\Phi$ as in~\eqref{eq:Phi},
\begin{equation}\label{eq:transfer-hypothesis}
\E\Bigl[\mathbf 1[f\in\mathcal G]\,\Bigl|\sum_{j\in\mathcal Y}p_{x(f)}(j)F(f,j)-\Phi(f)\Bigr|\Bigr]\le\eta.
\end{equation}
Let $q\ge0$ and $\tau\ge0$ be such that every classical algorithm that makes at most $q$ queries on every input and every run and outputs an element $J_{\mathcal A}$ of $\mathcal Y$ satisfies $|\E F(f,J_{\mathcal A})|\le\tau$ for uniformly random $f$. Let $\varepsilon\ge0$. If a classical algorithm makes at most $q$ queries on every input in $\mathcal G$ and every run, and its output distribution is within total variation distance $\varepsilon$ of $p_{x(f)}$ for every $f\in\mathcal G$, then, with $\Lambda_F$ and $\Phi_{\max}$ as in Lemma~\ref{lem:bias},
\begin{equation}\label{eq:transfer-conclusion}
|\Lambda_F|\le\eta+\tau+2\varepsilon+(2+\Phi_{\max})\delta_{\mathcal G}.
\end{equation}
\end{theorem}

\begin{proof}
By Lemma~\ref{lem:bias}, $\E\Phi(f)=\Lambda_F$, and every $f$ satisfies $|\Phi(f)|\le\Phi_{\max}$.

Truncating the algorithm after $q$ queries changes its behavior only outside $\mathcal G$, so we may assume that it makes at most $q$ queries on every input and every run. Let $J_{\mathcal A}$ be its output. Define $J_*$ to have distribution $p_{x(f)}$ when $f\in\mathcal G$ and to equal a fixed element of $\mathcal Y\setminus S$ otherwise. Then $\E F(f,J_*)=\E[\mathbf 1[f\in\mathcal G]\sum_jp_{x(f)}(j)F(f,j)]$, so~\eqref{eq:transfer-hypothesis} and $|\Phi|\le\Phi_{\max}$ imply
\[
|\E F(f,J_*)-\Lambda_F|\le\eta+\Phi_{\max}\delta_{\mathcal G}.
\]
For each $f$, the distributions of $J_{\mathcal A}$ and $J_*$ are within total variation distance $\varepsilon$ when $f\in\mathcal G$ and within distance one otherwise. Since $F(f,\cdot)$ takes values in an interval of length two, $|\E F(f,J_{\mathcal A})-\E F(f,J_*)|\le2(\varepsilon+\delta_{\mathcal G})$. Finally, $|\E F(f,J_{\mathcal A})|\le\tau$ by the condition on $q$ and $\tau$. Combining the three bounds proves~\eqref{eq:transfer-conclusion}.
\end{proof}

By Lemma~\ref{lem:transcript}, the condition on $q$ and $\tau$ in Theorem~\ref{thm:transfer} holds with $q=q_0(\xi,\delta)$ and $\tau=\tau_F(\xi,\delta)$, for all $\xi>0$ and $\delta\in(0,1)$.

In all our applications, the hypothesis~\eqref{eq:transfer-hypothesis} is verified with the following lemma. For a single Boolean function, it bounds the difference in~\eqref{eq:transfer-hypothesis} by two quantities: the distance between the coordinates of the target vector indexed by $S$ and the vector with coordinates $(\gamma_s+\beta Z_s(f))/\sqrt K$, and the deviation of the squared norm of the target vector from $d$.

\begin{lemma}\label{lem:two-conditions}
Let $F$ be the sine test or the cosine test, and put $F_\infty=1$ for the sine test and $F_\infty=2$ for the cosine test. Let $f$ be a Boolean function, let $x\in\mathbb C^{\mathcal Y}$ be nonzero, and let $\gamma\in\mathbb C^S$, $\beta\in\mathbb C$, and $d>0$. Let $x_S$ be the restriction of $x$ to $S$, and let $\bar x\in\mathbb C^S$ have the coordinates $\bar x_s=(\gamma_s+\beta Z_s(f))/\sqrt K$. Then, with $\Phi$ as in~\eqref{eq:Phi},
\[
\Bigl|\sum_{j\in\mathcal Y}p_x(j)F(f,j)-\Phi(f)\Bigr|\le F_\infty\Bigl(\frac{\|x_S-\bar x\|\,(\|x_S\|+\|\bar x\|)}{\|x\|^2}+\frac{\|\bar x\|^2\,\bigl|\|x\|^2-d\bigr|}{d\,\|x\|^2}\Bigr).
\]
\end{lemma}

\begin{proof}
Since $F(f,j)=0$ for $j\notin S$, the sum on the left equals $\sum_{s\in S}|x_s|^2F(f,s)/\|x\|^2$, and $\Phi(f)=\sum_{s\in S}|\bar x_s|^2F(f,s)/d$. The difference is therefore
\[
\sum_{s\in S}\frac{(|x_s|^2-|\bar x_s|^2)F(f,s)}{\|x\|^2}+\Bigl(\frac1{\|x\|^2}-\frac1d\Bigr)\sum_{s\in S}|\bar x_s|^2F(f,s).
\]
Since $\bigl||x_s|^2-|\bar x_s|^2\bigr|\le|x_s-\bar x_s|(|x_s|+|\bar x_s|)$ and $|F|\le F_\infty$, the Cauchy--Schwarz inequality and the triangle inequality bound the first sum by $F_\infty\|x_S-\bar x\|(\|x_S\|+\|\bar x\|)/\|x\|^2$. The second term is at most $F_\infty\|\bar x\|^2\,\bigl|d-\|x\|^2\bigr|/(d\|x\|^2)$.
\end{proof}

With $x=x(f)$, the expectation of $\mathbf 1[f\in\mathcal G]$ times the right side is a valid choice of $\eta$ in~\eqref{eq:transfer-hypothesis}. The first term is small when the coordinates of $x(f)$ indexed by $S$ are close to $\bar x$, and the second term is small when $\|x(f)\|^2$ is close to $d$.

The applications use the transfer theorem for a sequence of families with growing $M$ and fixed parameters. The next corollary states the resulting lower bound.

\begin{corollary}\label{cor:transfer-asymptotic}
Consider a sequence of instances of Theorem~\ref{thm:transfer}, indexed by $\ell$, with $M_\ell\to\infty$ and with the same test function $F$. Suppose that $\Lambda_{F,\ell}\to\Lambda_\infty$, that $\eta_\ell\le\eta_\infty$ for all sufficiently large $\ell$, and that $(2+\Phi_{\max,\ell})\delta_{\mathcal G,\ell}\to0$. Let $\varepsilon\ge0$ be a constant with $2\varepsilon<|\Lambda_\infty|-\eta_\infty$. Then there are $c>0$ and $\ell_0$ such that, for every $\ell\ge\ell_0$, every classical algorithm that samples within total variation distance $\varepsilon$ on $\mathcal G_\ell$ makes more than $cM_\ell/\log(2K_\ell)$ queries on some input in $\mathcal G_\ell$, and more than $cM_\ell/\log(2K_\ell)$ queries in expectation on some input in $\mathcal G_\ell$.
\end{corollary}

\begin{proof}
Put $\Delta=|\Lambda_\infty|-\eta_\infty-2\varepsilon>0$ and $\varepsilon'=\Delta/20$. Choose $\xi>0$ and $\delta\in(0,1)$ with $\tau_F(\xi,\delta)\le\Delta/10$, and choose $\ell_0$ such that, for $\ell\ge\ell_0$, $\eta_\ell\le\eta_\infty$, $|\Lambda_{F,\ell}|\ge|\Lambda_\infty|-\Delta/10$, and $(2+\Phi_{\max,\ell})\delta_{\mathcal G,\ell}\le\Delta/10$. By Lemma~\ref{lem:transcript}, Theorem~\ref{thm:transfer} applies with $q=q_0(\xi,\delta)$ and $\tau=\tau_F(\xi,\delta)$. For $\ell\ge\ell_0$, the right side of~\eqref{eq:transfer-conclusion} with error $\varepsilon+\varepsilon'$ is at most $\eta_\infty+2\varepsilon+3\Delta/10<|\Lambda_{F,\ell}|$. Hence no algorithm with error at most $\varepsilon+\varepsilon'$ on $\mathcal G_\ell$ makes at most $q_0(\xi,\delta)$ queries on every input in $\mathcal G_\ell$. An algorithm with error $\varepsilon$ and at most $\varepsilon'q_0(\xi,\delta)$ expected queries on every input in $\mathcal G_\ell$ could be stopped after $q_0(\xi,\delta)$ queries. By Markov's inequality, the stopped algorithm would have error at most $\varepsilon+\varepsilon'$, which is impossible. Finally, $\log(2K/\delta)\le(1+\log(1/\delta)/\log2)\log(2K)$, so $\varepsilon'q_0(\xi,\delta)\ge cM_\ell/\log(2K_\ell)$ for a suitable $c>0$ and all sufficiently large $\ell$.
\end{proof}

\subsection{Fourier sampling on a set of characters}\label{sec:fourier}

In all our matrix families, the output vector is close to an affine function of the Fourier coefficients on the coordinates indexed by a set $S$ of characters. The following sampling problem keeps only these coordinates, together with one fixed outcome. It is the simplest application of the transfer theorem.

\begin{definition}[Fourier sampling on a set of characters]\label{def:fourier-sampling}
Let $S\subseteq\{0,1\}^m$ be nonempty, put $K=|S|$, and let $\bot$ be a symbol that is not in $\{0,1\}^m$. For a Boolean function $f$, define the vector $z_f$ indexed by $S\cup\{\bot\}$ by $z_f(s)=Z_s(f)/\sqrt K$ for $s\in S$ and $z_f(\bot)=1$, and put $D_f=\|z_f\|^2=1+K^{-1}\sum_{s\in S}Z_s(f)^2$. For $\varepsilon\in[0,1)$, the task $\Samp_S(\varepsilon)$ is to output, given query access to $f$, a random element of $S\cup\{\bot\}$ whose distribution is within total variation distance $\varepsilon$ of $p_{z_f}$, for every Boolean function $f$.
\end{definition}

A sample from $p_{z_f}$ is $\bot$ with probability $p_{z_f}(\bot)=1/D_f$, and otherwise it is an element $s\in S$ drawn with probability proportional to $\widehat f(s)^2$, since $p_{z_f}(s)=Z_s(f)^2/(KD_f)$. The outcome $\bot$ makes the distribution well defined for every $f$, and it ensures that the quantum algorithm below has a success probability that is bounded below on every input. When $S=\{0,1\}^m$, $D_f=2$ by Parseval's identity, and $\Samp_S$ is Fourier sampling \cite{AaronsonChen2017} with probability $1/2$ and the outcome $\bot$ otherwise.

The proof of the classical lower bound uses the fact that $D_f$ is close to its mean $2$, which holds because distinct squared coefficients are negatively correlated.

\begin{lemma}\label{lem:normalization}
Let $f$ be uniformly random. For distinct $s,t\in\{0,1\}^m$, $\E[Z_s^2Z_t^2]=1-2/M$. Consequently, for every set $S$ of $K$ characters,
\[
\operatorname{Var}\Bigl(\sum_{s\in S}Z_s^2\Bigr)=2K\Bigl(1-\frac KM\Bigr).
\]
\end{lemma}

\begin{proof}
Write $\epsilon_\sigma=f(\sigma)$. For any $s,t$,
\[
\E[Z_s^2Z_t^2]=\frac1{M^2}\sum_{\sigma_1,\sigma_2,\sigma_3,\sigma_4}\chi_s(\sigma_1)\chi_s(\sigma_2)\chi_t(\sigma_3)\chi_t(\sigma_4)\,\E[\epsilon_{\sigma_1}\epsilon_{\sigma_2}\epsilon_{\sigma_3}\epsilon_{\sigma_4}].
\]
The expectation on the right equals one when every value among $\sigma_1,\sigma_2,\sigma_3,\sigma_4$ occurs an even number of times, and it is zero otherwise. The index tuples with this property form three disjoint classes: $\sigma_1=\sigma_2$ and $\sigma_3=\sigma_4$; $\sigma_1=\sigma_3\ne\sigma_2=\sigma_4$; and $\sigma_1=\sigma_4\ne\sigma_2=\sigma_3$. The first class contributes $M^2$. Since $\chi_s(\sigma)\chi_t(\sigma)=\chi_{s\oplus t}(\sigma)$, each of the other two classes contributes
\[
\sum_{\sigma_1\ne\sigma_2}\chi_{s\oplus t}(\sigma_1)\chi_{s\oplus t}(\sigma_2)=\Bigl(\sum_\sigma\chi_{s\oplus t}(\sigma)\Bigr)^2-M=M^2\,\mathbf 1[s=t]-M.
\]
Hence $\E[Z_s^2Z_t^2]=1+2\cdot\mathbf 1[s=t]-2/M$, which proves the first claim. Since $\E Z_s^2=1$, the covariance of $Z_s^2$ and $Z_t^2$ is $2\cdot\mathbf 1[s=t]-2/M$, and the sum of these covariances over $s,t\in S$ equals $2K-2K^2/M$.
\end{proof}

\begin{theorem}\label{thm:fourier-classical}
Let $K=|S|\ge128$ and $q_0=\lfloor M/(32\log(512K))\rfloor$. Every classical algorithm that makes at most $q_0$ queries on every input has, for some Boolean function $f$, an output distribution at total variation distance at least $31/1536$ from $p_{z_f}$. In particular, $\Samp_S(1/100)$ requires more than $q_0$ queries.
\end{theorem}

\begin{proof}
Apply Theorem~\ref{thm:transfer} with $\mathcal Y=S\cup\{\bot\}$, with $\mathcal G$ equal to the set of all Boolean functions, so that $\delta_{\mathcal G}=0$, and with the cosine test, $\gamma=0$, $\beta=1$, and $d=2$. The restriction of $z_f$ to $S$ equals the vector $\bar x$ of Lemma~\ref{lem:two-conditions}, and $\|\bar x\|^2=D_f-1\le D_f$. Lemma~\ref{lem:two-conditions}, with $F_\infty=2$, therefore bounds the difference in~\eqref{eq:transfer-hypothesis} by $|D_f-2|$. By Lemmas~\ref{lem:moments} and~\ref{lem:normalization}, $\E D_f=2$ and $\operatorname{Var}(D_f)\le2/K$, so~\eqref{eq:transfer-hypothesis} holds with $\eta=\sqrt{2/K}\le1/8$. By Lemma~\ref{lem:bias} and~\eqref{eq:moment-bounds}, $|\Lambda_{\cos}|=(c_M-a_M)/2\ge5/24$. Take $\xi=1/4$ and $\delta=1/256$, so that $q_0(\xi,\delta)=q_0$. Since $\log(512K)\ge\log(2^{16})>4$, we have $q_0/(2M)\le1/256$, and Lemma~\ref{lem:transcript} shows that every algorithm with at most $q_0$ queries on every input satisfies $|\E F_{\cos}(f,J_{\mathcal A})|\le\xi^2/2+q_0/(2M)+2\delta\le1/32+1/256+1/128=11/256$. If the error were less than $31/1536$, then~\eqref{eq:transfer-conclusion} with $q=q_0$ and $\tau=11/256$ would imply $5/24<1/8+11/256+31/768$, which is false.
\end{proof}

A quantum algorithm for $\Samp_S$ accesses $f$ through the oracle $O_f$ of Section~\ref{sec:known-magnitudes}, and each controlled phase query uses one call to $O_f$.

\begin{theorem}\label{thm:fourier-quantum}
For every nonempty $S$ and every $\varepsilon\in(0,1)$, a quantum algorithm solves $\Samp_S(\varepsilon)$ with $O(\sqrt{M/K}\,(1+\log(1/\varepsilon)))$ queries.
\end{theorem}

\begin{proof}
Use a flag qubit and an outcome register of $m+1$ qubits, in which $|0,s\rangle$ encodes $s\in\{0,1\}^m$ and $|1,0^m\rangle$ encodes $\bot$. Let $V$ be the following circuit, applied to the all-zero state. First rotate the flag to $\alpha_0|0\rangle+\alpha_1|1\rangle$, where $\alpha_0=\sqrt{M/(M+K)}$ and $\alpha_1=\sqrt{K/(M+K)}$. Controlled on the flag being $0$, apply $H^{\otimes m}$ to the last $m$ outcome qubits, a phase query, and $H^{\otimes m}$ again. Controlled on the flag being $1$, flip the first outcome qubit. Since $H^{\otimes m}\sum_\sigma M^{-1/2}f(\sigma)|\sigma\rangle=\sum_s\widehat f(s)|s\rangle$, the circuit $V$ uses one query and prepares
\[
V|0\rangle=\alpha_0|0\rangle\sum_{s\in\{0,1\}^m}\widehat f(s)|0,s\rangle+\alpha_1|1\rangle|1,0^m\rangle.
\]
Let $\Pi$ be the projection onto the span of the states $|0\rangle|0,s\rangle$ with $s\in S$ and the state $|1\rangle|1,0^m\rangle$. Then
\[
\|\Pi V|0\rangle\|^2=\alpha_0^2\sum_{s\in S}\widehat f(s)^2+\alpha_1^2=\frac K{M+K}\,D_f\ge\frac K{2M},
\]
since $D_f\ge1$ and $K\le M$. Measuring the outcome register in the normalized state $\Pi V|0\rangle/\|\Pi V|0\rangle\|$ returns $s\in S$ with probability $\alpha_0^2\widehat f(s)^2/\|\Pi V|0\rangle\|^2=p_{z_f}(s)$ and returns $\bot$ with probability $\alpha_1^2/\|\Pi V|0\rangle\|^2=p_{z_f}(\bot)$. By Lemma~\ref{lem:amplification} with $r=1$ and $p_0=K/(2M)$, this normalized state is prepared with probability at least $1-\varepsilon$ using $O(\sqrt{M/K}\,(1+\log(1/\varepsilon)))$ queries. The algorithm measures the outcome register of the prepared state, and it outputs $\bot$ if the preparation fails. Its output distribution is therefore within total variation distance $\varepsilon$ of $p_{z_f}$.
\end{proof}

The two theorems imply the following lower bound on the separation, as a function of the number of characters.

\begin{corollary}\label{cor:fourier-exponent}
Let $K\ge128$ and $K=M^\theta$ with $0<\theta<1$, and put $Q=M^{(1-\theta)/2}$. At error $1/100$, the quantum query complexity of $\Samp_S$ is $O(Q)$, and its classical query complexity is $\Omega(M/\log M)=\Omega(Q^{2/(1-\theta)}/\log M)$.
\end{corollary}

\begin{proof}
Apply Theorem~\ref{thm:fourier-quantum} with $\varepsilon=1/100$ and Theorem~\ref{thm:fourier-classical}. Since $K\le M$, $\log(512K)\le\log512+\log M$, and $M=Q^{2/(1-\theta)}$.
\end{proof}

For a constant number of characters, the power is two. For $K=\sqrt M$, the power is four; the linear systems of Section~\ref{sec:linear-systems} instantiate this case. For $K=M$, the state $V|0\rangle$ in the proof of Theorem~\ref{thm:fourier-quantum} lies in the range of $\Pi$, so one quantum query suffices. We do not know whether $\Omega(\sqrt{M/K})$ quantum queries are necessary for every set $S$ (Section~\ref{sec:open}).

For $K=M$, classical lower bounds for $\Samp_S$ follow from known results on Fourier sampling, the task of sampling $s\in\{0,1\}^m$ with probability $\widehat f(s)^2$ from query access to $f$. Aaronson and Ambainis proved that Fourier sampling requires $\Omega(M/\log M)$ classical queries \cite{AaronsonAmbainis2015}, and Aaronson and Chen \cite{AaronsonChen2017} and, independently, Montanaro \cite{Montanaro2019} proved that it requires $\Omega(M)$ classical queries at sufficiently small constant errors. These bounds also hold for $\Samp_S$ with $S=\{0,1\}^m$. Indeed, suppose that an algorithm solves $\Samp_S(\varepsilon)$ with $\varepsilon\le1/4$. Run it $T$ times, and return the first outcome different from $\bot$, or a fixed character if all outcomes equal $\bot$. Each run returns an outcome different from $\bot$ with probability at least $1/2-\varepsilon\ge1/4$. Conditioning the output distribution of a run and $p_{z_f}$ on this event multiplies their total variation distance by at most the inverse of the probability of the event under the first distribution. Since $p_{z_f}$ conditioned on this event is the distribution of Fourier sampling, the output of a run, conditioned on the event, is within total variation distance $\varepsilon/(1/2-\varepsilon)\le4\varepsilon$ of this distribution. The algorithm that returns the first such output therefore solves Fourier sampling within total variation distance $4\varepsilon+(3/4)^T$ and makes $T$ times as many queries. For $K=M$, Theorem~\ref{thm:fourier-classical} is thus weaker than the bound of Aaronson and Chen and of Montanaro by a factor $\log M$, and it is new only for $K<M$, the case that the families of Sections~\ref{sec:linear-systems} and~\ref{sec:positive-definite} require.

\section{Linear systems}\label{sec:linear-systems}

This section proves Theorem~\ref{thm:intro-linear}: for a family of linear systems with condition number bounded by a constant, sampling from the solution requires $\Omega(\rho^4/\log\rho)$ classical queries, while $\Theta(\rho)$ quantum queries suffice and are necessary. The family is defined in Section~\ref{sec:nonsymmetric-family}. It is block lower triangular, has a uniform right-hand side, and has two parameters: the parameter $\lambda$, which sets the size of the entries that depend on the unknown signs, and a scale parameter $\alpha$, which, together with an identity block, normalizes the operator norm to one. The solution has a closed form. Its third block is an affine function of the Fourier coefficients $Z_j$ of~\eqref{eq:hidden-notation}, with the linear term $\lambda^2Z_j/\sqrt n$, and its squared norm concentrates around its mean. The theorem is stated in Section~\ref{sec:nonsymmetric-theorem} and proved in three parts. The classical lower bound follows from the transfer theorem applied with the sine test, whose bias is of order $\lambda^2$. For the quantum upper bound, a circuit with one controlled phase query prepares a state whose projection onto a fixed subspace is proportional to the solution and has squared norm at least $1/(16n)$, and amplitude amplification then uses $O(\sqrt n)$ queries. For the quantum lower bound, we replace one row of $R=W/\sqrt n$ by the uniform vector, so that sampling from the solution solves unstructured search.

\subsection{The family and its solution}\label{sec:nonsymmetric-family}

The family consists of the matrices $A_W$ of Section~\ref{sec:overview}, rescaled and extended by an identity block. Fix rational constants $\lambda\in(0,1]$ and $\alpha\in(0,1/(1+8\lambda)]$. For a power of four $n$ and $W\in\{\pm1\}^{n\times n}$, with the notation of Section~\ref{sec:hidden-signs}, define
\begin{equation}\label{eq:nonsymmetric-family}
A_W=\begin{pmatrix}I&0&0\\-\lambda R&I&0\\0&-\lambda H&I\end{pmatrix},\qquad \widehat A_W=(\alpha A_W)\oplus I_n,\qquad \widehat b=\frac12\begin{pmatrix}u\\u\\u\\u\end{pmatrix}.
\end{equation}
The matrices have dimension $N=4n$, and $\widehat b$ is the uniform unit vector of dimension $4n$, the same for every $W$.

The scale $\alpha$ and the block $I_n$ only normalize the family, and the exponents in Theorem~\ref{thm:linear-systems} do not depend on $\alpha$. Together they make $\|\widehat A_W\|=1$ for every $W\in\mathcal G$ (Lemma~\ref{lem:nonsymmetric-parameters}): the block $I_n$ has norm one, and the condition $\alpha\le1/(1+8\lambda)$ ensures that $\|\alpha A_W\|\le1$. Under this normalization, the condition number of $\widehat A_W$ equals $\|\widehat A_W^{-1}\|$, and the solution of $\widehat A_Wx=\widehat b$ satisfies $1\le\|x\|\le\kappa(\widehat A_W)$; these bounds are used in Appendix~\ref{app:output-access}. By Lemma~\ref{lem:nonsymmetric-solution}, the first three blocks of the solution are proportional to $A_W^{-1}(u,u,u)$, and the fourth block does not depend on $W$. A smaller $\alpha$ reduces the share of the fourth block in the squared norm of the solution and therefore enlarges the range of errors in Theorem~\ref{thm:linear-systems}, but it increases the bound on the condition number.

\begin{lemma}\label{lem:nonsymmetric-parameters}
For every $W\in\mathcal G$, $\|\widehat A_W\|=1$, $\kappa(\widehat A_W)\le(1+8\lambda+8\lambda^2)/\alpha$, and $\rho(\widehat A_W)^2=n(1+\alpha^2(3+2\lambda^2))$. For $\alpha=1/(1+8\lambda)$, the condition number is at most $(1+8\lambda)(1+8\lambda+8\lambda^2)\le1+152\lambda$. The entries that depend on $W$ are $-\alpha\lambda W_{ab}/\sqrt n$, in row $n+a$ and column $b$, so the family satisfies the hypotheses of Lemma~\ref{lem:access} with $f(a,b)=W_{ab}$.
\end{lemma}

\begin{proof}
Write $A_W=I-\mathcal N$, where $\mathcal N$ is strictly block lower triangular with the blocks $\lambda R$ and $\lambda H$. For $w=(w_1,w_2,w_3)$, $\|\mathcal Nw\|^2=\lambda^2(\|Rw_1\|^2+\|Hw_2\|^2)$, so $\|\mathcal N\|=\lambda\max\{\|R\|,1\}\le8\lambda$. The matrix $\mathcal N^2$ has the single nonzero block $\lambda^2HR$, so $\|\mathcal N^2\|=\lambda^2\|R\|\le8\lambda^2$, and $\mathcal N^3=0$. Hence $\|A_W\|\le1+8\lambda$ and $A_W^{-1}=I+\mathcal N+\mathcal N^2$ has norm at most $1+8\lambda+8\lambda^2$. Since $\alpha(1+8\lambda)\le1$, the matrix $\widehat A_W$ obtained by adjoining $I_n$ satisfies $\|\widehat A_W\|=1$ and $\|\widehat A_W^{-1}\|\le(1+8\lambda+8\lambda^2)/\alpha$. The squared Frobenius norm is $\alpha^2(3n+\lambda^2\|R\|_F^2+\lambda^2\|H\|_F^2)+n=\alpha^2(3+2\lambda^2)n+n$. Finally, $(1+8\lambda)(1+8\lambda+8\lambda^2)=1+16\lambda+72\lambda^2+64\lambda^3\le1+152\lambda$ for $\lambda\le1$.
\end{proof}

We express the solution in the notation of~\eqref{eq:hidden-notation}: $v=Ru$ is the vector of normalized row sums of $W$, $y=Hv$ is its Walsh--Hadamard transform, with coordinates $y_j=Z_j/\sqrt n$, and $\nu=\|v\|^2$.
\begin{lemma}\label{lem:nonsymmetric-solution}
For every $W$, the solution is $\widehat A_W^{-1}\widehat b=x_W/(2\alpha)$, where
\begin{equation}\label{eq:nonsymmetric-solution}
x_W=\begin{pmatrix}u\\u+\lambda v\\u+\lambda e_0+\lambda^2y\\\alpha u\end{pmatrix}.
\end{equation}
Its squared norm $D_W=\|x_W\|^2$ satisfies $D_W\ge1$ and
\[
D_W=d_n+(\lambda^2+\lambda^4)(\nu-1)+2\lambda(1+\lambda^2)\omega+2\lambda^2v_{0^k},\qquad d_n=3+\alpha^2+2\lambda^2+\lambda^4+\frac{2\lambda}{\sqrt n},
\]
where $\omega=u^{\mathsf T}v$. For uniformly random $W$, $\E D_W=d_n$ and $\E|D_W-d_n|\le C_\lambda/\sqrt n$, where $C_\lambda=\sqrt2(\lambda^2+\lambda^4)+2\lambda(1+\lambda^2)+2\lambda^2$.
\end{lemma}

\begin{proof}
Solving $A_Wz=(u,u,u)$ by forward substitution, we obtain $z_1=u$, $z_2=u+\lambda Ru=u+\lambda v$, and $z_3=u+\lambda H(u+\lambda v)=u+\lambda e_0+\lambda^2y$. The last block of $\widehat A_W^{-1}\widehat b$ is $u/2$, which proves~\eqref{eq:nonsymmetric-solution}. The first block of $x_W$ has unit norm, so $D_W\ge1$. Expanding the block norms and using $\langle u,y\rangle=\langle Hu,v\rangle=v_{0^k}$ and $y_{0^k}=\langle u,v\rangle=\omega$, we obtain the formula for $D_W$. For uniformly random $W$, the coordinates $v_a$ are independent with mean zero and second moment $1/n$, so $\E\omega=\E v_{0^k}=0$, $\E\omega^2=\E v_{0^k}^2=1/n$, and $\operatorname{Var}(\nu)=2(n-1)/n^2$ by~\eqref{eq:nu-moments}. The Cauchy--Schwarz inequality then bounds $\E|D_W-d_n|$ by $C_\lambda/\sqrt n$.
\end{proof}

\subsection{Classical and quantum query complexity}\label{sec:nonsymmetric-theorem}

The following theorem states the classical and quantum query complexities of sampling from the solution. Its range of errors is determined by two quantities: the bias of the sine test, and the probability that the solution assigns to the coordinate of the modified row of $R$ for the inputs of Lemma~\ref{lem:search-instances}.

\begin{theorem}\label{thm:linear-systems}
Fix rational numbers $\lambda\in(0,1]$ and $\alpha\in(0,1/(1+8\lambda)]$, and put
\begin{equation}\label{eq:linear-error-range}
\varepsilon_{\mathrm{lin}}(\lambda,\alpha)=\min\Bigl\{\frac{\lambda^2}{\sqrt e\,(3+\alpha^2+2\lambda^2+\lambda^4)},\ \frac{\lambda^2}{3+\alpha^2+5\lambda^2+2\lambda^4}\Bigr\}.
\end{equation}
Consider the family $\{\widehat A_W:W\in\mathcal G\}$ with right-hand side $\widehat b$, for powers of four $n$. For every constant $\varepsilon\in(0,\varepsilon_{\mathrm{lin}}(\lambda,\alpha))$, there are $c>0$ and $n_0$ such that the following holds for every $n\ge n_0$.
\begin{enumerate}
\item Every classical algorithm that samples from $p_{\widehat A_W^{-1}\widehat b}$ within total variation distance $\varepsilon$ for every $W\in\mathcal G$ makes more than $cn^2/\log n$ queries on some $W\in\mathcal G$, and more than $cn^2/\log n$ queries in expectation on some $W\in\mathcal G$. Querying all $n^2$ unknown entries suffices to sample exactly.
\item A quantum algorithm samples within total variation distance $\varepsilon$ with $O(\sqrt n)$ queries and $\widetilde O(\sqrt n)$ gates, and every quantum algorithm that does so makes $\Omega(\sqrt n)$ queries on some $W\in\mathcal G$.
\end{enumerate}
Since $\rho(\widehat A_W)^2=\Theta(n)$, the classical and quantum query complexities are $\widetilde\Theta(\rho^4)$ and $\Theta(\rho)$, respectively.
\end{theorem}

By Lemma~\ref{lem:solution-access}, the classical lower bound of Theorem~\ref{thm:linear-systems} also applies to procedures that return sampling and query access to an approximate solution, counting the queries used to construct the description and to draw one sample from it.

Theorem~\ref{thm:intro-linear} follows with $\alpha=1/(1+8\lambda)$. Lemma~\ref{lem:nonsymmetric-parameters} bounds the condition number by $1+152\lambda$. For $\lambda,\alpha\le1$, the two denominators in~\eqref{eq:linear-error-range} are at most $7\sqrt e<12$ and $11$, so $\varepsilon_{\mathrm{lin}}(\lambda,\alpha)>\lambda^2/12$. An algorithm with error at most $\varepsilon\le\lambda^2/12$ also has error at most $\lambda^2/12$, so the constants $c$ and $n_0$ of Theorem~\ref{thm:linear-systems} for the error $\lambda^2/12$ apply to every smaller error. Table~\ref{tab:linear-examples} lists four choices of parameters. For each of them, explicit dimensions from which the classical lower bound holds are given in Appendix~\ref{app:regimes}.

\begin{table}[htbp]
\centering
\begin{tabular}{@{}cccc@{}}
\toprule
$\lambda$ & $\alpha$ & Condition number at most & $\varepsilon_{\mathrm{lin}}(\lambda,\alpha)$, rounded down \\
\midrule
$1$ & $1/16$ & $272$ & $0.0999$ \\
$1/4$ & $1/3$ & $21/2$ & $0.0116$ \\
$1/8$ & $1/2$ & $17/4$ & $0.0028$ \\
$1/32$ & $3/4$ & $161/96$ & $1.6\cdot10^{-4}$ \\
\bottomrule
\end{tabular}
\caption{Examples of parameters in Theorem~\ref{thm:linear-systems}. The condition-number bound is $(1+8\lambda+8\lambda^2)/\alpha$.}\label{tab:linear-examples}
\end{table}

\begin{proof}[Proof of the classical lower bound]
By Lemmas~\ref{lem:nonsymmetric-parameters} and~\ref{lem:access}, it suffices to consider algorithms that query the signs $W_{ab}$ directly. We apply Theorem~\ref{thm:transfer} and Corollary~\ref{cor:transfer-asymptotic} with $M=n^2$, $f(a,b)=W_{ab}$, $S=S_{\mathrm{row}}$, and $K=n$. The outcome set is $\mathcal Y=[4n]$, with the third block identified with $S_{\mathrm{row}}$; the promise set is $\mathcal G$, the target is $x_W$, and the test function is the sine test. By Lemma~\ref{lem:nonsymmetric-solution}, the third block of $x_W$ has coordinates
\[
(x_W)_{3,j}=\frac{\gamma_j+\lambda^2Z_j}{\sqrt n},\qquad \gamma_j=1+\lambda\sqrt n\,\mathbf 1[j=0^k].
\]
Take $\beta=\lambda^2$ and $d=d_n$. The third block of $x_W$ is then the vector $\bar x$ of Lemma~\ref{lem:two-conditions}, and $\|\bar x\|^2\le\|x_W\|^2=D_W$. Lemma~\ref{lem:two-conditions}, with $F_\infty=1$, therefore bounds the difference in~\eqref{eq:transfer-hypothesis} by $|D_W-d_n|/d_n$. Since $d_n\ge3$, Lemma~\ref{lem:nonsymmetric-solution} implies $\eta\le\E|D_W-d_n|/d_n\le C_\lambda/(3\sqrt n)$. By Lemma~\ref{lem:bias}, the bias of the sine test is
\[
\Lambda_{\sin}=\frac{2b_M}{d_nn}\sum_j\lambda^2\gamma_j=\frac{2\lambda^2b_M(1+\lambda/\sqrt n)}{d_n}\longrightarrow\frac{2\lambda^2e^{-1/2}}{3+\alpha^2+2\lambda^2+\lambda^4},
\]
where the limit follows from Lemma~\ref{lem:moments}.
Since $\|\gamma\|^2=n+2\lambda\sqrt n+\lambda^2n\le4n$, we have $\Phi_{\max}\le4(4n+n^2)/(3n)=O(n)$, and $\delta_{\mathcal G}\le\delta_n$ by Lemma~\ref{lem:spectral-tail}, so $(2+\Phi_{\max})\delta_{\mathcal G}\to0$. Every $\varepsilon<\varepsilon_{\mathrm{lin}}(\lambda,\alpha)$ satisfies $2\varepsilon<\lim|\Lambda_{\sin}|$. Since $\eta\le C_\lambda/(3\sqrt n)\to0$, we can choose $\eta_\infty>0$ with $2\varepsilon<\lim|\Lambda_{\sin}|-\eta_\infty$, and then $\eta\le\eta_\infty$ for all sufficiently large $n$. By Corollary~\ref{cor:transfer-asymptotic}, every classical algorithm with error $\varepsilon$ therefore makes $\Omega(M/\log(2K))=\Omega(n^2/\log n)$ queries. For the upper bound, querying all $n^2$ unknown entries determines $x_W$, after which exact sampling requires no further queries.
\end{proof}

\begin{proof}[Proof of the quantum upper bound]
The algorithm prepares $x_W/\|x_W\|$ by combining a part that does not depend on $W$ with a part that uses one controlled phase query to the Boolean function $f(a,b)=W_{ab}$. By Lemma~\ref{lem:access}, this query costs two reversible entry queries.

Start with uniform superpositions over $a\in\{0,1\}^k$ and $b\in\{0,1\}^k$, apply the phases $W_{ab}$ with one controlled phase query, and apply $H$ to the second register. Since $v_a=n^{-1}\sum_bW_{ab}$, the component in which the second register is $|0^k\rangle$ is $n^{-1/2}\sum_av_a|a\rangle|0^k\rangle$. Prepare in addition a block register in the state $(\lambda|2\rangle+\lambda^2|3\rangle)/\sqrt{\lambda^2+\lambda^4}$, and apply $H$ to the first register only in block $3$. The component in which the second register is $|0^k\rangle$ is then
\[
\frac{r_W}{\sqrt{(\lambda^2+\lambda^4)n}},\qquad r_W=\begin{pmatrix}0\\\lambda v\\\lambda^2y\\0\end{pmatrix}.
\]
The remaining part of the solution, $z_0=(u,u,u+\lambda e_0,\alpha u)$, does not depend on $W$. Its normalized version can be prepared with $O(\log n)$ gates apart from fixed rotations, because its entries are constant on each block apart from one known coordinate. Put $L_n=\|z_0\|+\sqrt{(\lambda^2+\lambda^4)n}$. The two parts are combined by a linear combination of two state preparations. Prepare a flag qubit with amplitudes $\sqrt{\|z_0\|/L_n}$ and $\sqrt{\sqrt{(\lambda^2+\lambda^4)n}/L_n}$. Controlled on the flag, prepare either $z_0/\|z_0\|$ with the second register in $|0^k\rangle$ or the state above, and then undo the preparation of the flag. The component in which the flag and the second register are both zero is $(z_0+r_W)/L_n=x_W/L_n$.

Its squared norm is $D_W/L_n^2\ge1/L_n^2\ge1/(16n)$. Indeed, $\|z_0\|^2=3+\alpha^2+\lambda^2+2\lambda/\sqrt n\le6$ and $\lambda^2+\lambda^4\le2$, so $L_n^2\le(\sqrt6+\sqrt{2n})^2\le16n$ for $n\ge4$. By Lemma~\ref{lem:amplification} with $p_0=1/(16n)$, the normalized state $x_W/\|x_W\|$ is prepared with probability at least $1-\varepsilon/2$ using $O(\sqrt n\,(1+\log(1/\varepsilon)))$ queries. Measuring this state produces a sample from $p_{x_W}$, and on failure the algorithm outputs a fixed index. The resulting circuit has $N_{\mathrm{rot}}=\widetilde O(\sqrt n)$ fixed rotations. If each of them is synthesized with operator-norm error at most $\varepsilon/(4N_{\mathrm{rot}})$, the output distribution changes by at most $\varepsilon/2$ in total variation distance, so the total error is at most $\varepsilon$. This synthesis costs $O(\log(N_{\mathrm{rot}}/\varepsilon))$ gates per rotation \cite{RossSelinger2016}, so the gate count is $\widetilde O(\sqrt n)$.
\end{proof}

\begin{proof}[Proof of the quantum lower bound]
We apply Lemma~\ref{lem:search} to the inputs $W^{(\emptyset)}$ and $W^{(a)}$ of Lemma~\ref{lem:search-instances}, with $j(a)$ equal to coordinate $a$ of the second block. By Lemmas~\ref{lem:nonsymmetric-parameters} and~\ref{lem:search-instances}, every coherent input operation on these inputs uses $O(1)$ queries to the search string. For $W^{(a)}$, $v=e_0+e_a$, and the second block of $x_W$ has the coordinate $(x_W)_{2,a}=n^{-1/2}+\lambda$. Expanding the block norms, we find $D_W=3+\alpha^2+5\lambda^2+2\lambda^4+(6\lambda+4\lambda^3)/\sqrt n$. The coordinate therefore has probability
\[
\frac{(\lambda+n^{-1/2})^2}{3+\alpha^2+5\lambda^2+2\lambda^4+(6\lambda+4\lambda^3)n^{-1/2}}\ge\frac{\lambda^2}{3+\alpha^2+5\lambda^2+2\lambda^4},
\]
because the left side increases with $n^{-1/2}\ge0$: its derivative has the sign of $2(3+\alpha^2+5\lambda^2+2\lambda^4)-(6\lambda+4\lambda^3)\lambda+(6\lambda+4\lambda^3)n^{-1/2}>0$. Every $\varepsilon<\varepsilon_{\mathrm{lin}}(\lambda,\alpha)$ is smaller than this probability, and the lower bound $\Omega(\sqrt n)$ follows from Lemma~\ref{lem:search}.
\end{proof}

\begin{remark}\label{rem:padding}
The lower bounds do not require the dimension to be $\Theta(\rho^2)$. For $N'\ge4n$, pad $\widehat A_W$ with zero rows and columns to an $N'\times N'$ matrix, and pad $\widehat b$ with zeros. The minimum-norm least-squares solution is $\widehat A_W^{-1}\widehat b$ followed by zeros, so the output distribution is unchanged. The padded matrix has the same nonzero singular values and the same $\rho$, and the ratio of its largest and smallest nonzero singular values is unchanged. The padded entries are zero and known, zero rows and columns have sampling probability zero, and conditional sampling in a zero row or column returns a fixed index. Every lower bound of this section therefore holds for least-squares instances of every dimension $N'\ge4n$. The same padding applies to the tasks of Section~\ref{sec:positive-definite}, with the inverse and the inverse square root replaced by the corresponding pseudo-inverses.
\end{remark}

\section{Positive-definite families}\label{sec:positive-definite}

This section proves the lower bounds for the tasks in Table~\ref{tab:results} other than linear systems and Fourier sampling on a set of characters: positive-definite linear systems and least squares; analytic matrix functions, including ridge regression, matrix powers, square roots, and exponentials; Hamiltonian evolution; and sampling from the top principal component and from a fixed row of a rank-one truncation. For each of these tasks and every fixed value of the parameter $t$ in the range of the corresponding theorem, sampling from the output with an error below a constant multiple of $t^4$, or of $t^2$ for Hamiltonian evolution, requires $\Omega(\rho^4/\log\rho)$ classical queries, while $\widetilde\Theta(\rho)$ quantum queries suffice and are necessary (for a matrix function $g$, when $g'(1)\ne0$). Several of these tasks, such as positive-definite systems, square roots, and Hamiltonian evolution, require a symmetric or positive-definite matrix, so we use a symmetric family for all of them.

The tasks use matrices built from a symmetric matrix $T_W$ that contains the blocks $R$, $R^{\mathsf T}$, and $H$ of Section~\ref{sec:hidden-signs}. Most of them use the positive-definite matrices $P_W=I-tT_W$, defined in Section~\ref{sec:pd-family}, where we also solve the linear system with right-hand side $b=(u,0,0)$ exactly. The third block of the solution equals $t^2y$ up to a relative error of order $t^2$, where, as in~\eqref{eq:hidden-notation}, $y=Hv$ has the coordinates $y_j=Z_j/\sqrt n$. The classical lower bounds are proved as in Theorem~\ref{thm:linear-systems}, with $M=n^2$, $f(a,b)=W_{ab}$, $S=S_{\mathrm{row}}$, and $K=n$. An expansion of the output vector shows that its third block is close to an affine function of the Fourier coefficients, and Lemma~\ref{lem:two-conditions} bounds the error $\eta$ in the hypothesis of the transfer theorem by the error of this approximation and by the deviation of the squared norm of the output from a fixed value. The quantum upper bounds follow from Lemma~\ref{lem:qsvt}, and the quantum lower bounds follow from Lemma~\ref{lem:search} with the sign matrices of Lemma~\ref{lem:search-instances}. By Lemma~\ref{lem:solution-access}, every classical lower bound of this section also applies to procedures that return sampling and query access to an approximate output vector.

Linear systems and least squares are treated in Section~\ref{sec:pd-systems}, and analytic matrix functions, including ridge regression, in Section~\ref{sec:functions}. For these tasks, the third block of the output has no component that is independent of the input, so we use the cosine test, whose bias is of order $t^4$. For Hamiltonian evolution, treated in Section~\ref{sec:hamiltonian}, we choose an initial vector for which the third block of the evolved vector contains a component that does not depend on the input. We then use the sine test, whose bias is of order $t^2$, so the lower bound holds for errors of order $t^2$. Finally, the lower bounds for the top principal component of a rank-one perturbation of the family, and for a fixed row of its rank-one truncation, are proved in Section~\ref{sec:pca}.

\subsection{The family and its solution}\label{sec:pd-family}

We first define the positive-definite family and solve its linear system exactly. Fix a rational number $t\in(0,1/64]$. For a power of four $n$ and $W\in\{\pm1\}^{n\times n}$, define the $3n\times3n$ matrices
\begin{equation}\label{eq:pd-family}
T_W=\begin{pmatrix}0&R^{\mathsf T}&0\\R&0&H\\0&H&0\end{pmatrix},\qquad P_W=I-tT_W,\qquad b=\begin{pmatrix}u\\0\\0\end{pmatrix}.
\end{equation}
The right-hand side $b$ is a fixed unit vector, the same for every $W$. The matrix $T_W$ is a symmetric analogue of the block structure of Section~\ref{sec:linear-systems}: it maps $b$ to $(0,v,0)$, and it maps $(0,v,0)$ to $(R^{\mathsf T}v,0,y)$. In the expansion $P_W^{-1}b=\sum_{k\ge0}t^kT_W^kb$, the term of order two therefore places $t^2y$ in the third block. The parameter $t$ is small, so that $P_W$ is well conditioned and the expansion converges (Lemma~\ref{lem:pd-parameters}).

\begin{lemma}\label{lem:pd-parameters}
For every $W\in\mathcal G$, $\|T_W\|^2=\|RR^{\mathsf T}+I\|\le65$. Consequently, $(1-9t)I\preceq P_W\preceq(1+9t)I$, so $P_W$ is positive definite with $\kappa(P_W)\le(1+9t)/(1-9t)\le73/55$. Moreover, $\|P_W\|_F^2=n(3+4t^2)$, so $\rho(P_W)=\Theta(\sqrt n)$. The entries that depend on $W$ are $-tW_{ab}/\sqrt n$, in two symmetric positions for each $(a,b)$, so the family satisfies the hypotheses of Lemma~\ref{lem:access} with $f(a,b)=W_{ab}$.
\end{lemma}

\begin{proof}
List the first and third blocks before the second. The matrix $T_W$ then has the off-diagonal blocks $B=(R^{\mathsf T};H)$ and $B^{\mathsf T}$, so $\|T_W\|^2=\|B^{\mathsf T}B\|=\|RR^{\mathsf T}+H^2\|=\|RR^{\mathsf T}+I\|\le65<81$. The diagonal blocks of $P_W$ contribute $3n$ to its squared Frobenius norm, and the four off-diagonal blocks contribute $t^2n$ each.
\end{proof}

\begin{lemma}\label{lem:pd-solution}
Put $C=RR^{\mathsf T}+I$, $\Gamma=I-t^2C$, $\theta_t=65t^2$, and $\tilde\theta_t=\theta_t/(1-\theta_t)$. For every $W\in\mathcal G$, the solution $x=P_W^{-1}b$ has blocks
\[
x_1=u+t^2R^{\mathsf T}\Gamma^{-1}v,\qquad x_2=t\Gamma^{-1}v,\qquad x_3=t^2H\Gamma^{-1}v.
\]
Its squared norm $D=\|x\|^2$ satisfies $D=1+t^2v^{\mathsf T}\Gamma^{-1}v+2t^2v^{\mathsf T}\Gamma^{-2}v$ and $1\le D\le1+4t^2\nu$. Moreover, $\|x_3-t^2y\|\le t^2\tilde\theta_t\sqrt\nu$ and $\|x_3\|\le t^2(1+\tilde\theta_t)\sqrt\nu$.
\end{lemma}

\begin{proof}
The first and third block equations are $x_1=u+tR^{\mathsf T}x_2$ and $x_3=tHx_2$. Substituting them into the second block equation and using $H^2=I$, we obtain $\Gamma x_2=tv$. Since $I\preceq C\preceq65I$, we have $(1-\theta_t)I\preceq\Gamma\preceq I$, which proves the formulas for the blocks and the bounds on $x_3$. Expanding the three squared block norms, we obtain
\[
D=1+2t^2v^{\mathsf T}\Gamma^{-1}v+t^2v^{\mathsf T}\Gamma^{-2}v+t^4v^{\mathsf T}C\Gamma^{-2}v,
\]
and substituting the identity $t^2C\Gamma^{-2}=\Gamma^{-2}-\Gamma^{-1}$ in the last term proves the stated expression. The upper bound follows from $(1-\theta_t)^{-1}+2(1-\theta_t)^{-2}<4$ for $t\le1/64$.
\end{proof}

\subsection{Positive-definite linear systems and least squares}\label{sec:pd-systems}

We first consider the linear system $P_Wx=b$. By Lemma~\ref{lem:pd-solution}, the third block of its solution is close to $t^2y$ and has no component that does not depend on $W$, so we use the cosine test.

\begin{theorem}\label{thm:pd-systems}
Fix a rational number $t\in(0,1/64]$. For every constant $\varepsilon\in(0,t^4/4)$, there are $c>0$ and $n_0$ such that, for every power of four $n\ge n_0$, every classical algorithm that samples from $p_{P_W^{-1}b}$ within total variation distance $\varepsilon$ for every $W\in\mathcal G$ makes more than $cn^2/\log n$ queries on some $W\in\mathcal G$, and more than $cn^2/\log n$ queries in expectation on some $W\in\mathcal G$. The quantum query complexity of this task is $\widetilde\Theta(\sqrt n)$, and the quantum algorithm uses $\widetilde O(\sqrt n)$ gates.
\end{theorem}

\begin{proof}
For the classical lower bound, apply Theorem~\ref{thm:transfer} and Corollary~\ref{cor:transfer-asymptotic} as in the proof of Theorem~\ref{thm:linear-systems}, now with the cosine test, $\gamma=0$, $\beta=t^2$, and $d=1$. The vector $\bar x$ of Lemma~\ref{lem:two-conditions} is then $t^2y$, and $\|t^2y\|^2=t^4\nu$. On $\mathcal G$, Lemma~\ref{lem:pd-solution} implies $\|x_3-t^2y\|(\|x_3\|+t^2\|y\|)\le t^4\tilde\theta_t(2+\tilde\theta_t)\nu$ and $1\le D\le1+4t^2\nu$. Lemma~\ref{lem:two-conditions}, with $F_\infty=2$, therefore bounds the difference in~\eqref{eq:transfer-hypothesis} by $2t^4\tilde\theta_t(2+\tilde\theta_t)\nu+8t^6\nu^2$, and by~\eqref{eq:nu-moments}, $\eta\le t^4(2\tilde\theta_t(2+\tilde\theta_t)+24t^2)$. For $t\le1/64$, $\tilde\theta_t\le65/4031$, so $\eta<0.1\,t^4$. By Lemma~\ref{lem:bias}, $\Lambda_{\cos}=t^4(a_M-c_M)$, which tends to $-t^4e^{-1/2}$ by Lemma~\ref{lem:moments}. Since $e^{-1/2}-0.1>1/2$, every $\varepsilon<t^4/4$ satisfies $2\varepsilon<|\Lambda_\infty|-\eta$. Finally, $\Phi_{\max}=4t^4n$, so $(2+\Phi_{\max})\delta_n\to0$.

For the quantum upper bound, apply Lemma~\ref{lem:qsvt} with $g(s)=1/s$, the interval $[1-9t,1+9t]$, and the vector $b$; the solution satisfies $\|P_W^{-1}b\|\ge1$ by Lemma~\ref{lem:pd-solution}. For the quantum lower bound, use the inputs $W^{(\emptyset)}$ and $W^{(a)}$ of Lemma~\ref{lem:search-instances}. For $W^{(a)}$, $v=e_0+e_a$ and $RR^{\mathsf T}v=2v$, so $Cv=3v$ and $x_2=tv/(1-3t^2)$. The coordinate $a$ of the second block has probability at least $t^2/D\ge t^2/(1+8t^2)\ge t^2/2$, which is larger than every $\varepsilon<t^4/4$. Lemma~\ref{lem:search} completes the proof.
\end{proof}

Since $P_W$ is invertible, the least-squares problem $\min_z\|P_Wz-b\|$ has the minimizer $P_W^{-1}b$, so the bounds of Theorem~\ref{thm:pd-systems} apply to least squares. Appending a zero row to $P_W$ gives an inconsistent problem with the same minimizer.

\begin{corollary}\label{cor:least-squares}
The bounds of Theorem~\ref{thm:pd-systems} hold for sampling from the minimizer of each of the following least-squares problems.
\begin{enumerate}
\item The consistent problem $\min_z\|P_Wz-b\|$, whose minimizer is $P_W^{-1}b$.
\item The inconsistent problem $\min_z\|\widetilde P_Wz-\widetilde b\|$, where $\widetilde P_W\in\mathbb R^{(3n+1)\times3n}$ is $P_W$ with an appended zero row and $\widetilde b=(b,1)$. Its minimizer is $P_W^{-1}b$, its optimal residual has norm one, and the projection of $\widetilde b/\|\widetilde b\|$ onto the column space of $\widetilde P_W$ has norm $1/\sqrt2$.
\end{enumerate}
\end{corollary}

\begin{proof}
The first statement is immediate. For the second, $\|\widetilde P_Wz-\widetilde b\|^2=\|P_Wz-b\|^2+1$, so the minimizer and the optimal residual are as stated. The nonzero singular values of $\widetilde P_W$ are those of $P_W$, so $\rho$ and the ratio of the extreme nonzero singular values are unchanged. The appended row has norm zero, so row sampling never returns it, and conditional sampling in it returns a fixed index. All other input operations are those of $P_W$.
\end{proof}

\subsection{Matrix functions and ridge regression}\label{sec:functions}

The lower bounds for the family~\eqref{eq:pd-family} extend to sampling from $g(P_W)b$, for every analytic function $g$ with $g(1)g''(1)\ne0$ and every sufficiently small $t$. Write $g(1+z)=\sum_{k\ge0}a_kz^k$, so that $g(P_W)b=\sum_{k\ge0}a_k(-t)^kT_W^kb$ when the series converges. The terms of even order are supported on the first and third blocks, and the terms of odd order on the second block (see the proof of Theorem~\ref{thm:functions}). The first three coefficients have the following roles. The term of order zero is $a_0b$, so $a_0=g(1)\ne0$ bounds the norm of the output from below. The term of order two contributes $a_2t^2y$ to the third block, so the third block is close to an affine function of the Fourier coefficients with $\beta=a_2t^2$, and $a_2=g''(1)/2\ne0$ makes the bias of the cosine test nonzero. The term of order one contributes $-a_1tv$ to the second block, and the quantum lower bound requires $a_1=g'(1)\ne0$.

To bound the remaining terms, let $g$ be analytic on the disk $\{z\in\mathbb C:|z-1|<r_g\}$ for some $r_g>0$. Fix $t_*\in(0,1/64]$ with $9t_*<r_g$, and define
\begin{align}
A_g&=\sum_{k\ge1}|a_{2k}|(65t_*^2)^{k-1},& B_g&=\sum_{k\ge2}|a_{2k}|65^{k-1}t_*^{2k-4},& L_g&=\sum_{k\ge0}|a_{2k+1}|(65t_*^2)^k,\nonumber\\
L_g'&=\sum_{k\ge1}|a_{2k+1}|3^kt_*^{2k-2},& K_g&=2|a_0|A_g+65t_*^2A_g^2+L_g^2,& c_g&=\frac{|a_2|^2}{|a_0|^2},\label{eq:function-constants}
\end{align}
and $M_g=4B_g(A_g+|a_2|)/|a_0|^2+12|a_2|^2K_g/|a_0|^4$. These series converge because $\sqrt{65}\,t_*<9t_*<r_g$. In the proof of Theorem~\ref{thm:functions}, $A_g$, $B_g$, and $L_g$ bound the terms of even order, the terms of even order beyond order two, and the terms of odd order, respectively. The constant $K_g$ bounds the deviation of the squared norm of the output from $|a_0|^2$, $M_g$ bounds the error $\eta$ in the hypothesis of the transfer theorem, $L_g'$ is used in the quantum lower bound, and $c_g$ is the factor in the bias of the cosine test.

\begin{theorem}\label{thm:functions}
Let $g$ be as above with $a_0\ne0$ and $a_2\ne0$, and let $t\in(0,t_*]$ be a rational number that satisfies
\begin{equation}\label{eq:function-conditions}
128K_gt^2\le|a_0|^2,\qquad M_gt^2\le\frac{c_g}{12}.
\end{equation}
For every constant $\varepsilon\in(0,c_gt^4/4)$, there are $c>0$ and $n_0$ such that, for every power of four $n\ge n_0$, every classical algorithm that samples from $p_{g(P_W)b}$ within total variation distance $\varepsilon$ for every $W\in\mathcal G$ makes more than $cn^2/\log n$ queries on some $W\in\mathcal G$, and more than $cn^2/\log n$ queries in expectation on some $W\in\mathcal G$. A quantum algorithm solves the task with $\widetilde O(\sqrt n)$ queries and gates. If moreover $a_1\ne0$ and
\begin{equation}\label{eq:function-search-conditions}
L_g't^2\le\frac{|a_1|}2,\qquad t^2\le\frac{2|a_1|^2}{5|a_2|^2},
\end{equation}
then every quantum algorithm for the task makes $\Omega(\sqrt n)$ queries.
\end{theorem}

\begin{proof}
For $W\in\mathcal G$, $\|tT_W\|\le9t<r_g$, so $g(P_W)=\sum_ka_k(-tT_W)^k$. Since $T_W(u,0,0)=(0,v,0)$, $T_W(0,w,0)=(R^{\mathsf T}w,0,Hw)$, and $T_W(p,0,r)=(0,Rp+Hr,0)$, induction on $k$ shows that, for $k\ge1$,
\[
T_W^{2k}b=\begin{pmatrix}R^{\mathsf T}C^{k-1}v\\0\\HC^{k-1}v\end{pmatrix},\qquad T_W^{2k-1}b=\begin{pmatrix}0\\C^{k-1}v\\0\end{pmatrix},
\]
with $C=RR^{\mathsf T}+I$. Hence $x=g(P_W)b$ has the blocks $a_0u+R^{\mathsf T}Ev$, $Ov$, and $HEv$, where $E=\sum_{k\ge1}a_{2k}t^{2k}C^{k-1}$ and $O=-\sum_{k\ge0}a_{2k+1}t^{2k+1}C^k$. Since $I\preceq C\preceq65I$ and $t\le t_*$,
\[
\|E\|\le A_gt^2,\qquad \|E-a_2t^2I\|\le B_gt^4,\qquad \|O\|\le L_gt.
\]
Expanding the block norms and using $\langle u,R^{\mathsf T}Ev\rangle=\langle v,Ev\rangle$ and $\|R\|\le8$, we obtain $|\|x\|^2-|a_0|^2|\le K_gt^2\nu$. Since $\nu\le\|R\|^2\le64$, the first condition in~\eqref{eq:function-conditions} implies $\|x\|^2\ge|a_0|^2/2$. The third block satisfies $\|x_3-a_2t^2y\|\le B_gt^4\sqrt\nu$ and $\|x_3\|\le A_gt^2\sqrt\nu$.

Apply Theorem~\ref{thm:transfer} and Corollary~\ref{cor:transfer-asymptotic} with the cosine test, $\gamma=0$, $\beta=a_2t^2$, and $d=|a_0|^2$. The vector $\bar x$ of Lemma~\ref{lem:two-conditions} is then $a_2t^2y$. On $\mathcal G$, we apply Lemma~\ref{lem:two-conditions} with $F_\infty=2$, the bounds above, and $\|x\|^2\ge|a_0|^2/2$. The difference in~\eqref{eq:transfer-hypothesis} is then at most
\[
\frac{4B_g(A_g+|a_2|)}{|a_0|^2}t^6\nu+\frac{4|a_2|^2K_g}{|a_0|^4}t^6\nu^2.
\]
By~\eqref{eq:nu-moments}, $\eta\le M_gt^6\le c_gt^4/12$. By Lemma~\ref{lem:bias}, $\Lambda_{\cos}=c_gt^4(a_M-c_M)$, which tends to $-c_gt^4e^{-1/2}$ by Lemma~\ref{lem:moments}. Since $e^{-1/2}-1/12>1/2$, every $\varepsilon<c_gt^4/4$ satisfies $2\varepsilon<|\Lambda_\infty|-\eta$. Since $\Phi_{\max}=4c_gt^4n$, the classical lower bound follows from Corollary~\ref{cor:transfer-asymptotic}, and reading all unknown entries shows that $n^2$ queries suffice.

For the quantum upper bound, Lemma~\ref{lem:qsvt} applies, because $g$ is analytic on a neighborhood of $[1-9t,1+9t]$ and $\|g(P_W)b\|\ge|a_0|/\sqrt2$. For the quantum lower bound, use the matrices $W^{(a)}$ of Lemma~\ref{lem:search-instances}, for which $\nu=2$ and $Cv=3v$. The second block of $x$ is then $o_tv$ with $o_t=-\sum_{k\ge0}a_{2k+1}t^{2k+1}3^k$, so the first condition in~\eqref{eq:function-search-conditions} implies $|o_t|\ge|a_1|t-L_g't^3\ge|a_1|t/2$. Also $\|x\|^2\le|a_0|^2+2K_gt^2\le\frac{65}{64}|a_0|^2$. The coordinate $a$ of the second block therefore has probability at least $\frac{16}{65}|a_1|^2t^2/|a_0|^2\ge|a_1|^2t^2/(5|a_0|^2)$. By the second condition in~\eqref{eq:function-search-conditions}, every $\varepsilon<c_gt^4/4$ is at most half of this probability, and Lemma~\ref{lem:search} completes the proof.
\end{proof}

Table~\ref{tab:functions} lists the consequences of Theorem~\ref{thm:functions} for standard functions. Ridge regression with regularization $\mu>0$ minimizes $\|P_Wz-b\|^2+\mu\|z\|^2$. Since $P_W$ is symmetric, its minimizer is $(P_W^2+\mu I)^{-1}P_Wb=g(P_W)b$ with $g(s)=s/(s^2+\mu)$. As in the other tasks, the algorithm has access to $P_W$, not to $P_W^2$. The table uses $\mu=1/2$, which is comparable to the squared singular values of $P_W$. For the exponentials, the table uses evolution time one; other fixed nonzero times and the powers $s^p$ with fixed $p\ge2$ satisfy the hypotheses of Theorem~\ref{thm:functions} for all sufficiently small $t$.

\begin{table}[htbp]
\centering
\begin{tabular}{@{}lllccc@{}}
\toprule
$g(s)$ & Task & $(a_0,a_1,a_2)$ & $c_g$ & \begin{tabular}[b]{@{}c@{}}Largest\\dyadic $t$\end{tabular} & \begin{tabular}[b]{@{}c@{}}$c_gt^4/4$,\\rounded down\end{tabular} \\
\midrule
$s^{-1}$ & Linear system & $(1,-1,1)$ & $1$ & $2^{-7}$ & $9.3\cdot10^{-10}$ \\
$s^2$ & Matrix power & $(1,2,1)$ & $1$ & $2^{-6}$ & $1.4\cdot10^{-8}$ \\
$s^3$ & Matrix power & $(1,3,3)$ & $9$ & $2^{-6}$ & $1.3\cdot10^{-7}$ \\
$s/(s^2+\frac12)$ & Ridge regression & $(\frac23,-\frac29,-\frac4{27})$ & $\frac4{81}$ & $2^{-7}$ & $4.5\cdot10^{-11}$ \\
$e^{-s}$ & Imaginary-time evolution & $e^{-1}(1,-1,\frac12)$ & $\frac14$ & $2^{-6}$ & $3.7\cdot10^{-9}$ \\
$e^{-is}$ & Real-time evolution & $e^{-i}(1,-i,-\frac12)$ & $\frac14$ & $2^{-6}$ & $3.7\cdot10^{-9}$ \\
$s^{1/2}$ & Square root & $(1,\frac12,-\frac18)$ & $\frac1{64}$ & $2^{-6}$ & $2.3\cdot10^{-10}$ \\
$s^{-1/2}$ & Inverse square root & $(1,-\frac12,\frac38)$ & $\frac9{64}$ & $2^{-7}$ & $1.3\cdot10^{-10}$ \\
\bottomrule
\end{tabular}
\caption{Applications of Theorem~\ref{thm:functions} with $t_*=1/64$. The fifth column is the largest $t=2^{-j}\le1/64$ that satisfies~\eqref{eq:function-conditions} and~\eqref{eq:function-search-conditions}, and the last column is the resulting range of errors. For the linear system, Theorem~\ref{thm:pd-systems} covers the larger range of errors below $t^4/4$ with $t=1/64$.}\label{tab:functions}
\end{table}

\subsection{Hamiltonian evolution}\label{sec:hamiltonian}

Theorem~\ref{thm:functions} covers evolution under $P_W$ for a fixed time at errors of order $t^4$. For Hamiltonian evolution, the range of errors improves to order $t^2$ when the initial vector is $\psi_0=(u,0,u)/\sqrt2$. For this initial vector, the third block of the evolved vector contains a component of norm close to $1/\sqrt2$ that does not depend on the input, so the bias of the sine test is of order $t^2$. In the proof of Theorem~\ref{thm:hamiltonian}, we expand the evolved vector to second order in $t$ and bound the remainder using $\|T_W\|\le9$.

\begin{theorem}\label{thm:hamiltonian}
Fix a rational number $t\in(0,1/64]$, let $P_W=P_W(t)$ be as in~\eqref{eq:pd-family}, and let $\psi_0=(u,0,u)/\sqrt2$. For every constant $\varepsilon\in(0,t^2/10)$, there are $c>0$ and $n_0$ such that, for every power of four $n\ge n_0$, every classical algorithm that samples from $p_{e^{-iP_W}\psi_0}$ within total variation distance $\varepsilon$ for every $W\in\mathcal G$ makes more than $cn^2/\log n$ queries on some $W\in\mathcal G$, and more than $cn^2/\log n$ queries in expectation on some $W\in\mathcal G$. The quantum query complexity of this task is $\widetilde\Theta(\sqrt n)$.
\end{theorem}

\begin{proof}
Since $e^{-iP_W}=e^{-i}e^{itT_W}$, the vector $\psi=e^{itT_W}\psi_0$ has the same output distribution. The matrix $T_W$ maps vectors supported on the first and third blocks to vectors supported on the second block, and conversely. The terms of even order in $\psi=\sum_k(it)^kT_W^k\psi_0/k!$ are therefore supported on the first and third blocks, and the terms of odd order on the second block. Since $T_W\psi_0=(0,v+e_0,0)/\sqrt2$ and $T_W^2\psi_0$ has third block $(y+u)/\sqrt2$, the third block of $\psi$ is $\psi_3=\bar\psi_3+r_3$ with
\[
\bar\psi_3=\frac1{\sqrt2}\Bigl(\Bigl(1-\frac{t^2}2\Bigr)u-\frac{t^2}2y\Bigr),\qquad \|r_3\|\le\sum_{k\ge4\text{ even}}\frac{(9t)^k}{k!}\le\frac{(9t)^4}{24}\cosh(9t)=:\varrho_t,
\]
on $\mathcal G$, where we used $\|T_W\|\le9$.

Apply Theorem~\ref{thm:transfer} and Corollary~\ref{cor:transfer-asymptotic} with the sine test, $\gamma_j=(1-t^2/2)/\sqrt2$, $\beta=-t^2/(2\sqrt2)$, and $d=1$. Since $(\bar\psi_3)_j=(\gamma_j+\beta Z_j)/\sqrt n$, the vector $\bar x$ of Lemma~\ref{lem:two-conditions} is $\bar\psi_3$, and since $\|\psi\|=1=d$, the second term in that lemma vanishes. On $\mathcal G$, $\|y\|=\|v\|\le8$, so $\|\bar\psi_3\|\le(1+7t^2/2)/\sqrt2$, and Lemma~\ref{lem:two-conditions}, with $F_\infty=1$, bounds the difference in~\eqref{eq:transfer-hypothesis} by
\[
\|r_3\|\bigl(\|\psi_3\|+\|\bar\psi_3\|\bigr)\le\|r_3\|(2\|\bar\psi_3\|+\|r_3\|)\le\varrho_t\bigl(\sqrt2(1+7t^2/2)+\varrho_t\bigr)=:\eta_t.
\]
For $t\le1/64$, $\varrho_t\le0.0675\,t^2$ and $\eta_t\le0.0956\,t^2$. By Lemma~\ref{lem:bias}, $\Lambda_{\sin}=-b_Mt^2(1-t^2/2)/2$, which tends to $-e^{-1/2}t^2(1-t^2/2)/2$ by Lemma~\ref{lem:moments}; its absolute value is at least $0.3032\,t^2$. Every $\varepsilon<t^2/10$ therefore satisfies $2\varepsilon<|\Lambda_\infty|-\eta_t$. Since $\Phi_{\max}=O(n)$, the classical lower bound follows from Corollary~\ref{cor:transfer-asymptotic}.

For the quantum upper bound, apply Lemma~\ref{lem:qsvt} with the entire function $g(s)=e^{-is}$ and the vector $\psi_0$, for which $\|g(P_W)\psi_0\|=1$. For the quantum lower bound, use the matrices $W^{(a)}$ of Lemma~\ref{lem:search-instances}, for which $v=e_0+e_a$. The second block of $\psi$ is $it(2e_0+e_a)/\sqrt2+r_2$ with $\|r_2\|\le(9t)^3\cosh(9t)/6\le0.031\,t$. The coordinate $a$ of the second block therefore has probability at least $(t/\sqrt2-0.031\,t)^2\ge t^2/4$, which exceeds every $\varepsilon<t^2/10$, and Lemma~\ref{lem:search} completes the proof.
\end{proof}

\subsection{Principal components}\label{sec:pca}

We now consider two spectral tasks on a symmetric matrix: sampling from its top principal component, and sampling from a fixed row of its rank-one spectral truncation, which keeps only the top eigenvalue and its eigenvector. Both are instances of singular value transformation by a function that removes the eigenvalues below a threshold. The first applies the corresponding projector to a fixed vector that overlaps the top eigenvector, and the second applies the truncation to a standard basis vector. Quantum and dequantized algorithms for singular value transformation apply to both tasks (Lemma~\ref{lem:qsvt} and \cite{BakshiTang2024}).

To obtain a constant spectral gap, we add a known rank-one term to a matrix built from $tT_W$. The third block of the top eigenvector is then still close to a multiple of $t^2y$. An additional coordinate gives the top eigenvector an entry of constant size, so the corresponding row of the rank-one truncation, which is proportional to the top eigenvector, has norm at least one. Theorem~\ref{thm:pca} also shows that the truncated row has at least a constant fraction of the norm of the original row, and that the threshold is at a constant distance from every eigenvalue. These facts show that the second task is not degenerate: an error that is small relative to the original row is also small relative to the truncated row, and small perturbations of the matrix change the truncation only slightly. The spectrum is bounded in Lemma~\ref{lem:pca-spectrum}, and the top eigenvector is computed by block elimination in Lemma~\ref{lem:pca-vector}. Theorem~\ref{thm:pca} then follows from the transfer theorem applied with the cosine test.

Fix a rational number $t\in(0,1/64]$. Adjoin a coordinate indexed by $0$ to the three blocks of~\eqref{eq:pd-family}, and define the $(3n+1)\times(3n+1)$ matrices
\begin{equation}\label{eq:pca-family}
\widetilde T_W=0\oplus T_W,\qquad \varphi=\frac1{\sqrt2}\begin{pmatrix}1\\u\\0\\0\end{pmatrix},\qquad G_W=I+\varphi\varphi^{\mathsf T}+t\widetilde T_W.
\end{equation}
Write $\lambda_1\ge\lambda_2\ge\cdots$ for the eigenvalues of $G_W$, and let $w_W$ be a unit eigenvector for $\lambda_1$ with $\varphi^{\mathsf T}w_W>0$. For $W\in\mathcal G$, this vector is unique, because $\lambda_1$ is simple and $\varphi^{\mathsf T}w_W\ne0$ (Lemmas~\ref{lem:pca-spectrum} and~\ref{lem:pca-vector}).

\begin{lemma}\label{lem:pca-spectrum}
For every $W\in\mathcal G$, $2\le\lambda_1\le2+9t$, $\lambda_2\le1+9t$, and the smallest eigenvalue is at least $1-9t$. Hence $\lambda_1$ is simple, $\lambda_1-\lambda_2\ge1-9t\ge55/64$, and $\kappa(G_W)\le(2+9t)/(1-9t)\le137/55<3$. Moreover, $\|G_W\|_F^2=3n+4+4t^2n$, so $\rho(G_W)=\Theta(\sqrt n)$, and the family satisfies the hypotheses of Lemma~\ref{lem:access}.
\end{lemma}

\begin{proof}
The matrix $I+\varphi\varphi^{\mathsf T}$ has eigenvalues $2,1,\ldots,1$, and $\|t\widetilde T_W\|\le9t$ by Lemma~\ref{lem:pd-parameters}. The Rayleigh quotient of $G_W$ at $\varphi$ equals $2$, because $\varphi$ is supported on coordinate $0$ and the first block, where $\widetilde T_W$ has a zero diagonal block. Weyl's inequality bounds $\lambda_2$ from above by $1+9t$ and all eigenvalues from below by $1-9t$. The supports of $\varphi\varphi^{\mathsf T}$ and of the nonzero blocks of $\widetilde T_W$ are disjoint, and $\operatorname{tr}\widetilde T_W=\varphi^{\mathsf T}\widetilde T_W\varphi=0$, so $\|G_W\|_F^2=\|I+\varphi\varphi^{\mathsf T}\|_F^2+t^2\|\widetilde T_W\|_F^2=(3n+4)+4t^2n$. Every entry of $G_W$ either is fixed or equals $tW_{ab}/\sqrt n$.
\end{proof}

\begin{lemma}\label{lem:pca-vector}
Put $\zeta=\lambda_1-1$, $z=w_W/(\varphi^{\mathsf T}w_W)$, $D_z=\|z\|^2$, $\omega_t=t^2/(1-65t^2)$, $L_t=1/(1-65t^2)$, $s_0=t^2/(1-t^2)$, and $b_t=s_0+64\omega_t/(1-t^2)$. Write $z_0$ for the coordinate of $z$ indexed by $0$, and $z_1,z_2,z_3$ for its three blocks. For every $W\in\mathcal G$, $z_0=1/(\sqrt2\zeta)$ and
\begin{gather*}
1\le\zeta\le1+\frac{\omega_t\nu}2,\qquad 1\le D_z\le1+t^2\nu,\\
\Bigl\|z_3-\frac{t^2}{\sqrt2}y\Bigr\|\le\frac{t^2}{\sqrt2}\Bigl(\frac32\omega_t\nu+b_t\Bigr)\sqrt\nu,\qquad \|z_3\|\le\frac{t^2}{\sqrt2}L_t\sqrt\nu.
\end{gather*}
\end{lemma}

\begin{proof}
By Lemma~\ref{lem:pca-spectrum}, $\zeta\ge1$, and a vector orthogonal to $\varphi$ has Rayleigh quotient at most $1+9t<2$, so $\varphi^{\mathsf T}w_W\ne0$. The eigenvalue equation becomes $(\zeta I-t\widetilde T_W)z=\varphi$. Put $s=t^2/(\zeta^2-t^2)$, which lies in $(0,s_0]$ because $\zeta\ge1$, and put $F_s=(I-sR^{\mathsf T}R)^{-1}$ and $r=(I-sRR^{\mathsf T})^{-1}v$. By block elimination and the normalization $\varphi^{\mathsf T}z=1$,
\[
z_0=\frac1{\sqrt2\zeta},\quad z_1=\frac{F_su}{\sqrt2\zeta},\quad z_2=\frac{t\,r}{\sqrt2(\zeta^2-t^2)},\quad z_3=\frac{t^2Hr}{\sqrt2\zeta(\zeta^2-t^2)},\quad \zeta=\frac{1+u^{\mathsf T}F_su}2.
\]
By the last identity, $\zeta-1=\frac s2v^{\mathsf T}(I-sRR^{\mathsf T})^{-1}v$. Since $1-64s_0=(1-65t^2)/(1-t^2)$, this is at most $\frac{s_0}{2(1-64s_0)}\nu=\frac{\omega_t}2\nu$. The same factor shows that $\|r\|\le(1-t^2)L_t\sqrt\nu$ and $\|r-v\|=\|sRR^{\mathsf T}(I-sRR^{\mathsf T})^{-1}v\|\le64\omega_t\sqrt\nu$. Since $1\le\zeta\le1+\omega_t\nu/2$,
\[
\Bigl|\frac1{\zeta(\zeta^2-t^2)}-1\Bigr|\le\max\Bigl\{s_0,\ 3(\zeta-1)\Bigr\}\le\frac32\omega_t\nu+s_0,
\]
and $1/(\zeta(\zeta^2-t^2))\le1/(1-t^2)$. Writing $z_3-\frac{t^2}{\sqrt2}Hv=\frac{t^2}{\sqrt2}H\bigl(\frac{r-v}{\zeta(\zeta^2-t^2)}+(\frac1{\zeta(\zeta^2-t^2)}-1)v\bigr)$ proves the bound on $z_3-\frac{t^2}{\sqrt2}y$, and the bound on $\|z_3\|$ follows in the same way.

For the norm, let $\varphi'=(1,u)/\sqrt2$ be the restriction of $\varphi$ to coordinate $0$ and the first block. The vector $X=(1,F_su)/\sqrt2$ has inner product $\zeta$ with the unit vector $\varphi'$, so $(z_0,z_1)=X/\zeta$ has inner product one with $\varphi'$. Hence $\|(z_0,z_1)\|^2=1+\|(z_0,z_1)-\varphi'\|^2\ge1$, and $(z_0,z_1)-\varphi'$ is the projection of $X/\zeta$ onto the orthogonal complement of $\varphi'$. Since $X-\varphi'=(0,F_su-u)/\sqrt2$ and $F_su-u=sR^{\mathsf T}r$, we get $\|(z_0,z_1)-\varphi'\|\le\|F_su-u\|/\sqrt2\le8s_0\|r\|/\sqrt2\le4\sqrt2\,\omega_t\sqrt\nu$. Together with $\|z_2\|^2\le t^2L_t^2\nu/2$ and $\|z_3\|^2\le t^4L_t^2\nu/2$,
\[
1\le D_z\le1+\Bigl(32\omega_t^2+\frac{t^2(1+t^2)L_t^2}2\Bigr)\nu\le1+t^2\nu,
\]
where the last inequality holds because $32t^2L_t^2+(1+t^2)L_t^2/2\le1$ for $t\le1/64$.
\end{proof}

\begin{theorem}\label{thm:pca}
Fix a rational number $t\in(0,1/64]$. For every constant $\varepsilon\in(0,t^4/8)$, there are $c>0$ and $n_0$ such that, for every power of four $n\ge n_0$, each of the following tasks on $\{G_W:W\in\mathcal G\}$ requires more than $cn^2/\log n$ classical queries on some input, both in the worst case and in expectation, and has quantum query complexity $\widetilde\Theta(\sqrt n)$:
\begin{enumerate}
\item sampling from $p_{w_W}$ within total variation distance $\varepsilon$, where $w_W$ is the top principal component;
\item sampling within total variation distance $\varepsilon$ from the squared entries of row $0$ of the rank-one spectral truncation $(G_W)_{\ge3/2}=\lambda_1w_Ww_W^{\mathsf T}$.
\end{enumerate}
The requested row is fixed independently of $W$. Its norm after truncation is at least one, its norm in $G_W$ is $\sqrt{5/2}$, and the threshold $3/2$ has a constant distance from every eigenvalue.
\end{theorem}

\begin{proof}
Since $w_W$ is proportional to $z$, $p_{w_W}=p_z$. Apply Theorem~\ref{thm:transfer} and Corollary~\ref{cor:transfer-asymptotic} with the cosine test on the third block, $\gamma=0$, $\beta=t^2/\sqrt2$, and $d=1$. The vector $\bar x$ of Lemma~\ref{lem:two-conditions} is then $t^2y/\sqrt2$. On $\mathcal G$, Lemma~\ref{lem:pca-vector} bounds $\|z_3-\bar x\|$ and $\|z_3\|$, and $1\le D_z\le1+t^2\nu$. Lemma~\ref{lem:two-conditions}, with $F_\infty=2$, therefore bounds the difference in~\eqref{eq:transfer-hypothesis} by
\[
t^4(L_t+1)\Bigl(\frac32\omega_t\nu+b_t\Bigr)\nu+t^6\nu^2.
\]
By~\eqref{eq:nu-moments}, $\eta\le t^4\bigl((L_t+1)(\frac92\omega_t+b_t)+3t^2\bigr)<t^4/24$ for $t\le1/64$. By Lemma~\ref{lem:bias}, $\Lambda_{\cos}=\frac{t^4}2(a_M-c_M)$, which tends to $-\frac{t^4}2e^{-1/2}$ by Lemma~\ref{lem:moments}. Since $\frac12e^{-1/2}-\frac1{24}>\frac14$, every $\varepsilon<t^4/8$ satisfies $2\varepsilon<|\Lambda_\infty|-\eta$. Since $\Phi_{\max}=2t^4n$, the classical lower bound for the first task follows.

By Lemma~\ref{lem:pca-spectrum}, $\lambda_1$ is the only eigenvalue above $3/2$, so $(G_W)_{\ge3/2}=\lambda_1w_Ww_W^{\mathsf T}$, whose row $0$ is $\lambda_1(w_W)_0w_W^{\mathsf T}$. By Lemma~\ref{lem:pca-vector}, $(w_W)_0=z_0/\sqrt{D_z}=1/(\sqrt2\zeta\sqrt{D_z})>0$, so the squared entries of row $0$ have distribution $p_{w_W}$, and the second task is equivalent to the first. The norm of row $0$ after truncation is $\lambda_1/(\sqrt2\zeta\sqrt{D_z})\ge1$, because $\lambda_1\ge2$, $\zeta\le1+9t\le73/64$, and $D_z\le1+64t^2\le65/64$. Row $0$ of $G_W$ is $(3/2,u^{\mathsf T}/2,0,0)$, with norm $\sqrt{5/2}$, and the eigenvalues satisfy $\lambda_1\ge2$ and $\lambda_2\le73/64$.

For the quantum upper bound, apply Lemma~\ref{lem:qsvt} to the projector onto the top eigenvector with the vector $\varphi$, whose overlap is $\varphi^{\mathsf T}w_W=1/\sqrt{D_z}\ge\sqrt{64/65}$. For the quantum lower bound, use the matrices $W^{(a)}$ of Lemma~\ref{lem:search-instances}, for which $v=e_0+e_a$ and $RR^{\mathsf T}v=2v$. Then $r=v/(1-2s)$ and $z_2=tv/(\sqrt2(\zeta^2-3t^2))$, with $1\le\zeta\le1+\omega_t$ and $D_z\le1+2t^2$ by Lemma~\ref{lem:pca-vector}. The coordinate $a$ of the second block therefore has probability at least $z_{2,a}^2/D_z\ge t^2/(2(1+\omega_t)^4(1+2t^2))\ge t^2/4$, which exceeds every $\varepsilon<t^4/8$, and Lemma~\ref{lem:search} completes the proof.
\end{proof}

\section{Discussion}\label{sec:discussion}

This section places the results in context. In Section~\ref{sec:larger-separations}, we show that the worst-case separation as a function of $\rho$ is quartic, and we compare it with a family whose quantum query complexity is constant. The ranges of errors in our lower bounds are discussed in Section~\ref{sec:error-ranges}. Settings that are not covered by our framework are discussed in Section~\ref{sec:not-covered}, and open problems are listed in Section~\ref{sec:open}.

\subsection{Separations as a function of \texorpdfstring{$\rho$}{rho}}\label{sec:larger-separations}

We compare classical and quantum algorithms through their worst-case query complexities as functions of $\rho$. Fix a bound $\kappa_0$ on the condition number and an error $\varepsilon$ to which Theorem~\ref{thm:linear-systems} applies. Let $C(\rho)$ and $Q(\rho)$ be the classical and quantum query complexities of sampling from $p_{A^{-1}b}$ within total variation distance $\varepsilon$, in the input models of Section~\ref{sec:input-models}, for the family of all invertible square matrices $A$ with normalized Frobenius norm at most $\rho$ and condition number at most $\kappa_0$, maximized over the fixed nonzero right-hand side $b$. Every singular value of such a matrix is at least $\|A\|/\kappa_0$, so an $N\times N$ matrix in this family satisfies $N\le\kappa_0^2\rho^2$, and reading all entries shows that $C(\rho)\le\kappa_0^4\rho^4$. Theorem~\ref{thm:linear-systems} shows that $C(\rho)=\Omega(\rho^4/\log\rho)$ and $Q(\rho)=\Omega(\rho)$, and $Q(\rho)=\widetilde O(\rho)$ by the standard quantum linear-system algorithms \cite{ChakrabortyEtAl2019,GilyenEtAl2019}. Hence $C(\rho)$ equals $Q(\rho)^4$ up to factors polylogarithmic in $\rho$: in the worst case, the separation is quartic. The dequantized algorithms cited in Section~\ref{sec:motivation} achieve the same dependence $\widetilde O(\rho^4)$ without a bound on the dimension \cite{ShaoMontanaro2022,BakshiTang2024,ChenEtAl2025}.

For particular families, the quantum query complexity can be much smaller than $\rho$. Fourier sampling over all characters also has a linear-system version. If the dense block $R$ is replaced by a diagonal matrix of unknown signs, the solution contains all Fourier coefficients of the signs. Proposition~\ref{prop:diagonal} in Appendix~\ref{app:diagonal} shows that, for the resulting $3N\times3N$ matrices with condition number at most six, one quantum query to the Boolean function given by the diagonal signs suffices to sample exactly from the solution, while every classical algorithm with error $1/100$ requires $\Omega(N/\log N)$ queries. At sufficiently small constant errors, $\Omega(N)$ classical queries are necessary, by the lower bounds for Fourier sampling of Aaronson and Chen \cite{AaronsonChen2017} and of Montanaro \cite{Montanaro2019} (Remark~\ref{rem:diagonal-fourier-sampling}). For this family, the classical query complexity is not bounded by any power of the quantum query complexity. Reading all $N$ unknown entries suffices classically, however, so at these errors the classical query complexity of this family is $\Theta(N)=\Theta(\rho^2)$, far below the worst case $C(\rho)=\widetilde\Theta(\rho^4)$.

\subsection{Ranges of errors}\label{sec:error-ranges}

Our classical lower bounds hold for every error below about half of the bias of the test function under exact sampling. The bias is of order $\lambda^2$ for the linear systems of Section~\ref{sec:linear-systems} and of order $t^2$ for Hamiltonian evolution, where the sine test is used, and of order $t^4$ for the other tasks, where the cosine test is used. A condition number close to one requires small values of $\lambda$ and $t$, and the error estimates for the families of Section~\ref{sec:positive-definite} require $t\le1/64$. The explicit errors are therefore small; for example, the lower bound for positive-definite linear systems holds for every error below $t^4/4$, and Table~\ref{tab:functions} lists the ranges for matrix functions. We do not know whether these tasks become classically easier at larger constant errors.

\subsection{Settings that are not covered by our framework}\label{sec:not-covered}

In the input models of Section~\ref{sec:input-models}, the transfer theorem applies to sampling tasks whose target vector is close, on a set of coordinates, to an affine function of Fourier coefficients of the unknown signs, as in Lemma~\ref{lem:two-conditions}. The following settings are not covered.
\begin{enumerate}
\item Matrix-vector multiplication corresponds to $g(s)=s$, for which $a_2=0$. Indeed, $P_Wb=(u,-tv,0)$ does not contain the Walsh--Hadamard transform $y=Hv$ of the normalized row sums $v=Ru$ in~\eqref{eq:hidden-notation}. A fourth-power lower bound for this task would require a different construction.
\item Our lower bounds concern sampling from vectors, not estimating scalar quantities such as eigenvalues, traces, determinants, or quadratic forms. For example, on the family of Section~\ref{sec:pca} with $t=1/64$, Lemma~\ref{lem:pca-vector} and the bound $\nu=\|Ru\|^2\le64$, which follows from~\eqref{eq:hidden-notation} and the promise~\eqref{eq:promise}, imply $2\le\lambda_1(G_W)\le2+32/4031<2+1/100$. The output $2$ therefore estimates the top eigenvalue to error $1/100$ without any query.
\item Theorem~\ref{thm:pca} concerns the top eigenvector and an exactly specified rank-one truncation. It does not cover low-rank approximations with a relative error in the Frobenius norm. For our full-rank matrices, the discarded part has Frobenius norm $\Theta(\sqrt n)$, so an approximation that satisfies such a guarantee can have an inaccurate top eigenvector.
\item Our lower bound for square roots concerns sampling from $P_W^{1/2}b$ for a fixed vector $b$. It does not apply to sampling from a Gaussian distribution with covariance $P_W$, which applies $P_W^{1/2}$ to a random vector, or to algorithms that receive copies of the density matrix $P_W/\operatorname{tr}P_W$ instead of the input operations of Section~\ref{sec:input-models}.
\end{enumerate}

\subsection{Open problems}\label{sec:open}

We conclude with several open problems.
\begin{enumerate}
\item Can the logarithmic factor in the classical lower bounds be removed? This factor comes from the union bound over the output coordinates in Lemma~\ref{lem:transcript}, which is needed because the output may depend on the previous queries: an algorithm can output the coordinate $j$ for which the queried signs contribute most to $Z_j$ in~\eqref{eq:hidden-notation}. Removing the factor therefore requires a different argument. For Fourier sampling over all characters, the logarithmic factor is not necessary \cite{AaronsonChen2017,Montanaro2019}.
\item Do the classical lower bounds of Section~\ref{sec:positive-definite} hold for all errors below a constant multiple of $t^2$, rather than of $t^4$? For Hamiltonian evolution they do, because the third block of the output contains a component that does not depend on the input, so the bias of the sine test is of order $t^2$ (Section~\ref{sec:error-ranges}). A positive-definite linear system whose solution has the same property might also have such a lower bound.
\item Does Fourier sampling on a set $S$ of $K$ characters require $\Omega(\sqrt{M/K})$ quantum queries for every $S$? By Theorem~\ref{thm:fourier-quantum}, $O(\sqrt{M/K})$ quantum queries suffice at every constant error.
\item How does the classical query complexity depend jointly on $\rho$ and on the accuracy? Our results hold at fixed accuracy, whereas the number of queries is polynomial in the inverse accuracy for dequantized algorithms and polylogarithmic in it for quantum algorithms.
\item Which other tasks of quantum linear algebra, such as matrix-vector multiplication and low-rank approximation with a relative error in the Frobenius norm, require $\widetilde\Omega(\rho^4)$ classical queries? Our constructions do not apply to these two tasks (Section~\ref{sec:not-covered}).
\end{enumerate}

\section*{Acknowledgment}
We thank Rob Otter and Ruslan Shaydulin for executive support and valuable feedback on this project. We thank Anupam Prakash for reading a near-complete version of the draft and giving helpful comments. We also acknowledge the support of our colleagues at the Global Technology Applied Research Center of JPMorganChase.

\section*{Disclaimer}
This paper was prepared for informational purposes by the Global Technology Applied Research center of JPMorgan Chase \& Co. This paper is not a product of the Research Department of JPMorgan Chase \& Co. or its affiliates. Neither JPMorgan Chase \& Co. nor any of its affiliates makes any explicit or implied representation or warranty and none of them accept any liability in connection with this paper, including, without limitation, with respect to the completeness, accuracy, or reliability of the information contained herein and the potential legal, compliance, tax, or accounting effects thereof. This document is not intended as investment research or investment advice, or as a recommendation, offer, or solicitation for the purchase or sale of any security, financial instrument, financial product or service, or to be used in any way for evaluating the merits of participating in any transaction.

\appendix

\section{Explicit dimension thresholds}\label{app:regimes}

Theorem~\ref{thm:linear-systems} is asymptotic. In this appendix we list explicit dimensions from which its classical lower bound holds, for several choices of the parameters. The thresholds are sufficient conditions from the proofs, not necessary conditions for hardness, and the numbers of excluded queries at the threshold dimensions are small.

The proof of the classical lower bound can be made explicit as follows. Let $n_0$ be a power of four, put $\overline C_\lambda=1.4143(\lambda^2+\lambda^4)+2\lambda(1+\lambda^2)+2\lambda^2\ge C_\lambda$ and $\overline d_0=3+\alpha^2+2\lambda^2+\lambda^4+2\lambda/\sqrt{n_0}$, and define
\[
L_0=\frac{2\lambda^2\cdot0.6065\,(1-1/(6n_0^2))-\overline C_\lambda/\sqrt{n_0}}{\overline d_0}.
\]
For $n\ge n_0$, we have $b_M\ge0.6065(1-1/(6n_0^2))$ by Lemma~\ref{lem:moments}, $\E|D_W-d_n|\le\overline C_\lambda/\sqrt{n_0}$ by Lemma~\ref{lem:nonsymmetric-solution}, and $d_n\le\overline d_0$. Put $U_W=\sum_j(x_W)_{3,j}^2\sin Z_j$, with $Z_j$ as in~\eqref{eq:hidden-notation}, so that the expectation of the sine test under exact sampling from $p_{x_W}$ is $U_W/D_W$. By Lemma~\ref{lem:bias}, $\E U_W=d_n\Lambda_{\sin}=2\lambda^2b_M(1+\lambda/\sqrt n)$. Since $|U_W|\le D_W$, we obtain $\E[U_W/D_W]\ge(\E U_W-\E|D_W-d_n|)/d_n\ge L_0$ whenever the numerator of $L_0$ is positive. Consider the sampler that samples exactly from $p_{x_W}$ under the promise and returns a fixed outcome outside the third block for $W\notin\mathcal G$. The expectation of the sine test at its output is at least $L_0-\delta_n$, where $\delta_n$ is the bound of Lemma~\ref{lem:spectral-tail} on the probability that the promise fails. For a classical algorithm with at most $q_0(\xi,\delta)=\lfloor\xi^2n^2/(2\log(2n/\delta))\rfloor$ queries, the expectation of the sine test at its output has absolute value at most $\xi+\delta$, by Lemma~\ref{lem:transcript}. As in the proof of Theorem~\ref{thm:transfer}, sampling within error $\varepsilon$ with at most $q_0(\xi,\delta)$ queries is impossible whenever
\begin{equation}\label{eq:explicit-condition}
\varepsilon<\frac{L_0-\xi-\delta}2-\frac32\delta_n.
\end{equation}
Table~\ref{tab:regimes} lists choices that satisfy this condition for every power of four $n\ge n_0$. In every row, $\delta_n<10^{-8}$, and the error is also smaller than the probability of the marked coordinate in the proof of the quantum lower bound. The number of excluded queries grows like a constant times $n^2/\log n$, and the last column lists its value at $n=n_0$.

\begin{table}[htbp]
\centering
\begin{tabular}{@{}cccccccc@{}}
\toprule
$\lambda$ & $\alpha$ & $\kappa\le$ & $\varepsilon$ & $n_0$ & $(\xi,\delta)$ & Right side of~\eqref{eq:explicit-condition} & $q_0(\xi,\delta)$ at $n_0$ \\
\midrule
$1$ & $1/16$ & $272$ & $1/100$ & $2^8$ & $(0.08,\,10^{-3})$ & $0.01344$ & $15$ \\
$1$ & $1/16$ & $272$ & $5/100$ & $2^{10}$ & $(0.04,\,10^{-3})$ & $0.05673$ & $57$ \\
$1$ & $1/16$ & $272$ & $0.092$ & $2^{16}$ & $(0.01,\,10^{-3})$ & $0.09251$ & $11489$ \\
$1/4$ & $1/4$ & $14$ & $1/100$ & $2^{14}$ & $(10^{-3},\,10^{-4})$ & $0.01039$ & $6$ \\
$1/8$ & $1/2$ & $17/4$ & $1/1000$ & $2^{14}$ & $(10^{-3},\,10^{-4})$ & $0.00197$ & $6$ \\
$1/32$ & $3/4$ & $161/96$ & $10^{-4}$ & $2^{18}$ & $(8\cdot10^{-5},\,2\cdot10^{-6})$ & $0.000107$ & $8$ \\
\bottomrule
\end{tabular}
\caption{Explicit dimension thresholds for the classical lower bound of Theorem~\ref{thm:linear-systems}. Each row holds for every power of four $n\ge n_0$. The right side of~\eqref{eq:explicit-condition} is rounded down.}\label{tab:regimes}
\end{table}

\section{Limits of communication-based lower bounds}\label{app:communication}

In this appendix we show that communication lower bounds in which a constant number of players hold the entries of the matrix according to a known partition cannot prove fourth-power lower bounds for sampling from the solutions of linear systems. Previous classical lower bounds for sampling from the solutions of dense linear systems use communication complexity \cite{MontanaroShao2024Communication,MandeShao2025}. These reductions divide the input among two or more players. After each player has sent the norm and the size of its part once, a coordinator can simulate each query to sampling and query access with one request and one reply of logarithmic length \cite{MandeShao2025}. A communication lower bound therefore implies a query lower bound that is smaller by a logarithmic factor, when the number of players is small compared with the number of queries. Communication complexity has also been studied for exact solutions. When $k$ players each hold some of the equations of a linear system in $N$ variables with $B$-bit coefficients, and a coordinator must output a solution, Vempala, Wang, and Woodruff show that the randomized communication complexity is $\widetilde\Theta(N^2B+kN)$ \cite{VempalaEtAl2020}. Sampling at a fixed accuracy is a weaker task. For a well-conditioned $N\times N$ system with $O(\log N)$-bit entries and a right-hand side known to all players, sampling at a fixed accuracy requires only $\widetilde O(kN)$ bits when $k$ players hold the entries of the matrix according to a partition known to all players. For a constant number of players and full-rank matrices, this is $\widetilde O(\rho^2)$. Proposition~\ref{prop:row-compression} concerns our linear systems when each row of $W$ belongs to one player, and the general upper bound is proved in Proposition~\ref{prop:communication-upper}.

A family can require many queries and still have a cheap communication protocol. For our linear systems, a player can compute the row sums of the rows it holds without communication, whereas a query algorithm must read the corresponding entries.

\begin{proposition}\label{prop:row-compression}
If each row of $W$ belongs to a single player and the assignment of rows to players is known to all players, exact sampling from the solution of the system~\eqref{eq:nonsymmetric-family} has a communication protocol with $O(n\log n)$ bits, for any number of players.
\end{proposition}

\begin{proof}
Each player sends the row sums of its rows, which are integers in $[-n,n]$. After receiving all $n$ sums, the coordinator computes $v=Ru$ and $y=Hv$ (see~\eqref{eq:hidden-notation}), then $x_W$ from~\eqref{eq:nonsymmetric-solution}, and samples locally.
\end{proof}

If $k$ players instead hold arbitrary disjoint subsets of the entries, each player can send its contribution to every row sum, which costs $O(kn\log n)$ bits. For fixed $k$, this rules out a communication lower bound of order $n^2$ for this family under every partition of the entries that is known to all players.

A general communication upper bound applies to all well-conditioned square systems at fixed sampling accuracy. It follows from gradient descent on the normal equations.

\begin{proposition}\label{prop:communication-upper}
Fix $c,\varepsilon\in(0,1)$. Let $A\in\mathbb R^{N\times N}$ have singular values in $[c,1]$, and let $b$ be a unit vector known to all players. Assume that $2^BA_{ij}$ and $2^{B_b}b_j$ are integers for all coordinates. If $k$ players hold the entries of $A$ according to a partition known to all players, a coordinator protocol samples within total variation distance $\varepsilon$ of $p_{A^{-1}b}$ using
\[
O_{c,\varepsilon}\bigl(kN(B+B_b+\log N)\bigr)
\]
bits of communication.
\end{proposition}

\begin{proof}
Let $A^{(i)}$ equal $A$ on the entries of player $i$ and zero elsewhere. Starting from $x_0=0$, use the iteration $x_{t+1}=x_t+A^{\mathsf T}(b-Ax_t)$. In each iteration, the coordinator sends $x_t$ to the players, receives $A^{(i)}x_t$ from each player, sends the residual $r_t=b-\sum_iA^{(i)}x_t$, and receives $(A^{(i)})^{\mathsf T}r_t$. Each iteration communicates $O(kN)$ numbers. For $x_*=A^{-1}b$, the error satisfies $x_{t+1}-x_*=(I-A^{\mathsf T}A)(x_t-x_*)$, with contraction factor at most $1-c^2$. Thus $T=O_c(\log(1/\varepsilon))$ iterations ensure $\|x_T-x_*\|\le(\varepsilon/2)\|x_*\|$, and by Lemma~\ref{lem:output-error}, a sample from $p_{x_T}$ has the required accuracy. For fixed $c$ and $\varepsilon$, $T$ is constant. Exact dyadic arithmetic during these iterations uses $O_{c,\varepsilon}(B+B_b+\log N)$ bits per number, because each matrix-vector product increases the lengths of numerators and denominators by at most $O(B+\log N)$. The coordinator samples locally from the explicit vector $x_T$.
\end{proof}

For a constant number of players, fixed condition number and accuracy, and $O(\log N)$-bit entries, the protocol of Proposition~\ref{prop:communication-upper} uses $\widetilde O(N)=\widetilde O(\rho^2)$ bits of communication. The proposition also applies to the linear systems of Sections~\ref{sec:linear-systems} and~\ref{sec:pd-systems}. Their entries are rational numbers with a common denominator of $O(\log N)$ bits, and multiplying the matrix by this denominator divided by a suitable power of two makes the entries dyadic and the singular values lie in $[c,1]$ for a constant $c>0$, without changing $p_{A^{-1}b}$. A fourth-power lower bound through communication complexity would therefore require a different distribution of the input, such as a number of players that grows with $N$.

\section{A linear system whose solution contains all Fourier coefficients}\label{app:diagonal}

In this appendix we prove Proposition~\ref{prop:diagonal}, which is discussed in Section~\ref{sec:larger-separations}. If the dense block $R$ of Section~\ref{sec:linear-systems} is replaced by a diagonal matrix of unknown signs, the solution contains all Fourier coefficients of the signs, so the resulting family instantiates the framework with $K=M$. As in the proof of Theorem~\ref{thm:fourier-classical}, the classical lower bound follows from the transfer theorem with the cosine test. One controlled phase query suffices to prepare the normalized solution. In Remark~\ref{rem:diagonal-fourier-sampling}, we derive the stronger classical lower bound $\Omega(N)$, at sufficiently small constant errors, from the lower bounds for Fourier sampling of Aaronson and Chen \cite{AaronsonChen2017} and of Montanaro \cite{Montanaro2019}.

\begin{proposition}\label{prop:diagonal}
Let $N\ge4$ be a power of four and $k=\log_2N$, and let $f\in\{\pm1\}^N$ be indexed by $\{0,1\}^k$. Let $H_N$ be the normalized Walsh--Hadamard matrix of order $N$, and let $u_N$ be the uniform unit vector of dimension $N$. Define the $3N\times3N$ matrix $B_f$ and the vector $r$ by
\[
B_f=\begin{pmatrix}I&0&0\\-\diag(f)&I&0\\0&-H_N&I\end{pmatrix},\qquad r=\begin{pmatrix}u_N\\0\\0\end{pmatrix}.
\]
\begin{enumerate}
\item The solution is $x_f=B_f^{-1}r=(u_N,N^{-1/2}f,\widehat f)$, and $\|x_f\|^2=3$ for every $f$.
\item One has $1\le\|B_f\|\le2$ and $\|B_f^{-1}\|\le3$, so $\kappa(B_f)\le6$, and $5N/4\le\rho(B_f)^2\le5N$.
\item Every classical algorithm that samples from $p_{x_f}$ within total variation distance $1/100$ for every $f$ makes more than $\lfloor N/(32\log(512N))\rfloor$ queries on some input.
\item One controlled phase query to $f$ suffices to sample exactly from $p_{x_f}$. In the coherent input model, this query uses $O(1)$ entry queries.
\end{enumerate}
\end{proposition}

\begin{proof}
By forward substitution, the blocks of $x_f$ are $u_N$, $\diag(f)u_N=N^{-1/2}f$, and $N^{-1/2}H_Nf=\widehat f$, and $\|x_f\|^2=3$ by Parseval's identity. Write $B_f=I-\mathcal N$, where $\mathcal N$ has the blocks $\diag(f)$ and $H_N$ below the diagonal. Then $\|\mathcal N\|\le1$, $\mathcal N^2$ has the single block $H_N\diag(f)$ of norm one, and $\mathcal N^3=0$, so $\|B_f\|\le2$ and $\|B_f^{-1}\|=\|I+\mathcal N+\mathcal N^2\|\le3$. The matrix $B_f$ fixes every vector supported on the third block, so $\|B_f\|\ge1$, and $\|B_f\|_F^2=3N+N+N=5N$.

For the third claim, by Lemma~\ref{lem:access}, it suffices to consider algorithms that query $f$ directly. Apply Theorem~\ref{thm:transfer} with $M=N$, $S=\{0,1\}^k$, $K=N$, the cosine test on the third block, $\gamma=0$, $\beta=1$, and $d=3$. Since $\|x_f\|^2=3$ for every $f$, the sum in~\eqref{eq:transfer-hypothesis} equals $\Phi$, so $\eta=0$, and $|\Lambda_{\cos}|=(c_N-a_N)/3\ge5/36$ by Lemma~\ref{lem:moments}. Take $\xi=1/4$ and $\delta=1/256$. As in the proof of Theorem~\ref{thm:fourier-classical}, since $\log(512N)>4$, Lemma~\ref{lem:transcript} shows that every algorithm with at most $\lfloor N/(32\log(512N))\rfloor$ queries satisfies $|\E F_{\cos}(f,J_{\mathcal A})|\le11/256$. Hence error $\varepsilon$ is impossible with this many queries whenever $5/36>11/256+2\varepsilon$, which includes $\varepsilon=1/100$.

For the fourth claim, prepare a block register in $(|1\rangle+|2\rangle+|3\rangle)/\sqrt3$ and a data register in the uniform state $u_N$. Apply a phase query controlled on the block register being $2$ or $3$, and then apply $H_N$ to the data register controlled on the block register being $3$. The result is $x_f/\|x_f\|$, and measuring both registers samples exactly from $p_{x_f}$. The entry of $B_f$ in row $N+\sigma$ and column $\sigma$ is $-f(\sigma)$, so two reversible entry queries implement the phase query.
\end{proof}

\begin{remark}\label{rem:diagonal-fourier-sampling}
At sufficiently small constant errors, the classical lower bound of Proposition~\ref{prop:diagonal} improves to $\Omega(N)$ queries, by the lower bounds for Fourier sampling of Aaronson and Chen \cite{AaronsonChen2017} and of Montanaro \cite{Montanaro2019} (Section~\ref{sec:fourier}). By the first claim, the blocks of $x_f$ are $u_N$, $N^{-1/2}f$, and the vector $\widehat f$ of Fourier coefficients of $f$, and each of them has norm one, the last by Parseval's identity. A sample from $p_{x_f}$ therefore lies in the third block with probability $1/3$ and, conditioned on this event, equals coordinate $s$ of this block with probability $\widehat f(s)^2$. By Lemma~\ref{lem:access}, it suffices to consider algorithms that query $f$ directly. Suppose that such an algorithm samples from $p_{x_f}$ within total variation distance $\varepsilon\le1/12$ for every $f$. Run it $T$ times, and return the first outcome in the third block, or a fixed coordinate of this block if there is none. As in the argument for $\Samp_S$ with $S=\{0,1\}^m$ in Section~\ref{sec:fourier}, each run returns an outcome in the third block with probability at least $1/3-\varepsilon\ge1/4$, and, conditioned on this event, its output is within total variation distance $\varepsilon/(1/3-\varepsilon)\le4\varepsilon$ of the distribution that assigns probability $\widehat f(s)^2$ to coordinate $s$. The algorithm that returns the first such output therefore solves Fourier sampling for $f$ within total variation distance $4\varepsilon+(3/4)^T$ and makes $T$ times as many queries. With a sufficiently large constant $T$ and a sufficiently small constant $\varepsilon$, the lower bound $\Omega(N)$ for Fourier sampling implies the same lower bound for sampling from $p_{x_f}$.
\end{remark}

\section{Output access with constant success probability}\label{app:output-access}

Lemma~\ref{lem:solution-access} applies to procedures whose output fails to approximate the solution with probability smaller than the error $\varepsilon$. In this appendix we treat procedures whose output approximates the solution only with probability $2/3$. Such a procedure requires an additional step that amplifies the success probability, and for this step we assume exact access to the returned vector: for a nonzero vector $w$, $\SQ(w)$ consists of entry queries, exact norm queries, and independent samples from $p_w$. In Lemma~\ref{lem:confidence}, an accurate candidate is selected from a constant number of runs by estimating the distances between candidates. Corollary~\ref{cor:solution-access-amplified} then follows from this lemma and Theorem~\ref{thm:linear-systems}. In the lemma, every run returns correct access operations to some vector, whether or not that vector approximates the solution; only the approximation guarantee is probabilistic.

\begin{lemma}\label{lem:confidence}
Let $x\in\mathbb C^N$, $\kappa_0\ge1$, $1\le\|x\|\le\kappa_0$, and $0<\eta<1/2$. Suppose independent runs of a procedure return $\SQ$ access to nonzero vectors $w_i\in\mathbb C^N$ with $\Prb(\|w_i-x\|\le\eta\|x\|)\ge2/3$, and all returned access operations are correct. For every $\delta\in(0,1/2)$, $O(\log(1/\delta))$ runs and a number of access operations polynomial in $\eta^{-1}$ and $\log(1/\delta)$ suffice to select a candidate $w$ with
\[
\Prb(\|w-x\|\le4\eta\kappa_0)\ge1-\delta.
\]
The number of access operations is independent of the dimension $N$.
\end{lemma}

\begin{proof}
Take an odd number $r=O(\log(1/\delta))$ of independent candidates. By a Chernoff bound, more than half of them are good approximations, except with probability at most $\delta/2$. Reject candidates with norm greater than $2\kappa_0$; a good candidate is not rejected. Put $\Delta=\eta\kappa_0$. Two good candidates have distance at most $2\Delta$.

For each pair of retained candidates $w,w'$, estimate the squared distance $\|w-w'\|^2$ to additive error at most $\Delta^2$. To do so, sample $j$ from $p_w$ and put $X=\overline{w_j}w'_j/p_w(j)$. Then $\E X=w^\dagger w'$ and $\E|X|^2\le\|w\|^2\|w'\|^2\le16\kappa_0^4$. By Chebyshev's inequality and a Chernoff bound, the median of $O(\log(r/\delta))$ independent averages, each of $O(\eta^{-4})$ independent copies of $\operatorname{Re}X$, estimates $\operatorname{Re}(w^\dagger w')$ to additive error $\Delta^2/2$, except with probability at most $\delta/r^2$. With exact norm queries and the identity $\|w-w'\|^2=\|w\|^2+\|w'\|^2-2\operatorname{Re}(w^\dagger w')$, this estimate determines the squared distance to error $\Delta^2$. By a union bound over the pairs, all pair estimates are correct, except with probability at most $\delta/2$.

Select any retained candidate whose estimated squared distance is at most $5\Delta^2$ from more than $r/2$ retained candidates, counting itself. Such a candidate exists when more than half of the candidates are good. Since more than half of all candidates are good, one of these retained candidates is good, and its true distance from the selected candidate is at most $\sqrt6\,\Delta$, so the selected candidate has distance at most $(1+\sqrt6)\Delta<4\Delta$ from $x$. If no candidate is selected, the output is arbitrary. The two exceptional events have total probability at most $\delta$.
\end{proof}

\begin{corollary}\label{cor:solution-access-amplified}
Consider a family in Theorem~\ref{thm:linear-systems}, with condition-number bound $\kappa_0$ and a constant error $\varepsilon>0$ in its range, and put $x=\widehat A_W^{-1}\widehat b$ and $\eta_*=\varepsilon/(64\kappa_0)$. Suppose a classical procedure uses the input operations of Definition~\ref{def:sq} and returns exact $\SQ(w)$ access to a nonzero vector $w$ on every run, with $\Prb(\|w-x\|\le\eta_*\|x\|)\ge2/3$. If the construction uses at most $T$ input queries and each access operation uses at most $S'$ input queries, then $T+S'=\Omega(n^2/\log n)=\Omega(\rho^4/\log\rho)$.
\end{corollary}

\begin{proof}
The solution satisfies $1\le\|x\|\le\kappa_0$, because $\|\widehat A_W\|=\|\widehat b\|=1$ and $\|\widehat A_W^{-1}\|\le\kappa_0$. Apply Lemma~\ref{lem:confidence} with $\eta=\eta_*$ and $\delta=\varepsilon/4$. Except with probability $\delta$, the selected vector $w$ satisfies $\|w-x\|\le4\eta_*\kappa_0=\varepsilon/16\le(\varepsilon/16)\|x\|$. One exact sample from $p_w$ then has total variation error at most $\varepsilon/8+\varepsilon/4<\varepsilon$ by Lemma~\ref{lem:output-error}. All parameters of the amplification are independent of $n$, so the resulting sampler uses $O(T+S')$ input queries, and the claim follows from Theorem~\ref{thm:linear-systems}.
\end{proof}

The corollary counts both the construction and the use of the output. It does not bound the construction cost alone, because an access operation may perform input queries. The same argument applies to the families of Section~\ref{sec:positive-definite}, because their target vectors have norms bounded above and below by constants.

\section{A memory separation in a streaming model}\label{app:streaming}

In this appendix we show that Theorem~\ref{thm:linear-systems} implies a separation in memory between quantum and classical algorithms that sample from the solutions of the linear systems of Section~\ref{sec:nonsymmetric-family}, of dimension $N=4n$, in a streaming model of Zhao et al.\ \cite{ZhaoEtAl2026}. In this model, each unknown sign $W_{ab}$ is encoded by two random strings of $O(\log N)$ bits, called shares and numbered $0$ and $1$. Each sample contains a uniformly random position $(a,b)$ and one of the two shares at this position, together with its number. The number of this share is the same for all samples in a block of $n^2$ consecutive samples and is drawn uniformly at random for each block, so the two shares of a sign never appear in the same block. For every fixed error below $\varepsilon_{\mathrm{lin}}(\lambda,\alpha)$ in~\eqref{eq:linear-error-range}, a quantum streaming algorithm uses $O(\log N)$ qubits of memory and $O(N^3\log^2N)$ samples, whereas every classical streaming algorithm with $O(N^3\log^2N)$ samples requires $\Omega(N/\log^2N)$ bits of memory (Corollary~\ref{cor:streaming}). For the quantum part, each of the $O(\sqrt N)$ queries of the quantum algorithm of Theorem~\ref{thm:linear-systems} is implemented with $O(\log N)$ queries to the bits of the shares, and each of these is implemented from $O(N^{5/2}\log N)$ samples by the quantum oracle sketching of Zhao et al.\ (Lemma~\ref{lem:streaming-compiler}). For the classical part, a simulation of Zhao et al.\ (Lemma~\ref{lem:streaming-simulation}) turns a classical streaming algorithm with $B$ bits of memory and $N_{\mathrm{samp}}$ samples, extended by $O(\log N)$ bits of memory and $O(N^2)$ samples, into a classical algorithm that samples from nearly the same distribution with $O((N_{\mathrm{samp}}+N^2)(B+\log N)/(N^2\log N))$ queries to the signs, while Theorem~\ref{thm:linear-systems} requires $\Omega(N^2/\log N)$ queries. Proposition~\ref{prop:streaming-limits} shows that $O(\log N)$ bits of classical memory suffice under sampling and query access, with independent samples of the signs or of the nonzero entries, or in the streaming model with $O(N^5)$ samples. Citations of Zhao et al.\ use the numbering of version~2 of their paper, so numbers such as Theorem~E.24 in these citations refer to their paper and not to this appendix.

\paragraph{The streaming model.} Let $n$ be a power of four (Section~\ref{sec:hidden-signs}), and let $W\in\mathcal G$, the set of sign matrices that satisfy the promise~\eqref{eq:promise}. As in Section~\ref{sec:test-functions}, we identify the $M=n^2$ positions $(a,b)$ of the signs with $\{0,1\}^m$, where $m=2k$ and $k=\log_2n$, and put $f(a,b)=W_{ab}$. The target distribution is $p_{x_W}=p_{\widehat A_W^{-1}\widehat b}$, with $x_W$ as in~\eqref{eq:nonsymmetric-solution}. Each sign $f(\sigma)$ is encoded by two shares $Y^{(0)}_\sigma,Y^{(1)}_\sigma\in\{0,1\}^w$, where $w=\lceil c_*\log_2M\rceil$, and $c_*=3460\log_23460$ is the constant in the hypothesis of Lemma~\ref{lem:streaming-simulation} below; the value of $c_*$ is explained after that lemma. For each position $\sigma$, independently, the pair $(Y^{(0)}_\sigma,Y^{(1)}_\sigma)$ is uniformly random among the pairs of strings in $\{0,1\}^w$ whose inner product modulo two equals $(1-f(\sigma))/2$, so that $f(\sigma)=(-1)^{Y^{(0)}_\sigma\cdot Y^{(1)}_\sigma}$. The shares are drawn once and do not change. The samples form consecutive blocks of $M$ samples, and each block has a context $0$ or $1$ that is uniformly random and independent of the shares and of the other blocks. In a block with context $0$, each sample is $(X,Y^{(0)}_X,0)$, and in a block with context $1$, it is $(X,Y^{(1)}_X,1)$, where the position $X$ is a uniformly random element of $\{0,1\}^m$, independent of the shares, the contexts, and the other positions. A sample therefore never contains an entry of $\widehat A_W$, and the two shares that determine a sign never appear in the same block. This is the noisy data generation process of Zhao et al.\ \cite[Appendix~E.1]{ZhaoEtAl2026} with blocks of length $M$ and with the inner product modulo two as the encoding function, the function that maps the two shares of a position to the encoded bit $(1-f(\sigma))/2$. Zhao et al.\ also call these groups of samples blocks, and they call the contexts situations.

A streaming algorithm receives the samples one at a time. Its memory is the information that it keeps between two samples, and its computation between two samples is unrestricted. After each sample, a classical algorithm with $B$ bits of memory replaces the content of its memory by a function of this content, the sample, and the index of the sample. Its random bits are not counted as memory, because a randomized algorithm is a probability distribution over deterministic ones. A quantum algorithm, whose memory is counted in qubits, applies to its qubits a channel that depends on the sample and its index; this channel may include measurements whose outcomes are kept in the memory. The number of samples is fixed in advance, and after the last sample the algorithm outputs a random coordinate $J\in[4n]$, which is a function of its memory or, for a quantum algorithm, the outcome of a measurement of its memory. These are the learning algorithms of Zhao et al.\ \cite[Appendices~C.2 and~C.3]{ZhaoEtAl2026}. A streaming algorithm samples from the solution with error $\varepsilon$ if, for every $W\in\mathcal G$, the distribution of $J$ over the shares, the contexts, the positions, and the randomness and measurements of the algorithm is within total variation distance $\varepsilon$ of $p_{x_W}$.

The classical lower bound rests on the low discrepancy of the inner product. For independent and uniformly random $U,U'\in\{0,1\}^w$, the discrepancy of the inner product \cite[Definition~E.8]{ZhaoEtAl2026} is the largest value of $|\E[(-1)^{U\cdot U'}\mathbf 1[U\in\mathcal R]\,\mathbf 1[U'\in\mathcal R']]|$ over all sets $\mathcal R,\mathcal R'\subseteq\{0,1\}^w$. When it is small, information about $U$ and information about $U'$, obtained separately, reveal little about $U\cdot U'$. The discrepancy of the inner product is at most $2^{-w/2}$ \cite[Lemma~E.41]{ZhaoEtAl2026}. Indeed, the expectation above is $2^{-2w}$ times the bilinear form of the matrix with entries $(-1)^{\sigma\cdot\sigma'}$, for $\sigma,\sigma'\in\{0,1\}^w$, at the indicator vectors of $\mathcal R$ and $\mathcal R'$. This matrix is $2^{w/2}$ times the normalized Walsh--Hadamard matrix of order $2^w$, so its norm is $2^{w/2}$, and the indicator vectors have norms at most $2^{w/2}$.

\paragraph{Two results of Zhao et al.} The first result, quantum oracle sketching, implements a phase query to a Boolean function from samples of its values. We state it for a single query, the case used in the proof of Corollary~\ref{cor:streaming}. For two channels $\mathcal E$ and $\mathcal E'$, the diamond norm $\|\mathcal E-\mathcal E'\|_\diamond$ does not increase under composition with other channels, and it bounds the total variation distance between the outcome distributions of a measurement applied after $\mathcal E$ and after $\mathcal E'$.

\begin{lemma}[Quantum oracle sketching {\cite[Theorem~D.13]{ZhaoEtAl2026}}]\label{lem:streaming-compiler}
Let $g$ be a Boolean function on $\{0,1\}^m$, let $\mathcal U$ be the unitary channel of the phase query $|\sigma\rangle\mapsto g(\sigma)|\sigma\rangle$ or of a controlled phase query to $g$ (Section~\ref{sec:known-magnitudes}), and let $\eta>0$. There is a quantum algorithm on the qubits of $\mathcal U$ that processes $\lceil2\pi^2M/\eta\rceil$ independent samples $(X,(1-g(X))/2)$, with uniformly random positions $X$, one at a time, and implements a random channel $\mathcal V$ whose average over the samples satisfies $\|\E[\mathcal V]-\mathcal U\|_\diamond\le\eta$.
\end{lemma}

In Theorem~D.13 of Zhao et al., $Q$ queries to the unitary $\sum_\sigma e^{itp(\sigma)(1-g(\sigma))/2}|\sigma\rangle\langle\sigma|$, to its inverse, and to their controlled versions, where $p$ is a probability distribution on $\{0,1\}^m$ and $t>0$, are implemented from $Q\lceil2t^2Q\max_\sigma p(\sigma)/\eta\rceil$ independent samples $(X,(1-g(X))/2)$ whose positions $X$ have the distribution $p$, and the channel of the algorithm, averaged over the samples, changes by at most $\eta$ in diamond norm. Lemma~\ref{lem:streaming-compiler} is its case $Q=1$, with $p$ uniform and $t=\pi M$, for which this unitary is the phase query to $g$ and equals its inverse.

The second result concerns classical streaming algorithms with one output bit. The transcript of such an algorithm is the sequence of its memory states immediately before the samples whose context differs from that of the previous sample, followed by its output \cite[Appendix~C.2]{ZhaoEtAl2026}. It contains all the information that the algorithm keeps from one context to the next. This transcript, defined by Zhao et al., is different from the transcript of queries in the proof of Lemma~\ref{lem:transcript}.

\begin{lemma}[Simulation of the transcript {\cite[Theorem~E.24]{ZhaoEtAl2026}}]\label{lem:streaming-simulation}
There are absolute constants $C_{\mathrm{sim}}$ and $M_*$ such that the following holds for $M\ge M_*$ in the streaming model of Appendix~\ref{app:streaming}, whose shares have length $w\ge c_*\log_2M$. For every randomized classical streaming algorithm $\mathcal L$ with $B$ bits of memory, $N_{\mathrm{samp}}\le2M^2w$ samples, and one output bit, there is a randomized classical algorithm that queries values of $f$, makes at most $C_{\mathrm{sim}}(N_{\mathrm{samp}}B/(Mw)+1)$ queries on every input and every run, and, for every sign matrix $W$, outputs a bit string whose distribution is within total variation distance $2^{-w/16}$ of that of the transcript of $\mathcal L$.
\end{lemma}

Theorem~E.24 of Zhao et al.\ applies to every encoding function whose discrepancy is at most $2^{-\theta w}$ for a constant $\theta\in(0,2]$, with shares of length $w\ge(865/\theta^2)\log_2(865/\theta^2)\log_2M$, and its bound on the total variation distance is $2^{-\theta w/8}$. For the inner product, this condition holds with $\theta=1/2$ by the bound on its discrepancy above, so that $(865/\theta^2)\log_2(865/\theta^2)=c_*$ and $2^{-\theta w/8}=2^{-w/16}$, and Lemma~\ref{lem:streaming-simulation} is this case of Theorem~E.24, with blocks of length $M$ and with the oracle $(1-f)/2\in\{0,1\}^M$, as Zhao et al.\ call the string of encoded bits. Their simulation stops before its number of queries would exceed a bound of the order of $N_{\mathrm{samp}}B/(Mw)+1$ \cite[Algorithm~3]{ZhaoEtAl2026}, so the bound in Lemma~\ref{lem:streaming-simulation} holds on every input and every run.

\begin{corollary}\label{cor:streaming}
Fix rational numbers $\lambda\in(0,1]$ and $\alpha\in(0,1/(1+8\lambda)]$, an error $\varepsilon\in(0,\varepsilon_{\mathrm{lin}}(\lambda,\alpha))$, with $\varepsilon_{\mathrm{lin}}$ as in~\eqref{eq:linear-error-range}, and a constant $C_1>0$. In the streaming model of Appendix~\ref{app:streaming}, for powers of four $n$ and $N=4n$, the following holds.
\begin{enumerate}
\item There is a quantum streaming algorithm that samples from the solution $\widehat A_W^{-1}\widehat b$ with error $\varepsilon$, using $O(\log N)$ qubits of memory and $O(N^3\log^2N)$ samples.
\item Every classical streaming algorithm that samples from the solution with error $\varepsilon$, using at most $C_1N^3\log^2N$ samples, requires $\Omega(N/\log^2N)$ bits of memory.
\end{enumerate}
The constants in the asymptotic notation depend only on $\lambda$, $\alpha$, $\varepsilon$, and $C_1$.
\end{corollary}

\begin{proof}[Proof of the quantum part]
Let $\mathcal B$ be the quantum algorithm in the proof of the quantum upper bound of Theorem~\ref{thm:linear-systems} for the error $\varepsilon/2$. The streaming algorithm implements queries one at a time with Lemma~\ref{lem:streaming-compiler}, so we first transform $\mathcal B$ into an algorithm whose sequence of queries is fixed in advance and whose queries are to functions with values in the samples. The algorithm $\mathcal B$ applies Lemma~\ref{lem:amplification} to a circuit $V$ with one phase query to $f$ controlled by a flag qubit. A controlled phase query is its own inverse, so $V^\dagger$ contains the same query as $V$, and every query of $\mathcal B$ is a phase query to $f$ controlled by this qubit. The number of queries of $\mathcal B$ is random, because each trial in the proof of Lemma~\ref{lem:amplification} applies a random power of the operator $G$ defined there, and $\mathcal B$ stops after the first successful trial. We give each trial as many query steps as the largest number of queries of a trial, so that, by the proof of Lemma~\ref{lem:amplification}, there are $Q_{\max}=O(\sqrt n)$ query steps in total. At each step, the algorithm $\mathcal B'$ computes into an ancilla qubit the conjunction of the flag qubit and a classical bit that equals one if and only if $\mathcal B$ queries at this step, applies a phase query to $f$ controlled by the ancilla, and uncomputes the ancilla. This bit, like the choice of the other operations of $\mathcal B$, is determined by the random choices of $\mathcal B$ so far, including the power of $G$ chosen in the current trial, and by the measurement outcomes of its earlier trials, which $\mathcal B$ keeps in classical registers. A query whose control is zero acts as the identity, so $\mathcal B'$ implements the channel of $\mathcal B$ with exactly $Q_{\max}$ controlled phase queries to $f$, separated by channels that do not depend on $W$.

Next, condition on the shares, and let $f^{(0)}_i(\sigma)=(-1)^{(Y^{(0)}_\sigma)_i}$ and $f^{(1)}_i(\sigma)=(-1)^{(Y^{(1)}_\sigma)_i}$ for $i\in\{1,\ldots,w\}$. Since $f(\sigma)=\prod_i(-1)^{(Y^{(0)}_\sigma)_i(Y^{(1)}_\sigma)_i}$, a controlled phase query to $f$ is implemented by writing $Y^{(0)}_\sigma$ and $Y^{(1)}_\sigma$ into two registers of $w$ qubits, initially zero, with one call to each oracle $O_{f^{(0)}_i}$ and $O_{f^{(1)}_i}$ of Section~\ref{sec:known-magnitudes}, applying for each $i$ a $Z$ gate to the control qubit controlled by qubit $i$ of both registers, and resetting the registers with $2w$ further calls. A call to an oracle $O_g$ is a phase query to $g$ controlled by the target qubit of $O_g$ and conjugated by Hadamard gates on this qubit. The resulting algorithm $\mathcal B''$ implements the channel of $\mathcal B$ with $Q'=4wQ_{\max}$ controlled phase queries to the functions $f^{(0)}_i$ and $f^{(1)}_i$, in an order fixed in advance.

It remains to implement the queries of $\mathcal B''$ from the samples. Put $\eta=\varepsilon/(4Q')$ and $r=\lceil2\pi^2/\eta\rceil$, and divide the samples into $Q'$ consecutive segments of $8r$ blocks. The streaming algorithm replaces the $j$th query $\mathcal U_j$ of $\mathcal B''$, to some $f^{(0)}_i$ or $f^{(1)}_i$, by the random channel $\mathcal V_j$ of Lemma~\ref{lem:streaming-compiler}, implemented from the first $\lceil2\pi^2M/\eta\rceil$ samples of segment $j$ with context $0$ or $1$, respectively; in these samples, bit $i$ of the share is $(1-f^{(0)}_i(X))/2$ or $(1-f^{(1)}_i(X))/2$. It ignores the other samples and applies the other operations of $\mathcal B''$ between segments, so the set of samples that it uses depends only on the contexts and not on the memory. Condition also on the contexts, and suppose that each segment has at least $r$ blocks with the required context, hence at least $rM\ge\lceil2\pi^2M/\eta\rceil$ samples with this context. The positions of these samples are independent and uniformly random, so the channels $\mathcal V_j$ are independent and satisfy $\|\E[\mathcal V_j]-\mathcal U_j\|_\diamond\le\eta$. The channel of the streaming algorithm, averaged over the positions, is therefore that of $\mathcal B''$ with each $\mathcal U_j$ replaced by $\E[\mathcal V_j]$, and by a telescoping sum it is within diamond norm $Q'\eta=\varepsilon/4$ of the channel of $\mathcal B''$. By Hoeffding's inequality, a segment has fewer than $r$ of its $8r$ blocks with the required context with probability at most $e^{-2(3r)^2/(8r)}\le e^{-r}\le\varepsilon/(4Q')$, since $r\ge8\pi^2Q'/\varepsilon\ge\log(4Q'/\varepsilon)$, so some segment has too few such blocks with probability at most $\varepsilon/4$. Since $\mathcal B$ has error $\varepsilon/2$, the output distribution is therefore within total variation distance $\varepsilon/2+\varepsilon/4+\varepsilon/4=\varepsilon$ of $p_{x_W}$ for every value of the shares, and hence also on average over the shares.

The algorithm uses $8rMQ'=O(n^2w^2Q_{\max}^2)=O(n^3\log^2n)$ samples. Its memory consists of the registers of $\mathcal B$ (two registers of $k$ qubits, a register that selects a block of $x_W$, the flag qubit, $O(\log n)$ ancilla qubits, and classical registers of $O(\log n)$ bits for the trials), one ancilla qubit, $2w$ qubits for the shares, and a counter of $O(\log n)$ bits for the number of samples of the current segment that it has used. The model counts classical registers as memory, so the memory is $O(\log N)$ qubits.
\end{proof}

\begin{proof}[Proof of the classical part]
Let $\mathcal L$ be a classical streaming algorithm that samples from the solution with error $\varepsilon$, using $B$ bits of memory and $N_{\mathrm{samp}}\le C_1N^3\log^2N$ samples. Lemma~\ref{lem:streaming-simulation} concerns algorithms with one output bit, and the algorithm that it provides reproduces their transcripts, not their outputs, so we first extend $\mathcal L$ to an algorithm whose transcript determines the output of $\mathcal L$ with high probability. Put $\varepsilon'=(\varepsilon+\varepsilon_{\mathrm{lin}}(\lambda,\alpha))/2$ and $N_{\mathrm{ext}}=M\lceil\log_2(4/(\varepsilon'-\varepsilon))\rceil$. The algorithm $\mathcal L'$ runs $\mathcal L$, stores its output $J$ and a flag bit equal to one, processes $N_{\mathrm{ext}}$ further samples without changing its memory, and outputs zero. It has $B'=B+\lceil\log_2(4n)\rceil+1$ bits of memory, and its flag is zero until it stores $J$. At least $N_{\mathrm{ext}}/M-1$ of the further samples are the first sample of a block that has a previous block, and the context of each of these blocks differs from that of the previous block with probability $1/2$, independently. Hence, with probability at least $1-2^{1-N_{\mathrm{ext}}/M}\ge1-(\varepsilon'-\varepsilon)/2$, the last memory state in the transcript of $\mathcal L'$ was recorded after $\mathcal L'$ stored $J$. On this event, the function that maps a transcript to the value of $J$ in its last memory state if the flag of this state is one, and to a fixed coordinate otherwise, returns $J$.

We now apply Lemma~\ref{lem:streaming-simulation} to $\mathcal L'$. Let $n$ be so large that $M\ge M_*$, $2^{-w/16}\le(\varepsilon'-\varepsilon)/2$, and $N_{\mathrm{samp}}+N_{\mathrm{ext}}\le2M^2w$; the last condition holds for large $n$ because $N_{\mathrm{samp}}=O(n^3\log^2n)$ and $w=\Theta(\log n)$. Applying the function that recovers $J$ from a transcript to the output of the algorithm of Lemma~\ref{lem:streaming-simulation} for $\mathcal L'$, we obtain a classical algorithm that makes at most $C_{\mathrm{sim}}((N_{\mathrm{samp}}+N_{\mathrm{ext}})B'/(Mw)+1)$ queries to $f$ on every input and every run. Since a function does not increase the total variation distance, this algorithm samples from $p_{x_W}$ with error at most $\varepsilon+(\varepsilon'-\varepsilon)/2+2^{-w/16}\le\varepsilon'$ for every $W\in\mathcal G$. Its queries to $f$ are entry queries of $\widehat A_W$ (Lemma~\ref{lem:nonsymmetric-parameters}), so by item~1 of Theorem~\ref{thm:linear-systems} for the error $\varepsilon'$, there are $c>0$ and $n_0$ such that, for $n\ge n_0$, this bound on the number of queries exceeds $cn^2/\log n$. If also $cn^2\ge2C_{\mathrm{sim}}\log n$, then, since $Mw\ge2c_*n^2\log_2n$, it follows that $(N_{\mathrm{samp}}+N_{\mathrm{ext}})B'\ge cc_*n^4/(C_{\mathrm{sim}}\log2)$. Since $N_{\mathrm{samp}}+N_{\mathrm{ext}}=O(n^3\log^2n)$ and $B'=B+O(\log n)$, we obtain $B=\Omega(n/\log^2n)=\Omega(N/\log^2N)$. The constants are chosen in this order: $\varepsilon'$ and $N_{\mathrm{ext}}/M$ from $\lambda$, $\alpha$, and $\varepsilon$; then $c$ and $n_0$ from Theorem~\ref{thm:linear-systems}; then, for the given $C_1$, all $n$ large enough for the conditions above.
\end{proof}

\paragraph{Limits of the separation.} The classical lower bound of Corollary~\ref{cor:streaming} depends on three features of the streaming model, as the following proposition shows: the input is not available after it has been read, the two shares of a sign arrive in blocks with different contexts, and the number of samples is $O(N^3\log^2N)$.

\begin{proposition}\label{prop:streaming-limits}
Fix rational numbers $\lambda\in(0,1]$ and $\alpha\in(0,1/(1+8\lambda)]$, let $n\ge4$ be a power of four, and put $N=4n$. In each of the following settings, there is a classical algorithm with $O(\log N)$ bits of memory, not counting random bits, that samples from $p_{x_W}$ as stated for every $W\in\mathcal G$. The constants depend only on $\lambda$ and $\alpha$, and in item~2 also on $\varepsilon$.
\begin{enumerate}
\item With the sampling and query access of Definition~\ref{def:sq} to $\widehat A_W$, where the memory is the information kept between queries: exactly, with $n^3+n^2$ queries.
\item With independent samples $(X,f(X))$ with uniformly random positions $X$, or independent samples of uniformly random nonzero entries of $\widehat A_W$ with their positions: with any fixed error $\varepsilon\in(0,1)$, from $O(N^3/\varepsilon^2)$ samples.
\item In the streaming model of Appendix~\ref{app:streaming}: with error at most $e^{-(n^3+n^2)/2}$, from $16(n^3+n^2)M=O(N^5)$ samples.
\end{enumerate}
\end{proposition}

\begin{proof}
Each algorithm computes the coordinates of $x_W$, or estimates of them, one at a time, and samples an index by weighted reservoir sampling: it keeps the sum of their squares so far and one selected index, which it replaces by the index $j$ of a new coordinate with probability equal to the square of this coordinate divided by the new sum. By induction, the final selected index has the length-squared distribution of the whole vector. Since $\lambda$ and $\alpha$ are rational and $\sqrt n$ is an integer, the squares of the coordinates of $x_W$ in~\eqref{eq:nonsymmetric-solution}, and of the estimates below, have numerators of $O(\log n)$ bits and denominators that divide an integer of $O(\log n)$ bits that does not depend on $W$, so weighted reservoir sampling keeps $O(\log n)$ bits.

For item~1, the first and fourth blocks of $x_W$ do not depend on $W$, and by~\eqref{eq:hidden-notation}, coordinate $a$ of the second block and coordinate $j$ of the third block are determined by $\sum_bW_{ab}$ and $\sum_{a,b}h_{ja}W_{ab}$. The algorithm accumulates these sums term by term and reads $W_{ab}$ with one entry query in row $n+a$ and column $b$ (Lemma~\ref{lem:nonsymmetric-parameters}), so it makes $n\cdot n+n\cdot n^2$ queries.

For item~2, write $X=(X_1,X_2)$ with $X_1,X_2\in\{0,1\}^k$. The parts $\lambda v_a$ and $\lambda^2Z_j/\sqrt n$ of the coordinates of the second and third blocks that depend on $W$, with $v$ and $Z_j$ as in~\eqref{eq:hidden-notation}, are the expectations of $\lambda n\,\mathbf 1[X_1=a]f(X)$ and $\lambda^2\sqrt n\,h_{jX_1}f(X)$, whose second moments are $\lambda^2n$ and $\lambda^4n$. The algorithm estimates each of these parts by an average over $\lceil4(\lambda^2+\lambda^4)n^2/\varepsilon^2\rceil$ new samples. Let $\widetilde x$ be the vector whose second and third blocks consist of the estimated coordinates and whose other blocks equal those of $x_W$; then $\E\|\widetilde x-x_W\|^2\le\varepsilon^2/4$, and $\widetilde x\ne0$ because its first block is $u$. If $\|\widetilde x-x_W\|<\|x_W\|$, then $\TV(p_{\widetilde x},p_{x_W})\le2\|\widetilde x-x_W\|$ by Lemma~\ref{lem:output-error}, since $\|x_W\|\ge1$; otherwise, this bound holds because $\|\widetilde x-x_W\|\ge1$ and a total variation distance is at most one. The output distribution $\E[p_{\widetilde x}]$ is therefore within total variation distance $2(\E\|\widetilde x-x_W\|^2)^{1/2}\le\varepsilon$ of $p_{x_W}$. For samples of nonzero entries, the same algorithm works with three times as many samples, because the entries $-\alpha\lambda W_{ab}/\sqrt n$ in row $n+a$ and column $b$ form a fraction $n/(2n+4)\ge1/3$ of the $2n^2+4n$ nonzero entries, and the two functions above, divided by this fraction on these entries and set to zero on the others, keep their expectations and at most triple their second moments.

For item~3, the algorithm runs the algorithm of item~1 and obtains each sign $f(\sigma)$ in two steps, each starting with a new block of samples: it waits for a sample with position $\sigma$ in a block with context $0$ and stores $Y^{(0)}_\sigma$, and then for a sample with position $\sigma$ in a block with context $1$, and computes $f(\sigma)=(-1)^{Y^{(0)}_\sigma\cdot Y^{(1)}_\sigma}$, with $O(w+\log n)$ additional bits of memory. Whatever the step in progress, a block has the required context and contains a sample with the required position with probability $\frac12(1-(1-1/M)^M)\ge1/4$, independently of the earlier blocks. By Hoeffding's inequality, all $2(n^3+n^2)$ steps therefore end within $16(n^3+n^2)$ blocks, except with probability at most $e^{-(n^3+n^2)/2}$, in which case the algorithm outputs a fixed coordinate.
\end{proof}

Item~1 concerns the sampling and query access of Section~\ref{sec:input-models}, under which the input remains available and its data structures are not counted as memory. By item~1 and the count of registers in the proof of Corollary~\ref{cor:streaming}, $O(\log N)$ bits suffice for a classical algorithm and $O(\log N)$ qubits suffice for the quantum algorithm of Theorem~\ref{thm:linear-systems}, so in that model the separation is in the number of queries and not in memory. The algorithm of item~2 also works when each sample contains both shares of its position, since they determine $f(X)$, so the classical lower bound depends on the separation of the two shares in time. By item~3, $O(\log N)$ bits suffice for a classical algorithm with $O(N^5)$ samples and a number of operations polynomial in $N$. Corollary~\ref{cor:streaming} is therefore a separation in memory for classical algorithms with at most a constant multiple of $N^3\log^2N$ samples, and not a superpolynomial separation in the number of samples or in time.

\paragraph{Comparison with Zhao et al.} The classical part of Corollary~\ref{cor:streaming} is the analogue, for a task whose output is a sample, of the sample-space lower bound of Zhao et al.\ \cite[Theorem~E.2]{ZhaoEtAl2026}. That theorem concerns noisy oracle property estimation, in which a classical streaming algorithm receives samples as in our model and must output the value of a given function of the encoded bits. Under the hypotheses of Lemma~\ref{lem:streaming-simulation}, it states that every such algorithm with $N_{\mathrm{samp}}$ samples and $B$ bits of memory that errs with probability at most $\delta-2^{-w/16}$, for every string of encoded bits on which the function is defined, satisfies $N_{\mathrm{samp}}B=\Omega(Q_\delta Mw)$, where $Q_\delta$ is the $\delta$-error randomized query complexity of the function, provided that $Q_\delta$ exceeds a sufficiently large constant. That theorem and the classical part of Corollary~\ref{cor:streaming} are both proved with the simulation of Lemma~\ref{lem:streaming-simulation}, and the bound $(N_{\mathrm{samp}}+N_{\mathrm{ext}})B'=\Omega(n^4)$ in the proof of the corollary has the same form, with the query complexity $\Omega(n^2/\log n)$ of Theorem~\ref{thm:linear-systems} in place of $Q_\delta$. The memory lower bounds of Zhao et al.\ for linear systems and binary classification \cite[Theorems~F.7 and~F.17]{ZhaoEtAl2026} are proved instead from a hardness result for a task in which the encoded bits change over time while the bit to be computed stays fixed \cite[Theorem~E.26]{ZhaoEtAl2026}. The proof of this result uses a learning XOR lemma, which bounds the probability of predicting the parity of the values of the function at several strings of encoded bits \cite[Lemma~E.29]{ZhaoEtAl2026}. The streaming model, quantum oracle sketching, and the simulation of the transcript are due to Zhao et al.

\printbibliography
\end{document}